\documentclass[letterpaper,english,reprint,aps,notitlepage,superscriptaddress,twocolumn]{revtex4-2}

\usepackage[T1]{fontenc}
\usepackage[utf8]{inputenc}
\usepackage{amsmath}
\usepackage{amssymb}
\usepackage[dvipsnames]{xcolor}

\colorlet{mylinkcolor}{RoyalPurple}
\colorlet{mycitecolor}{RoyalPurple}
\colorlet{myurlcolor}{RoyalPurple}

\usepackage{hyperref}
\hypersetup{
	linkcolor  = mylinkcolor,
	citecolor  = mycitecolor,
	urlcolor   = myurlcolor,
	colorlinks = true,
	breaklinks = true
}

\makeatletter

\pdfpageheight\paperheight
\pdfpagewidth\paperwidth

\makeatother

\usepackage{graphicx}
\usepackage{babel}
\usepackage{physics}

\begin{document}
\title{Continuous wideband microwave-to-optical converter based on room-temperature Rydberg atoms}

\author{Sebastian Borówka}
\affiliation{Centre for Quantum Optical Technologies, Centre of New Technologies, University of Warsaw, Banacha 2c, 02-097 Warsaw, Poland}
\affiliation{Faculty of Physics, University of Warsaw, Pasteura 5, 02-093 Warsaw, Poland}
\author{Uliana Pylypenko}
\affiliation{Centre for Quantum Optical Technologies, Centre of New Technologies, University of Warsaw, Banacha 2c, 02-097 Warsaw, Poland}
\affiliation{Faculty of Physics, University of Warsaw, Pasteura 5, 02-093 Warsaw, Poland}
\author{Mateusz Mazelanik}
\affiliation{Centre for Quantum Optical Technologies, Centre of New Technologies, University of Warsaw, Banacha 2c, 02-097 Warsaw, Poland}
\affiliation{Faculty of Physics, University of Warsaw, Pasteura 5, 02-093 Warsaw, Poland}
\author{Michał Parniak}
\email{m.parniak@cent.uw.edu.pl}
\affiliation{Centre for Quantum Optical Technologies, Centre of New Technologies, University of Warsaw, Banacha 2c, 02-097 Warsaw, Poland}
\affiliation{Niels Bohr Institute, University of Copenhagen, Blegdamsvej 17, 2100 Copenhagen, Denmark}

\begin{abstract}
The coupling of microwave and optical systems presents an immense challenge due to the natural incompatibility of energies, but potential applications range from optical interconnects for quantum computers to next-generation quantum microwave sensors, detectors or coherent imagers. Several engineered platforms have emerged that are constrained by specific conditions, such as cryogenic environments, impulse protocols, or narrowband fields. Here we employ Rydberg atoms that allow for the natural wideband coupling of optical and microwave photons even at room temperature and with the use of a modest setup. We present continuous-wave conversion of a $13.9\ \mathrm{GHz}$ field to a near-infrared optical signal using an ensemble of Rydberg atoms via a free-space six-wave mixing process, designed to minimize noise interference from any nearby frequencies. The Rydberg photonic converter exhibits an unprecedented conversion dynamic range of $57\ \mathrm{dB}$ and a wide conversion bandwidth of $16\ \mathrm{MHz}$. Using photon counting, we demonstrate the readout of photons of free-space $300\ \mathrm{K}$ thermal background radiation at $1.59\ \mathrm{nV}\mathrm{cm}^{-1}(\mathrm{rad}/\mathrm{s})^{-1/2}$ ($3.98 \ \mathrm{nV}\mathrm{cm}^{-1}\mathrm{Hz}^{-1/2}$) with the sensitivity down to $3.8\ \mathrm{K}$ of noise-equivalent temperature, allowing us to observe Hanbury Brown and Twiss interference of microwave photons. 
\end{abstract}
\maketitle

\begin{figure*}
\includegraphics[width=\textwidth]{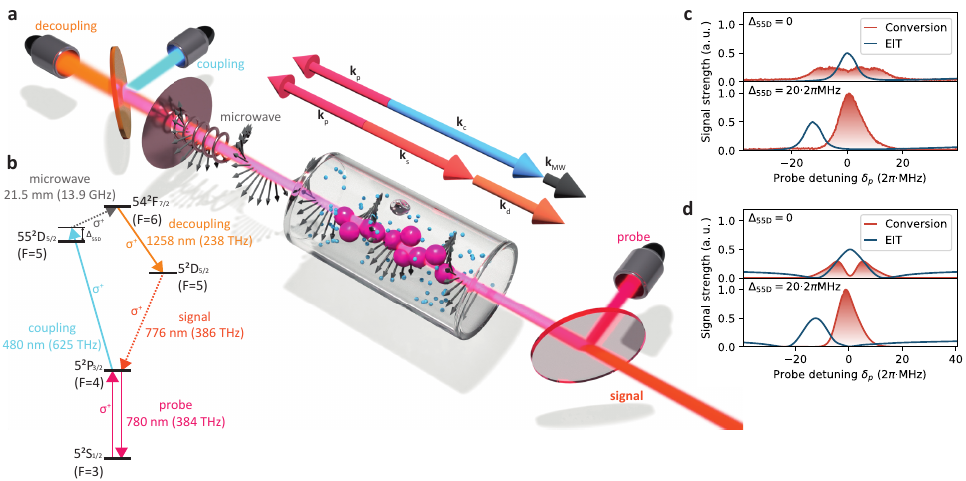}
\caption{\textbf{Room-temperature microwave-to-optical converter.} \textbf{a},~Concept illustration of a warm vapor Rydberg converter: a circularly polarized MW field enters a rubidium vapor cell, Rydberg-state atoms, excited in an interaction region defined by three laser beams, contribute to the conversion process and generate a \emph{signal} beam. The arrows represent the wavevectors of the interacting fields obeying phase-matching principles. Constant supply of ground-state atoms is assured by the Maxwellian distribution of velocities, resulting in a continuous process without the need for atomic trapping or repumping. \textbf{b},~${}^{85} \mathrm{Rb}$ energy level structure employed in the conversion process. Three strong fields ($780\ \mathrm{nm}$: \emph{probe}, $480\ \mathrm{nm}$: \emph{coupling} and $1258\ \mathrm{nm}$: \emph{decoupling}) are applied to the atomic medium in the near-resonant scheme. Introducing MW field ($13.9\ \mathrm{GHz}$) results in a converted emission (\emph{signal}) at the $776\ \mathrm{nm}$ transition. Optimal transduction is assured with sign-matched $\sigma$ transitions (between the states of maximal $F$ and $m_F$ quantum numbers). \textbf{c},~Comparison of measured EIT and conversion in the domain of probe field detuning for resonant and $55^2 \mathrm{D}_{5/2}$ level detuned case. \textbf{d},~The same EIT and conversion relation predicted in the numerical simulation.}
\label{qura}
\end{figure*}

Coherent conversion between energetically separated domains of microwave (MW) and optical radiation demonstrates a demanding issue at the frontier of photonics and quantum science. With the recent progress in the field of quantum computing \cite{Arute_2019}, the most disruptive invention would be a realization of optically connected qubits, giving rise to the perspective of hybrid quantum networks \cite{Muralidharan_2016} and the quantum internet \cite{Awschalom_2021}. Nevertheless, several other rapidly developing fields would greatly benefit from MW-photonic conversion with near-term and still noisy performance, including fundamental \cite{Riechers_2022,Pankratov_2022,Komatsu_2022} and observational radio-astronomy \cite{Greaves_2018,Gorski_2021,Kou_2022}, THz imaging \cite{Wade_2016} and next-generation MW sensing \cite{Xia_2020}.

To this day various approaches to the MW conversion task have been presented, utilizing piezo-optomechanics \cite{Bochmann_2013,Forsch_2019,Jiang_2020,Mirhosseini_2020,Honl_2022,Stockill_2022}, electro-optomechanics \cite{Andrews_2014,Peairs_2020,Arnold_2020,Delaney_2022}, magneto-optics \cite{Hisatomi_2016,Bartholomew_2020,Zhu_2020}, electro-optics \cite{Rueda_2016,Witmer_2020,McKenna_2020,Xu_2021,Sahu_2022,Wang_2022}, vacancy centers \cite{Lekavicius_2017} and Rydberg excitons \cite{Gallagher_2022}. Notably, several works presented realize MW conversion in Rydberg alkali atoms \cite{Han_2018,Vogt_2019,Tu_2022,Kumar_2022}. This flexible medium provides a multitude of possible conversion modes with resonant transition frequency scaling as $1/n^3$ for consecutive principal quantum numbers $n$, and transition dipole moment scaling as $n^2$, thus allowing for excellent sensitivity and in turn efficiency of conversion. Owing to their MW transitions, Rydberg atoms also carry a history of extremely impactful experiments in MW cavity QED (quantum electrodynamics) with atomic beams \cite{Goy_1980,Raimond_1982,Meschede_1985}.
Up to date all of the works realizing MW-to-optical conversion require carefully engineered systems, either in the form of developed transducer structure, cryogenic environment or laser-cooled atoms. Optical interconnection of remote qubits will certainly necessitate a system of uncompromising performance, but many applications could benefit from near-term and noisy converters. In this case, it is of particular benefit to employ a simple system, preferably operating at room temperature.
Recent works show that many of the quantum devices can be based on hot atomic vapors, including quantum memories \cite{Finkelstein_2018,Kaczmarek_2018}, single-photon sources \cite{Ripka_2018, Dideriksen_2021} and photonic isolators \cite{Dong_2021}. With astronomical MW measurements in mind, it is also worth remembering that atomic vapors already play an invaluable role in satellite navigation as a part of atomic clocks \cite{Steigenberger_2015}.

Here we employ a hot-atom system for the task of MW-to-optical upconversion. We demonstrate as a proof-of-concept that the Doppler-broadened rubidium energy level structure is well-suited to produce atomic coherence resulting in a microwave-to-optical converted field. Furthermore, we take advantage of the straightforward nature of the hot-atom approach to present a continuous-wave (CW) realization of free-space MW field conversion, in opposition to the previous approaches \cite{Han_2018,Vogt_2019,Tu_2022} in cold atomic media, which had to work in impulse regime due to magneto-optical trap (MOT) operational sequence. Despite the simplicity, we achieve an unprecedented MW conversion dynamic range, descending down to thermal limit, where we are able to measure the autocorrelation function of free-space MW thermal photons to support this claim. The conversion bandwidth is on par with the best results in employing other wideband conversion media \cite{Jiang_2020,McKenna_2020,Zhu_2020,Sahu_2022,Stockill_2022}. We present the means to tune the conversion band beyond this, and we discuss the methods to extend the dynamic range, surpassing the thermal limit.

Our idea is based on extending the robust two-photon Rydberg excitation scheme in rubidium, utilized in the measurement of Autler-Townes (A-T) splitting of electromagnetically induced transparency (EIT) in MW field electrometry \cite{Sedlacek_2012, Jing_2020} and atomic receivers \cite{Deb_2018, Meyer_2018, Bor_wka_2022}. We introduce an additional near-infrared (at the higher limit of telecom O-band) laser field, transferring atomic population from Rydberg state and enabling the emission at the $5^2\mathrm{D}_{5/2} \rightarrow 5^2\mathrm{P}_{3/2}$ ($776\ \mathrm{nm}$) transition. This scheme allows for robust conversion -- free from both noise interference and from strong classical fields spectrally close to signal field. Remarkably, the realization is all-optical and no ancillary MW field is necessary, further simplifying the setup, as well as broadening its potential applications. We also avoid using ultraviolet (UV) fields (employed in the recent remarkable work of \emph{Kumar et al.} \cite{Kumar_2022}), which are particularly problematic both in the generation of narrowband CW laser beams and in their destructive impact on optical elements and thus the lifespan of the device.

\section*{Results}

\begin{figure*}
\includegraphics[width=\textwidth]{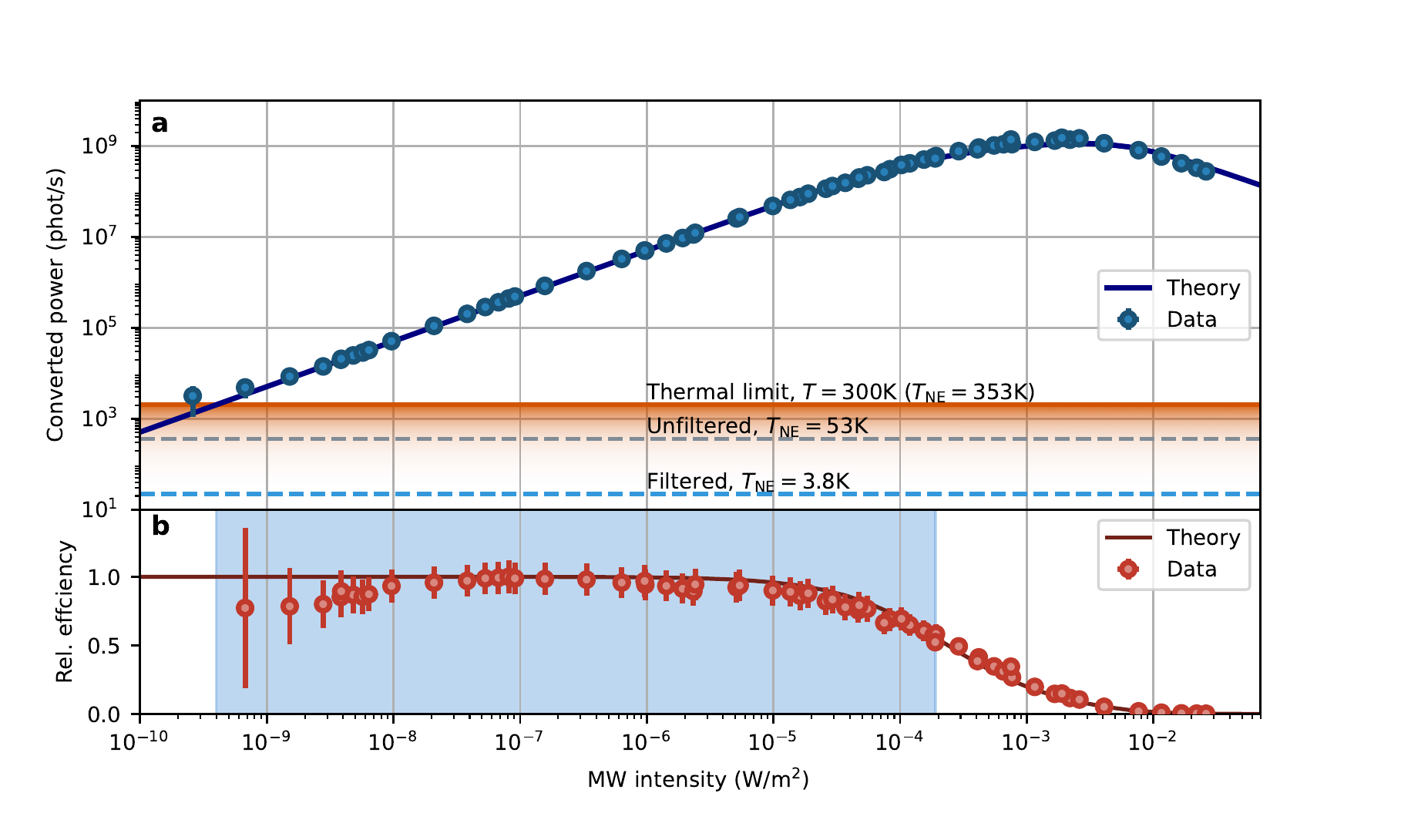}
\caption{\textbf{Microwave-to-optical conversion with 57 dB dynamic range.} \textbf{a},~Photon counting yields conversion response to applied MW field intensity, from thermal noise level ($2.1 {\cdot} 10^3\ \mathrm{phot} / \mathrm{s}$, including thermal radiation and added non-thermal noise) to conversion saturation limit ($1.5 {\cdot} 10^9\ \mathrm{phot} / \mathrm{s}$). The observed non-thermal added noise is at the level of noise-equivalent temperature $T_{\mathrm{NE}} {=} 53\ \mathrm{K}$ and can be further filtered down to $T_{\mathrm{NE}} {=} 3.8\ \mathrm{K}$. \textbf{b},~Relative conversion efficiency (thermal noise subtracted) with ${>}1$ SNR and ${>}0.5$ relative efficiency achieved over $57\ \mathrm{dB}$ dynamic range of MW intensity -- from $4.0 {\cdot} 10^{-10}$ to $1.9 {\cdot} 10^{-4}\ \mathrm{W} / \mathrm{m}^2$. The efficiency is normalized to the highest measured value. In both cases we compare the results with a theoretical curve, where the overall efficiency is the only free parameter. Notably, the saturation intensity is well reproduced along with the shape of the curve. The error bars presented on \textbf{a} and \textbf{b} are calculated from photon-counting standard deviation, $\sqrt{\bar{n}}$ (with thermal radiation treated as added noise), where in each case the $\bar{n}$ is explicitly noted as photon rates on the Y axis, and from the propagation of the calibration's standard deviation.}
\label{sateff}
\end{figure*}

\paragraph{Conversion in warm Rydberg vapors} 
Room-temperature atoms in vapor cells are easy to harness experimentally, yet they provide limited options for full quantum control due to Doppler broadening and collisions, which may particularly influence Rydberg atoms. We therefore start by showing that the room-temperature atomic vapors are adequate to facilitate the Rydberg-assisted conversion process. In our demonstration, a cylindrical vapor cell serves as a hot-atom Rydberg converter (Fig.~\ref{qura}\textbf{a}). Three focused optical fields determine an interaction region, where ground-state atoms are supplied from the remainder of the cell via thermal atomic motion, assuring that neither depletion nor other long-term time-dependence occur, thus satisfying CW operational framework. These fields combined with a MW field realize coherent (for the measurements of coherence see Supplement S.4) six-wave mixing (Fig.~\ref{qura}\textbf{c}) with emission at $776\ \mathrm{nm}$. We utilize sign-matched $\sigma$ transitions with the largest dipole moments (maximal $F$ and $m_F$ quantum numbers) driven by circularly-polarized fields. The conversion scheme constitutes a closed cycle without the need for repumping. The conservation of energy and momentum, governing the wave mixing process, assure that temporal and spatial properties of converted photons are preserved. In contrast to previously presented conversion schemes \cite{Han_2018,Vogt_2019,Tu_2022}, this scheme does not require additional MW field, and all the introduced optical fields are spectrally different well enough to be separated with the use of free-space optics, even in a collinear configuration.

Considering field detunings from the energy level structure, we observe that similarly as in \cite{Tu_2022} the best conversion efficiency is achieved for off-resonant realization, as shown in the Fig.~\ref{qura}\textbf{d} in relation to the EIT effect. Here the detuning applies to Doppler-broadened level structure, involving a range of atomic velocity classes. We particularly find that the detuning from level $55^2 \mathrm{D}_{5/2}$ plays an important role in increasing the efficiency of conversion, yielding a nearly fivefold improvement in efficiency at $\Delta_{55\mathrm{D}} {=} 20 {\cdot} 2 \pi\ \mathrm{MHz}$ in comparison to the resonant case.

\begin{figure*}
\includegraphics[width=\textwidth]{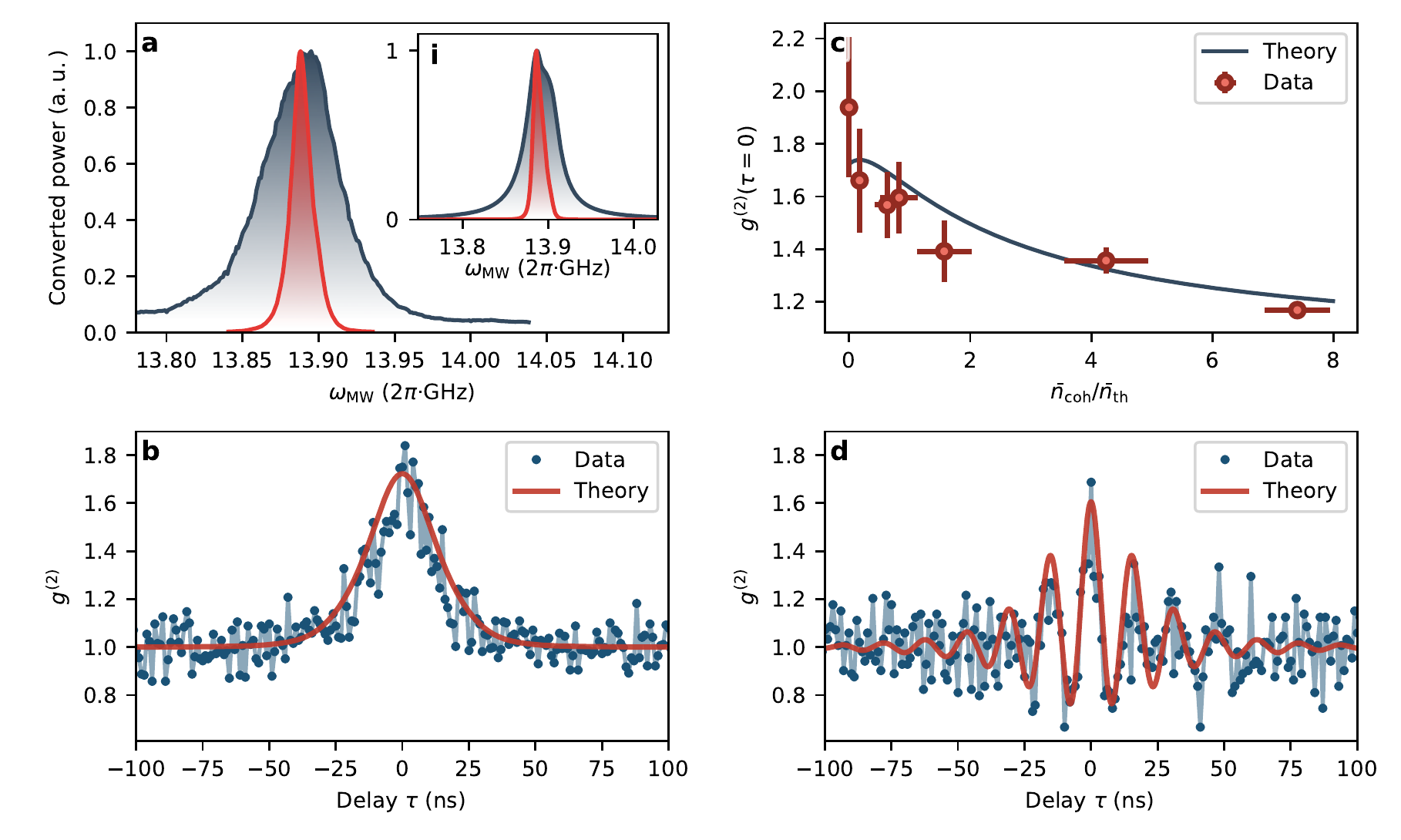}
\caption{\textbf{Wideband conversion of microwave thermal radiation at the single-photon level.} \textbf{a},~Conversion's dependence on MW field detuning (red curve), revealing a bandwidth of $\Gamma_{\mathrm{con}} {=} 16 {\cdot} 2 \pi\ \mathrm{MHz}$ FWHM. Interdependence on the detunings of MW and decoupling fields shows that with commensurate detuning of decoupling field the tunable bandwidth of conversion (dark curve) can be broadened to $59 {\cdot} 2 \pi\ \mathrm{MHz}$ FWHM. This is in agreement with the general theoretical prediction,~\textbf{i}~(numerical simulation). \textbf{b},~Photon autocorrelation (homodyne) measurement yields second order autocorrelation function of thermal MW photons, confirming the conversion of quantum thermal state. This experimental result agrees perfectly with parameter-free theory prediction based on the conversion band and measured noise properties. \textbf{c},~The decreasing of $g^{(2)} (0)$ with the introduction of coherent state photons rate $\bar{n}_{\mathrm{coh}}$, in agreement with the theory. The error bars presented are calculated from photon-counting standard deviation, $\sqrt{\bar{n}}$, where $\bar{n}$ denote photon rates, and from the propagation of the calibration's standard deviation in relation to the thermal level. \textbf{d},~Introducing far detuned ($\Delta \omega {=} 4 \Gamma_{\mathrm{con}} {=} 64 {\cdot} 2 \pi\ \mathrm{MHz}$) coherent MW field with rate $\bar{n}_{\mathrm{coh}} {\approx} \bar{n}_{\mathrm{th}}$ induces interference in the autocorrelation function showing beat modulation frequency equal to the detuning.}
\label{results}
\end{figure*}

\paragraph{Continuous-wave conversion}
Atomic conversion processes arise from atomic coherences, which in the quantum mechanical approach can be derived from state density matrices $\hat{\rho}$. With the incident weak MW field $E_\mathrm{MW}$ and strong drive fields we obtain an optical coherence $\rho_s$ and thus expect emission of the signal field as $E_s\sim\rho_s=\mathrm{Tr}(|5^2D_{5/2}\rangle\langle5^2P_{3/2}|\hat{\rho})\sim E_\mathrm{MW}$. The time evolution of the atomic state is governed by Gorini--Kossakowski--Sudarshan--Lindblad (GKSL) equation,
\begin{equation}\label{GKSL}
    \partial_t \hat{\rho} = \frac{1}{i \hbar} [\hat{H}, \hat{\rho}] + \mathcal{L}[\hat{\rho}],
\end{equation}
where by $\hat{H}$ we denote Hamiltonian and by $\mathcal{L}[{\cdot}]$ a superoperator responsible for spontaneous emission and other sources of decoherence. In our approach for conversion, we consider the density matrix in steady state, $\partial_t \hat{\rho} (t) = 0$, similarly to EIT-based Rydberg electrometry systems. We employ the steady-state solution of the GKSL equation as the basis for the parametric numerical simulation utilized as an aid to interpret the experimental results we present. We consider a range of atomic velocity classes contributing to the conversion process and take into consideration the shape of the interaction volume. We find that the conversion is facilitated by the steady-state coherence $\rho_s$. Remarkably, the steady-state coherence arises partially due to decoherence and influx of ground-state atoms into the interaction region.

Our approach can be compared with the pulsed approach to conversion which requires tailoring of the Hamiltonian $\hat{H} (t)$ (e.g.~by sequentially turning on the driving fields and enabling the conversion for short time only) to maximize the coherence and in consequence the conversion output. This approach, showcased particularly in \cite{Tu_2022}, can be implemented in Doppler-free cold-atom system and leads to high, local conversion efficiency, albeit it is practically limited to operating for as short as $1/1000$ of $10\ \mathrm{ms}$ time sequence. The potential applications are thus narrowed down to the instances where the converter can be triggered to a short signal or where we aim to convert a strong classical field. Our antipodal CW approach addresses this gap, enabling uniform, trigger-free conversion of weak or asynchronous signals.

\paragraph{Dynamic range and efficiency}
With the use of a single-photon counter, we measure the converter's response to MW field intensity. As shown in the Fig.~\ref{sateff}\textbf{a}, the converted signal ranges from $2.1 {\cdot} 10^3$ to $1.5 {\cdot} 10^9\ \mathrm{phot} / \mathrm{s}$. We identify the upper bound as saturation from energy levels shift due to A-T splitting, and the lower limit as free-space thermal MW photons coupling to the converter. We adjust for other sources of noise, directional and polarization coupling to the converter, and the measured conversion band. The electric field spectral density level of observed thermal radiation, $1.59\ \mathrm{nV}\mathrm{cm}^{-1}(\mathrm{rad}/\mathrm{s})^{-1/2}$ (isotropic, both polarizations), agrees very well with the theoretical prediction of $1.64\ \mathrm{nV}\mathrm{cm}^{-1}(\mathrm{rad}/\mathrm{s})^{-1/2}$ (see Methods for derivation). We note, however, that such a remarkable agreement occurs for the employed model that is simpler than the reality of the experiment, i.e.~the presence of MW antenna near the converter is not accounted for. Nevertheless, we are able to confirm the matching of two fundamental references for MW field intensity, namely the thermal bath and the Autler-Townes splitting.

We consider the efficiency of the implemented conversion: Fig.~\ref{sateff}\textbf{b} shows relative conversion efficiency ${>}0.5$ and signal-to-noise ratio (SNR) ${>}1$ (in relation to thermal noise) for $57\ \mathrm{dB}$ of MW intensities (from $4.0 {\cdot} 10^{-10}$ to $1.9 {\cdot} 10^{-4}\ \mathrm{W} / \mathrm{m}^2$), which we identify as the converter's dynamic range. The data is presented with normalization to the highest measured value. The converter's response to MW field and conversion efficiency are well-predicted by the numerical simulation, where only the overall efficiency is parametrically matched. 

As far as the absolute efficiency is concerned, several teams adopted an approach where the reference MW photon rate is taken from the intensity multiplied by the area of the interaction medium  \cite{Han_2018,Vogt_2019,Tu_2022}. Care must be taken when interpreting this area-normalized efficiency, since the interaction region is significantly sub-wavelength for the MW fields. By adopting the same approach, we obtain a $3.1{\pm}0.4\%$ atomic efficiency (see Supplement S.1 for estimation). We are also able to use our full model to give a theoretical prediction (assuming no signal depletion) for the area-normalized efficiency as $2.8{\pm}1.6\%$ (see Supplement S.1 for calculation), which matches the observed value. We expect that for our case as well as the previous free-space experiments this efficiency would be achieved if the MW mode was confined to the interaction volume, e.g.~by means of a waveguide. In port-coupled or cavity-based conversion cases \cite{Kumar_2022} the absolute efficiency can be estimated unambiguously.

\paragraph{Conversion band}
With the use of single photon counter we measure the conversion's dependence on MW frequency in the linear response regime (Fig.~\ref{results}\textbf{a}), arriving at a bandwidth $\Gamma_{\mathrm{con}} = 16 {\cdot} 2 \pi\ \mathrm{MHz}$ FWHM (full width at half maximum). Further explorations reveal interdependence on MW and decoupling ($1258\ \mathrm{nm}$) fields detunings, that can be utilized as a means to fine-tune the converter to incoming MW field. We show that with such a procedure we are able to widen the tunable conversion bandwidth up to $59 {\cdot} 2 \pi\ \mathrm{MHz}$ FWHM.

\paragraph{Photonic conversion of thermal radiation}
We perform single-photon autocorrelation measurement (Fig.~\ref{results}\textbf{b}) on the thermal radiation signal, confirming the Hanbury Brown and Twiss (HBT) effect of thermal photon bunching ($g^{(2)} (0) {>} 1$) for MW photons. The results obtained experimentally agree perfectly with the parameter-free theory drawn from the Wiener-Khinchin theorem,
\begin{equation}\label{WK}
    g^{(1)}_{\mathrm{th}} (\tau) = \frac{1}{2 \pi} \int_{- \infty}^{\infty} \left| S(\omega) \right|^2 e^{- i \omega \tau} \mathrm{d} \omega,
\end{equation}
where $|S(\omega)|^2$ is normalized ($\int_{- \infty}^{\infty} | S(\omega) |^2 \mathrm{d} \omega {=} 2 \pi$) power spectral density, here measured in the Fig.~\ref{results}\textbf{a} with the assumption that thermal radiation locally has white noise characteristics. Naturally following, $g^{(2)}_{\mathrm{th}} (\tau) = 1 + | g^{(1)}_{\mathrm{th}} (\tau) |^2$. To arrive with the theoretical results presented in the Fig.~\ref{results}\textbf{b}, we consider both thermal radiation and non-thermal noise present in the system (see Methods for details). An alternative parametric fitting of exponent function, $g^{(2)} (\tau) {=} 1 {+} ( g^{(2)} (0) {-} 1 ) e^{-2 \tau / \tau_0}$, yields $g^{(2)} (0) = 1.868 {\pm} 0.035$ and coherence time $\tau_0 = 26.0 {\pm} 1.5\ \mathrm{ns}$.

Next we introduce a resonant coherent MW field of strength comparable to the thermal field -- in terms of converted photons rates $\bar{n}_{\mathrm{coh}}$ for coherent photons and $\bar{n}_{\mathrm{th}}$ for thermal photons. We estimate the $g^{(2)} (0)$ value for consecutive measurements (Fig.~\ref{results}\textbf{c}), in agreement with the theory. Interestingly, the introduction of a stronger but $\Delta \omega$ far-detuned coherent field, while keeping $\bar{n}_{\mathrm{coh}} {\approx} \bar{n}_{\mathrm{th}}$, leads to a beat modulation of the autocorrelation function (Fig.~\ref{results}\textbf{d}). Specifically in this case
\begin{equation}
    g^{(2)} (\tau) = 1 + \frac{1}{4} \left( \left| g^{(1)}_{\mathrm{th}} (\tau) + e^{- i \omega \tau} \right|^2 - 1 \right),
\end{equation}
where $g^{(1)}_{\mathrm{th}} (\tau) {=} A(\tau) e^{- i \omega_0 \tau}$ and $\omega_0$ is the mean frequency of the conversion band. The beat modulation is at the frequency of the detuning $\Delta \omega {=} \omega {-} \omega_0$.

\paragraph{Noise figure}
With no external field applied to the converter, we observe photons, which by disabling each field and cross-correlation analysis we identify as consisting of $85\%$ thermal radiation, $13\%$ fluorescence of optical elements induced by $480\ \mathrm{nm}$ laser field and $2\%$ other noise. This proportion results in an overall $5.7{:}1$ signal-to-noise ratio, and as the signal is thermal in this case, we can translate it straightforwardly to a noise-equivalent temperature $T_{\mathrm{NE}} = 53\ \mathrm{K}$. As the added noise is wideband, we utilize additional cavity-assisted filtering of thermal signal, improving the ratio to $77{:}1$, thus $T_{\mathrm{NE}} = 3.8\ \mathrm{K}$. We present these values in the Fig.~\ref{sateff}\textbf{a} as a reference to the converter's response characteristics. See Supplement S.2 for cross-correlation analysis and comparison of different sources of noise.

\begin{figure}
\includegraphics[width=.48\textwidth]{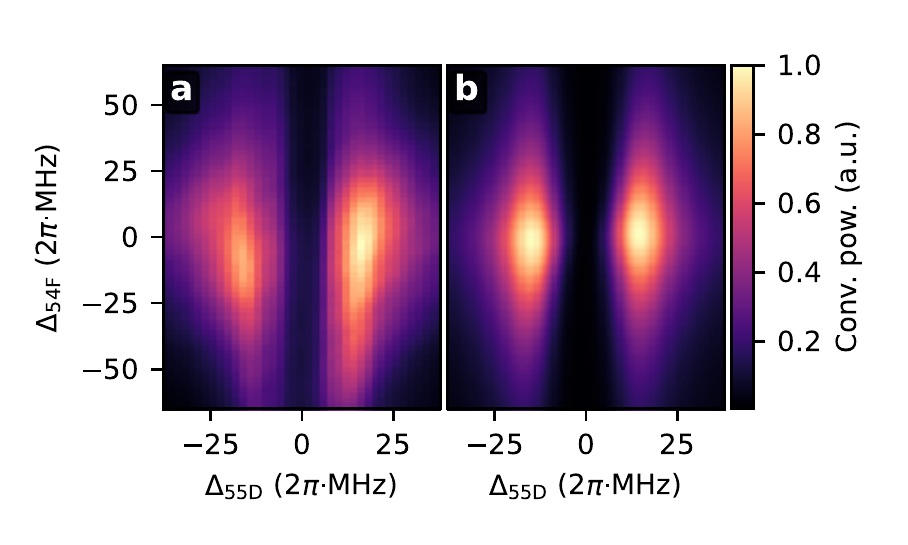}
\caption{\textbf{Atomic bright resonances enabling the conversion process.}
\textbf{a},~Measured conversion in the domain of bare-state detunings from energy levels $55^2\mathrm{D}_{5/2}$ and $54^2\mathrm{F}_{7/2}$ (other detunings being near-zero), unveiling the underlying bright resonances $\ket{B+}$ and $\ket{B-}$. From this measurement we obtain the optimal level detunings $\Delta_{55\mathrm{D}} {=} 16{\cdot} 2 \pi\ \mathrm{MHz}$, $\Delta_{54\mathrm{F}} {=} 0$. \textbf{b},~The same conversion interdependence predicted in the numerical simulation.}
\label{levmap}
\end{figure}

\paragraph{Observation of bright resonances}
Following the short demonstration of the level-detuned working regime in the Fig.~\ref{qura}\textbf{d}, we further explore the conversion's dependence on the detunings from bare atomic levels -- realized with proportional detunings of the fields. We find this domain the most convenient to operate in, as it naturally obeys the conservation of energy in six-wave mixing. In general, with a hot-atom system, the Doppler effect prevents interpretations as simple as in the cold-atom case. However, in our case, we find that the Doppler effect is mostly canceled via the selection of velocity class in the two-photon process, i.e.~with probe and coupling fields.  We expect that the conversion loop is actually composed of two dressed-state subsystems: the three-level system dressed by coupling and probe fields, and the two-level system dressed by the decoupling field. Indeed, by scanning the $\Delta_{55\mathrm{D}}$ level detuning, we observe resonances corresponding to two bright states, that correspond to the simplified and unnormalized eigenstates given by $\ket{B\pm} = \Omega_p^*\ket{5^2\mathrm{S}_{1/2}} \pm \sqrt{|\Omega_p|^2{+}|\Omega_c|^2}\ket{5^2\mathrm{P}_{3/2}} + \Omega_c\ket{55^2\mathrm{D}_{5/2}}$ split by roughly $\Omega_c$ (as seen in the Fig.~\ref{levmap}), where by $\Omega_p$ and $\Omega_c$ we denote the Rabi frequencies of probe and coupling fields respectively. The dark state $\ket{D}=\Omega_c^*\ket{5^2\mathrm{S}_{1/2}}-\Omega_p\ket{55^2\mathrm{D}_{5/2}}$ does not take part in the conversion process, but could be accessed in a cavity-enhanced system to yield conversion with lower loss due to $5\mathrm{P}_{3/2}$ state decay. On the other hand, as we sweep the detuning $\Delta_{54\mathrm{F}}$, we observe broadening rather than clear splitting into two dressed states $\ket{\pm}=\ket{54^2\mathrm{F}_{7/2}}\pm\ket{5^2\mathrm{D}_{5/2}}$, which is due to particular interference of different velocity classes. Via numerical studies we found that this splitting is observed only at much higher decoupling Rabi frequencies in different scans. We also note that the blue-detuned resonance results in a slightly stronger conversion than the red-detuned, which we attribute to the interfering effect of other energy levels (that may take part in wave-mixing process) on the latter. Our observation confirms that bright resonances $\ket{B\pm}$ are essential in the conversion process. The conversion process is optimized by accessing one of the resonances through detuning from the two-photon resonance (and thus also avoiding the dark resonance $\ket{D}$), as seen in the Fig.~\ref{qura}\textbf{c} and as observed in other free-space experiments \cite{Tu_2022}.

\section*{Discussion}

The presented proof-of-concept warm-atomic converter exhibits flexibility in terms of high dynamic range and wide bandwidth, which supports its potential for applications. We believe the fields of contemporary astronomy could benefit from conversion-based MW photonic measurements \cite{Riechers_2022,Pankratov_2022,Komatsu_2022,Greaves_2018,Gorski_2021,Kou_2022}, where Rydberg atoms excel in simplicity, adaptability and presented low noise figures. As the conversion scheme utilized here is all-optical, it can be applied in similar systems (i.e.~cold trapped atoms and superheterodyne Rydberg electrometry) to avoid the introduction of noise by spectrally close fields. We highlight many desirable properties of the presented converter. The very ability to perform photon counting of microwave radiation at room temperature along with observation of the HBT interference and coherent-thermal interference is remarkable, as those feats typically require deeply cryogenic conditions \cite{PhysRevLett.107.217401}. The converter presents an extremely large dynamic range and excellent bandwidth, along with many options for both fine and very coarse tuning to different Rydberg levels. The all-optical realization gives further prospects for applications as even strong electromagnetic interference would not damage the device. This could be important for microwave-based communication, where another advantage may come from avoiding shot noise of homodyne detection via photon counting.

There are numerous issues that so far escaped the scope of this work and could be explored in more detail to the benefit of increasing the conversion's range and efficiency. Because thermal photons are spread over the whole conversion band, the limit approached in the Fig.~\ref{sateff}\textbf{a} is not fundamental, as efficient narrow spectral filtering or phase-locking measurement would push it down, enabling detection at the single-photon level. The non-thermal noise may be decreased further e.g.~by utilizing different optical elements in the setup (specifically addressing the $480\ \mathrm{nm}$ fluorescence) or by exploring non-collinear configuration of the laser fields. Various optimization efforts may also be performed to find working points with greater conversion efficiency, though the operational regime of warm atomic vapors and strong probe field does not allow for simple theoretical predictions. The converter's response to MW pulses is yet to be investigated, as well as pulse-control of laser fields or coherent repumping of atomic population, though the complexity of the setup would then inevitably increase. Additionally, the operation mode of the Rydberg converter in principle can be thoroughly shifted to adapt to emission at the telecom C-band wavelength -- specifically, $4^2\mathrm{D}_{5/2} \rightarrow 5^2\mathrm{P}_{3/2}$ transition occurs at $1530\ \mathrm{nm}$.

The all-optical realization could be retained while designing a setup that exploits the Doppler effect cancellation in a wider scope (i.e.~by using different transitions), enabling more atoms to take part in the conversion process. Conversely, a range of atomic velocities may be seen as the means to further widen the tunable conversion band. We believe that with commensurate detunings of all-optical field, the tunable conversion bandwidth can be extended to as much as $600\ \mathrm{MHz}$ (the width of Doppler-broadened probe absorption line), thus with the consideration of the neighboring Rydberg transitions, the converter could efficiently cover the full range of MW frequencies up to $50\ \mathrm{GHz}$, only with the tuning of laser fields. To our knowledge, such an ultrawide adaptable conversion band is possible only with warm Rydberg atomic vapors. Finally, we also expect that the reverse process of optical-to-microwave conversion should be facilitated by our setup via coupling of atoms to microwave resonators (see Supplement S.6 for details).

Thus we envisage that further progress will be centered around introducing warm Rydberg atomic vapors to MW cavity systems, as it will enhance the coupling between MW field and atomic interaction volume. Our all-optical scheme is well-suited for that task; in comparison, the six-wave mixing schemes involving two MW transitions would require a doubly-resonant cavity with close-by but non-degenerate resonances. The design principles for hot-atom MW-cavity systems can be drawn from cavity-enhanced atomic clocks. Even with moderate finesse, we expect the conversion efficiency to be significantly enhanced, likely even leading to radiative cooling of the cavity mode via conversion. At the same time, a lot of potential lies in recent progresses in microfabrication of atomic devices \cite{Kitching_2018}, such as micro- and nanoscale vapor cells \cite{Cutler_2020,Lucivero_2022} and hollow-core photonic bandgap fibers filled with alkali atoms \cite{Peters_2020}. These instruments may enable all-fiber, reproducible applications of the presented converter's model, paving the way to next-generation MW converting sensors.

\section*{Methods}

\paragraph{Density matrix calculation}
As indicated in the main text, we solve the five-level GKSL equation in the steady state, i.e.~$\partial_t\hat{\rho}=0$. The Lindbladian is constructed from the Hamiltonian and the jump operators using the QuantumOptics.jl package \cite{kramer2018quantumoptics}. We next solve for the steady-state using standard linear algebra methods finding the zero eigenvalue.  We employ jump operators $\hat{J}_n$ for spontaneous decay and for transit-time decay, i.e.~atoms exiting the interaction region with ground-state atoms entering instead. Each solution is for a given set of Rabi frequencies, detuning and a specific velocity class $v$ (along the longitudinal direction). The detunings are modified due to the Doppler effect as $\Delta_n^v {=} \Delta_n {\pm} k_n v$ with $n {\in} \{p,c,\mathrm{MW},d\}$. We next perform averaging of the resulting state-state density matrix $\hat{\rho}$ over the Doppler velocity profile with weighting function $f(v) {=} \sqrt{m/2\pi k_\mathrm{B} T} \exp(-mv^2/k_\mathrm{B} T)$ (with $m$ being the $^{85}$Rb atomic mass) and over the Gaussian profile of the beams due to changing Rabi frequencies. Unless noted otherwise, the MW Rabi frequency is taken well below the saturation point. Finally, we extract the generated signal, always plotted as intensity, as $|E_s|^2 {\sim} |\mathrm{Tr}(\ket{5^2D_{5/2}}\bra{5^2P_{3/2}}\hat{\rho})|^2$ or the EIT signal as $\mathrm{Im}(\mathrm{Tr}(\ket{5^2S_{1/2}}\bra{5^2P_{3/2}}\hat{\rho}))$.

\paragraph{Phase matching}
Efficient conversion requires all atoms interacting with the MW field to emit in phase. This renders the phase-matching condition that determines the conversion efficiency depending on the MW field mode. Thus we introduce the phase-matching factor $\eta_\mathrm{phm}(\theta)$ for a plane-wave MW field $E_\mathrm{MW}(\theta)$ entering the medium at angle $\theta$ to the propagation axis ($z$). The factor is then calculated by projecting the generated electric susceptibility onto the detection mode $u_s$ being a Gaussian beam focused in the center of the cell ($z_0{=}0$) with $w_0{=}100\ \mathrm{\mu m}$. To obtain the projection we assume the spatial dependence of the susceptibility in the form of the product of all interacting fields. This is a simplification compared with the full model, as in general the steady-state density matrix and, in turn, the coherence $\rho_s$ have a more complex dependence on strong drive field amplitudes. Nevertheless, for the purpose of spatial calculation, the simplified approach yields proper results and then the susceptibility takes on a form:
\begin{equation}
    \chi_\theta = E_pE_p^*E_cE_d^{*}E_\mathrm{MW}(\theta),
\end{equation}
with the optical fields ($E_p$, $E_c$, $E_d$) taken to be Gaussian beams with the same $z_0$ and $w_0$ as the detection mode $u_s$. Additionally, we account for the relatively strong absorption of the probe ($E_p$) field by multiplying its amplitude by the exponential decay $\exp(-\alpha z)$ with (measured) $\alpha{=}19\ \mathrm{m}^{-1}$. Finally, the coefficient is calculated as the following integral:
\begin{equation}
\eta_\mathrm{phm}(\theta)=\int_{-L/2}^{L/2}\mathrm{d}z \int_0^{2\pi}\mathrm{d}\phi\int_0^{\infty}\rho\mathrm{d}\rho \chi_\theta u_s^{*},
\end{equation}
where $L{=}50\ \mathrm{mm}$ is the length of the glass cell.

\begin{figure}
    \includegraphics[width=1\columnwidth]{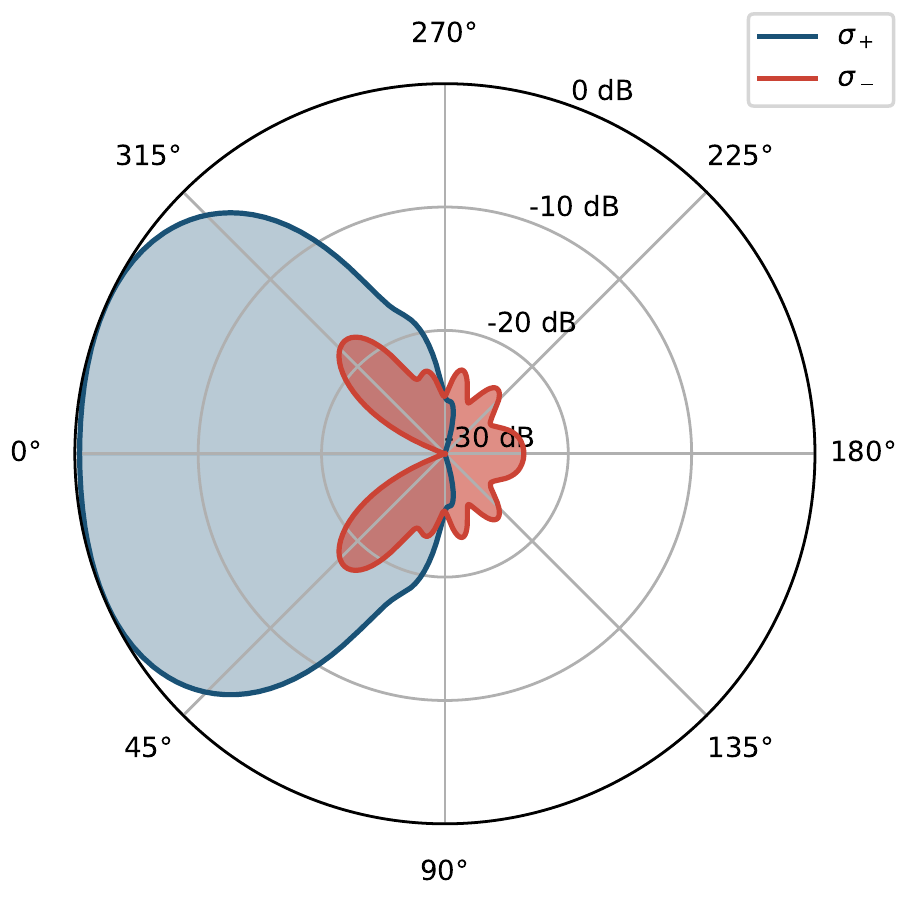}
    \caption{\textbf{Converter's spatial reception pattern.} Calculated angular dependence of the MW conversion efficiency for two circular polarizations: $\sigma_+$ (antenna polarization) and orthogonal $\sigma_-$. The scale for both plots represents efficiency relative to the maximum of the $\sigma_+$ pattern in $\mathrm{dB}$. The slight drop in efficiency at $0^\circ$ is due to the Gouy phase of optical beams.}
    \label{fig:qra}
\end{figure}

\begin{figure*}
\includegraphics[width=\textwidth]{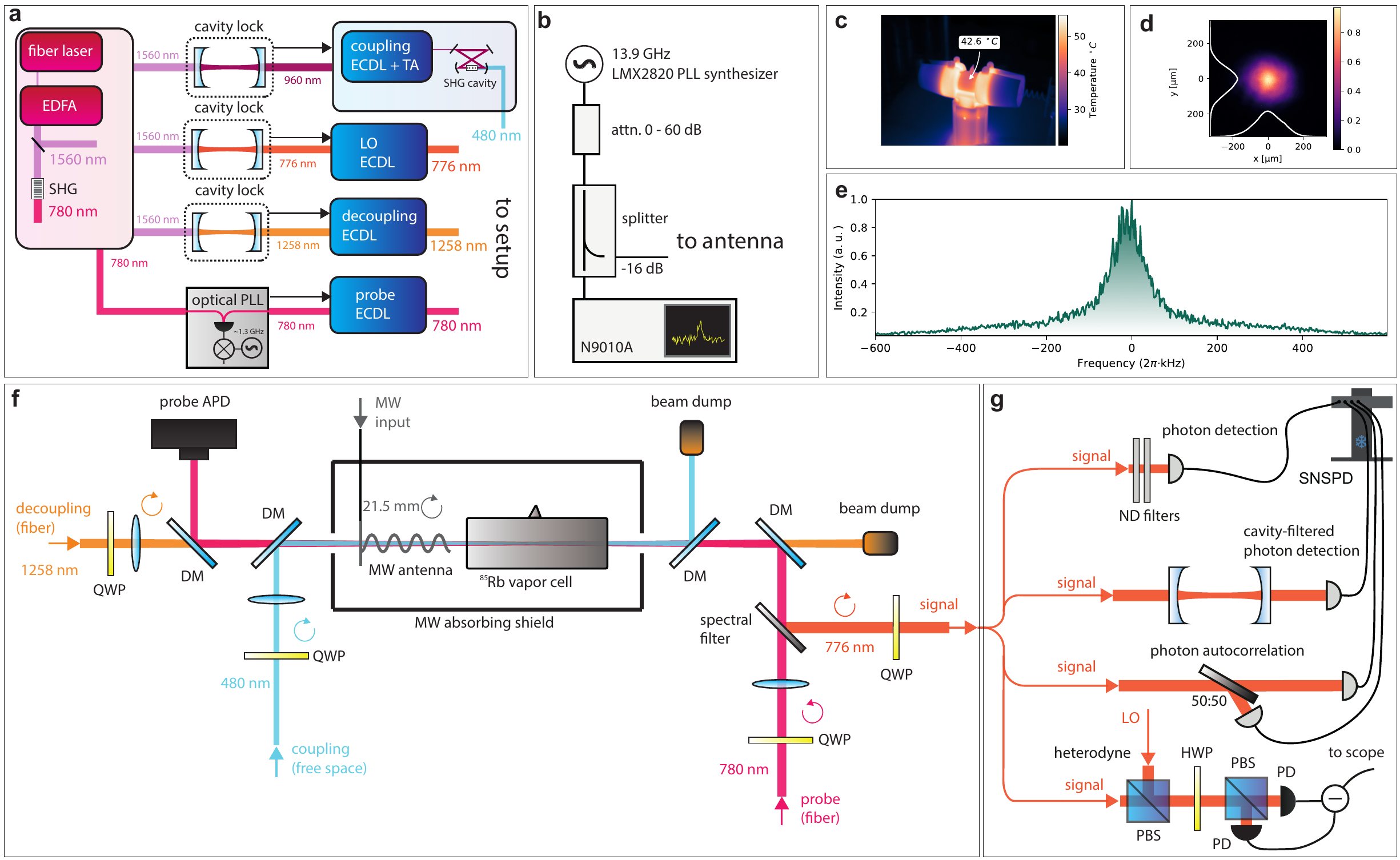}
\caption{
\textbf{Details of the entire experimental setup.}
\textbf{a},~Laser system. A narrowband fiber laser at $1560\ \mathrm{nm}$ serves as a frequency reference for the entire system. The laser is amplified via erbium-doped fiber amplifier (EDFA) and the frequency is doubled using a second-harmonic generation (SHG) process to $780\ \mathrm{nm}$. The $960\ \mathrm{nm}$ external-cavity diode laser (ECDL) is amplified using a tapered amplifier (TA) and frequency-doubled via cavity-enhanced SHG to yield a coupling field at $480\ \mathrm{nm}$. The $960\ \mathrm{nm}$ laser is stabilized to the master laser of $1560\ \mathrm{nm}$ using a common cavity. A similar independent system is introduced for the stabilization of decoupling ($1258\ \mathrm{nm}$) laser, likewise for local oscillator (LO) $776\ \mathrm{nm}$ laser. The second harmonic of the $1560\ \mathrm{nm}$ fiber laser serves as a reference for offset-locking of the probe laser in an optical phase-locked loop (PLL).
\textbf{b},~Microwave generation system. The generated microwaves are attenuated adequately and split into the spectrum analyzer for power and frequency references and to the antenna.
\textbf{c},~Thermal image of the cell and cell holder. The constant temperature of the cell is assured with hot-air heating via hollow channels in the 3D-printed cell holder.
\textbf{d},~The generated signal light exits the converter in a beam with a Gaussian profile shown in the image.
\textbf{e},~Heterodyne measurement with LO yields spectrum of the converted signal: the residual broadening of $\Gamma_{\mathrm{sig}} {=} 86 {\cdot} 2 \pi\ \mathrm{kHz}$ FWHM is due to collective laser-locking phase noise.
\textbf{f},~Scheme of the experimental setup: probe, coupling and decoupling laser beams are combined with dichroic mirrors (DM) into a collinear configuration, and focused inside a rubidium vapor cell of length $50\ \mathrm{mm}$ and diameter $25\ \mathrm{mm}$ with a MW helical antenna pointed there. The circular polarization of lasers is assured with quarter-wave plates (QWP). The probe ($780\ \mathrm{nm}$) signal is registered by an avalanche photodiode (APD), which enables laser calibration by observation of EIT features.   Converted $776\ \mathrm{nm}$ signal is spectrally separated and coupled into a single-mode fiber.
\textbf{g},~Different setups are used for detection, with single setup being used at a time. The detection setups include direct photon counting with optional attenuation, cavity-filtered photon counting, photon autocorrelation with two channels simultaneously, and heterodyne detection. For photon counting, a multichannel superconducting nanowire single-photon detector (SNSPD) is used. For heterodyne measurement we combine the signal with LO using a polarization beam splitter (PBS) and split the combined signal $50{:}50$ with a half-wave plate (HWP) and a second PBS. The signal is then registered on a differential photodiode (PD).
}
\label{setup}
\end{figure*}

\paragraph{Thermal modes consideration}
Let us now estimate the effective field root-mean-square amplitude due to thermal black-body radiation in our given conversion band, which takes into account that our converter is both polarization and wavevector sensitive. In order to do so, we follow a standard derivation of Planck's law, however, when considering all electromagnetic modes we account for both the polarization-dependence and phase matching during conversion. The effective mean-square field amplitude at the temperature $T$ may be written as follows:
\begin{align}
    \langle E^2_{\mathrm{eff}} \rangle &=\frac{\omega^2\langle \mathcal{E}\rangle}{\pi^2c^3\varepsilon_0} \frac{1}{4\pi}\int_0^{2\pi}\mathrm{d}\phi\int_0^\pi\mathrm{d}\theta\sin(\theta) |\eta(\theta)|^2, \\
    \langle \mathcal{E}\rangle &= \frac{\hbar \omega}{e^{\hbar \omega/k_\mathrm{B} T}-1},
\end{align}
where $\omega$ is the frequency of the MW field, $c$ is the speed of light, $\varepsilon_0$ is vacuum permittivity and $k_\mathrm{B}$ is the Boltzmann constant. Importantly, $\eta(\theta)$ is the field conversion efficiency coefficient that arises from projecting the MW field in the plane-wave modes at the given angle $\theta$ onto circular polarization along the conversion axis and multiplying it by the phase matching factor: 
\begin{equation}
    |\eta(\theta)|^2=\left(\cos\left(\frac{\theta}{2}\right)^4 + \sin\left(\frac{\theta}{2}\right)^4\right)|\eta_\mathrm{phm}(\theta)|^2.
\end{equation}
In fact the efficiency $|\eta(\theta)|^2$ represents the converter reception pattern, which we plot in the Fig.~\ref{fig:qra}. From the pattern we see that the contribution to the effective noise-equivalent field at temperature $T$ comes mostly from the phase-matched microwaves coming from an almost right-angled cone with the correct circular polarization. Finally, we refer the effective field convertible by the atoms $\sqrt{\langle E_\mathrm{eff}^2\rangle}$ to total RMS field fluctuations $\sqrt{\langle E^2\rangle}$, which can be calculated by plugging in $|\eta(\theta)|^2 {=} 2$ (because of two polarizations).

In the Fig.~\ref{sateff}\textbf{a} we find that the thermal radiation level (without added non-thermal noise) corresponds to $3.41{\cdot}10^{-10}\ \mathrm{W}/\mathrm{m}^2$ or $5.07\ \mathrm{\mu V}/\mathrm{cm}$, as calibrated from the A-T splitting at large fields. To calculate spectral field density, we consider the conversion bandwidth using an integral measure $\Tilde{\Gamma}_{\mathrm{con}} = 1/\max(\left| S(\omega) \right|^2) \int_{- \infty}^{\infty} \left| S(\omega) \right|^2 \mathrm{d} \omega = 17.8 {\cdot} 2 \pi\ \mathrm{MHz}$, to arrive with $\sqrt{\langle E_\mathrm{eff}^2\rangle} = 480\ \mathrm{pV}\mathrm{cm}^{-1}(\mathrm{rad}/\mathrm{s})^{-1/2}$. This is indeed highly consistent with the model prediction of $\sqrt{\langle E_\mathrm{eff}^2\rangle} = 495\ \mathrm{pV}\mathrm{cm}^{-1}(\mathrm{rad}/\mathrm{s})^{-1/2}$. When referred to the total field fluctuations, we obtain the results quoted in the text, i.e.~ $\sqrt{\langle E^2\rangle} = 1.59\ \mathrm{nV}\mathrm{cm}^{-1}(\mathrm{rad}/\mathrm{s})^{-1/2}$ measured and $\sqrt{\langle E^2\rangle}=1.64\ \mathrm{nV}\mathrm{cm}^{-1}(\mathrm{rad}/\mathrm{s})^{-1/2}$ predicted. (For detailed analysis of photon rates and electric field densities see Supplement S.3.)

\paragraph{Thermal and coherent states autocorrelation function}
We consider the autocorrelation measurements with three different sources of photons, defined by rates: $\bar{n}_{\mathrm{th}}$ for thermal state photons, $\bar{n}_{\mathrm{coh}}$ for coherent state photons and $\bar{n}_{\mathrm{noise}}$ for photons coming from wideband, non-interfering sources (e.g.~various forms of fluorescence). We follow the derivation presented in \cite{Marian_1993} (and later showcased in \cite{Forsch_2019,Kumar_2022}) to arrive with the following general form of second order autocorrelation function
\begin{equation}
    g^{(2)} (\tau) = 1 + \frac{ \left| \bar{n}_{\mathrm{th}} g^{(1)}_{\mathrm{th}} (\tau) + \bar{n}_{\mathrm{coh}} e^{- i \omega \tau} \right|^2 - \bar{n}_{\mathrm{coh}}^2}{\left( \bar{n}_{\mathrm{th}} + \bar{n}_{\mathrm{coh}} + \bar{n}_{\mathrm{noise}} \right)^2},
\end{equation}
where $\omega$ is the coherent state frequency. In the Fig.~\ref{results}\textbf{b},\textbf{c},\textbf{d} $\bar{n}_{\mathrm{noise}} / \bar{n}_{\mathrm{th}} {=} 15/85$, in the Fig.~\ref{results}\textbf{d} $\bar{n}_{\mathrm{coh}} / \bar{n}_{\mathrm{th}} {=} 100/85$.

The $g_{\mathrm{th}}^{(1)}$ correlation function is obtained from power spectral density in the Fig.~\ref{results}\textbf{a} via the Wiener-Khinchin theorem, Eq.~(\ref{WK}), and then used to calculate $g^{(2)}$ for comparison with the experiment. The conversion bandwidth is measured via photon-counting detection in the domain of MW field detuning -- we directly change the MW frequency fed to the antenna. The next crucial step to equate the measured conversion bandwidth with the power spectral density of thermal radiation coupling to the converter $|S(\omega)|^2$ is to assume that thermal radiation has a white noise characteristics within our band of interest (i.e.~for free uncoupled radiation $|S(\omega)|^2 = \mathrm{const}$). Then we arrive with the function presented in the Fig.~\ref{results}\textbf{b}.

\paragraph{Laser fields parameters}
The laser beams are focused to equal Gaussian waists of $w_0 {=} 100\ \mathrm{\mu m}$ and combined with the use of dichroic mirrors (Fig.~\ref{setup}\textbf{f}). The probe beam counterpropagates with respect to the other fields and its transmission through ${}^{85} \mathrm{Rb}$ medium is then utilized as a means for calibration via EIT effects (Fig.~\ref{qura}\textbf{d},\textbf{e}). The length of vapor cell is $50\ \mathrm{mm}$. The collinear laser fields configuration is assured with $4 f$ optical systems (not shown on the scheme) with mirrors at focal planes, enabling independent control of every beam's position an propagation angle. The employed laser beam effective peak Rabi frequencies for $780\ \mathrm{nm}$, $480\ \mathrm{nm}$ and $1258\ \mathrm{nm}$ fields are derived from the numerical simulation as $\Omega_p{=}8 {\cdot} 2 \pi\ \mathrm{MHz}$, $\Omega_c{=}22 {\cdot} 2 \pi\ \mathrm{MHz}$ and $\Omega_d{=}17 {\cdot} 2 \pi\ \mathrm{MHz}$ respectively. The optimal dominant detuning from the $55 \mathrm{D}$ level is measured as $\Delta_{55 \mathrm{D}} = 16 {\cdot} 2 \pi\ \mathrm{MHz}$ (Fig.~\ref{levmap}\textbf{a}) and is used as a working point for measurements, with the other detunings being near-zero. For the Fig.~\ref{qura} and \ref{setup}\textbf{e} the MW Rabi frequency is set to about $\Omega_\mathrm{MW}{=}8 {\cdot} 2 \pi\ \mathrm{MHz}$, while for the Figs. \ref{results} and \ref{levmap}\textbf{a} it is deeply in the unsaturated regime. For the Fig.~\ref{levmap}\textbf{a} the decoupling intensity is increased to about $\Omega_d{=}25 {\cdot} 2 \pi\ \mathrm{MHz}$ peak. The calibration of detuning zero-points takes place at the ${}^{85} \mathrm{Rb}$ working temperature to account for pressure shifts of Rydberg energy levels.

\paragraph{Temperature stabilization and microwave shielding}
The measured optimal ${}^{85} \mathrm{Rb}$ working temperature, $T {=} 42 {}^\circ \mathrm{C}$, is ensured via hot air heating, as this method introduces little interference to MW fields propagating through the vapor cell. The air is heated and pumped through a specially designed hollow 3D-printed resin cell holder (Fig.~\ref{setup}\textbf{c}), where the heat exchange takes place. The cell is enclosed inside a MW absorbing shield, made from a material with ${>}30\ \mathrm{dB}$ loss at $14\ \mathrm{GHz}$ (LeaderTech EA-LF500), with sub-MW-wavelength apertures for optical beams -- the shield also provides additional thermal isolation, reducing fluctuations of temperature. The temperature of the shield inside, for the reference to black-body radiation, is measured as $26{-}27{}^\circ \mathrm{C}$. The MW helical antenna is placed inside the shield and the collinear propagation is assured with optical fields passing through an aperture at the backplate of the antenna (Fig.~\ref{setup}\textbf{f}). 

\paragraph{Frequency stabilization and calibration}
The lasers in the experiment are stabilized (Fig.~\ref{setup}\textbf{a}) to a narrowband, frequency-doubled fiber laser (NKT Photonics, $1560\ \mathrm{nm}$), which is itself stabilized to a Rb cell via a modulation-transfer lock. The probe laser at $780\ \mathrm{nm}$ is stabilized to the frequency-doubled reference via an optical phase-locked loop (PLL). The $1258\ \mathrm{nm}$ laser (decoupling), the $960\ \mathrm{nm}$ laser (frequency-doubled to yield coupling light at $480\ \mathrm{nm}$, Toptica DL-SHG pro) and the $776\ \mathrm{nm}$ laser acting as LO, are all stabilized to the reference at $1560\ \mathrm{nm}$ using independent cavities via transfer locks. Laser fields are calibrated in low-intensity regimes with the use of EIT effects observed in the probe field transmission, registered on the control avalanche photodiode (Thorlabs APD120A, Fig.~\ref{setup}\textbf{f}). MW field, generated via LMX2820 PLL frequency synthesizer, is coupled to the antenna and spectrum analyzer (Agilent N9010A EXA) used for relative reference of field amplitude (Fig.~\ref{setup}\textbf{b}). Absolute MW field amplitude is calibrated with a standard measurement of A-T splitting \cite{Sedlacek_2012}. As a method to quantify the converter's ability to distinguish photons incoming from spectrally different sources, we perform a heterodyne measurement, yielding the converter's signal width $\Gamma_{\mathrm{sig}} = 86 {\cdot} 2 \pi \mathrm{kHz}$ FWHM (Fig.~\ref{setup}\textbf{e}). This value may be interpreted as a measure of collective laser-locking phase noise.

\paragraph{Measurement techniques}
Initially, the $776\ \mathrm{nm}$ signal is reflected from a bandpass filter and then passed through a series of free-space spectral filters (highpass, lowpass, $1.2\ \mathrm{nm}$ spectral width bandpass) and coupled to a fiber. Due to the Gaussian characteristics of the beam (Fig.~\ref{setup}\textbf{d}), coupling losses are negligible in this case (${<}25\%$). We apply various detection methods pictured in the Fig.~\ref{setup}\textbf{g}. Additional filtering in the measurement of noise-equivalent temperature is performed with an optical cavity ($160\ \mathrm{MHz}$ spectral width, $80$ finesse). The measurement of photon counting is performed with a single-photon detector (ID Quantique ID281 superconducting nanowire single-photon detector) with $85\%$ quantum efficiency, ${<}1\ \mathrm{Hz}$ dark count rate and $35\ \mathrm{ns}$ recovery time. As the detector's reliable response to incoming photons is at the order of ${<}10^7\ \mathrm{phot} / \mathrm{s}$, calibrated neutral density filters are applied to the signal above this value. The autocorrelation measurement is performed on two channels of the single-photon detector after splitting the signal $50{:}50$ with a fiber splitter. The heterodyne measurement is performed using a custom-made differential photodiode with the use of $776\ \mathrm{nm}$ LO laser, $20\ \mathrm{mW}$ power. The image of the beam profile (Fig.~\ref{setup}\textbf{d}) is taken on a ${>}50\%$ quantum efficiency CMOS camera (Basler acA2500-14gm). The supplementary measurements of temperature (including Fig.~\ref{setup}\textbf{c}) are performed with a calibrated thermal camera (Testo 883).

\begin{acknowledgments}
We thank K.~Banaszek for the generous support and W.~Wasilewski for discussions and support with cavity and offset locking systems. We also thank J.~Kołodyński, M.~Papaj, M.~Lipka and G.~Santamaría Botello for their comments and discussion regarding the manuscript. The “Quantum Optical Technologies” (MAB/2018/4) project is carried out within the International Research Agendas programme of the Foundation for Polish Science co-financed by the European Union under the European Regional Development Fund.  This research was funded in whole or in part by National Science Centre, Poland grant no. 2021/43/D/ST2/03114.
\end{acknowledgments}

\section*{Author contributions}
S.B.~and U.P.~built the optical and microwave setups. S.B.~took the measurements and analyzed the data assisted by other authors. S.B, M.M.~and M.P.~prepared the figures and the manuscript. M.M.~and M.P.~developed the theory assisted by S.B.~who facilitated comparison with experiments. M.P.~led the project assisted by~M.M.

\section*{Competing interests}
We declare that none of the authors have competing interests.

\section*{Data availability}
Data underlying the results presented in this paper are available in the Ref.~\cite{Data23}.

\section*{Code availability}
The codes used for the numerical simulation and the analysis of experimental data are available from the authors upon request.

\bibliographystyle{apsrev4-2}

\bibliography{refs}

\begin{thebibliography}{58}%
\makeatletter
\providecommand \@ifxundefined [1]{%
 \@ifx{#1\undefined}
}%
\providecommand \@ifnum [1]{%
 \ifnum #1\expandafter \@firstoftwo
 \else \expandafter \@secondoftwo
 \fi
}%
\providecommand \@ifx [1]{%
 \ifx #1\expandafter \@firstoftwo
 \else \expandafter \@secondoftwo
 \fi
}%
\providecommand \natexlab [1]{#1}%
\providecommand \enquote  [1]{``#1''}%
\providecommand \bibnamefont  [1]{#1}%
\providecommand \bibfnamefont [1]{#1}%
\providecommand \citenamefont [1]{#1}%
\providecommand \href@noop [0]{\@secondoftwo}%
\providecommand \href [0]{\begingroup \@sanitize@url \@href}%
\providecommand \@href[1]{\@@startlink{#1}\@@href}%
\providecommand \@@href[1]{\endgroup#1\@@endlink}%
\providecommand \@sanitize@url [0]{\catcode `\\12\catcode `\$12\catcode
  `\&12\catcode `\#12\catcode `\^12\catcode `\_12\catcode `\%12\relax}%
\providecommand \@@startlink[1]{}%
\providecommand \@@endlink[0]{}%
\providecommand \url  [0]{\begingroup\@sanitize@url \@url }%
\providecommand \@url [1]{\endgroup\@href {#1}{\urlprefix }}%
\providecommand \urlprefix  [0]{URL }%
\providecommand \Eprint [0]{\href }%
\providecommand \doibase [0]{https://doi.org/}%
\providecommand \selectlanguage [0]{\@gobble}%
\providecommand \bibinfo  [0]{\@secondoftwo}%
\providecommand \bibfield  [0]{\@secondoftwo}%
\providecommand \translation [1]{[#1]}%
\providecommand \BibitemOpen [0]{}%
\providecommand \bibitemStop [0]{}%
\providecommand \bibitemNoStop [0]{.\EOS\space}%
\providecommand \EOS [0]{\spacefactor3000\relax}%
\providecommand \BibitemShut  [1]{\csname bibitem#1\endcsname}%
\let\auto@bib@innerbib\@empty
\bibitem [{\citenamefont {Arute}\ \emph {et~al.}(2019)\citenamefont {Arute},
  \citenamefont {Arya}, \citenamefont {Babbush}, \citenamefont {Bacon},
  \citenamefont {Bardin}, \citenamefont {Barends}, \citenamefont {Biswas},
  \citenamefont {Boixo}, \citenamefont {Brandao}, \citenamefont {Buell},
  \citenamefont {Burkett}, \citenamefont {Chen}, \citenamefont {Chen},
  \citenamefont {Chiaro}, \citenamefont {Collins}, \citenamefont {Courtney},
  \citenamefont {Dunsworth}, \citenamefont {Farhi}, \citenamefont {Foxen},
  \citenamefont {Fowler}, \citenamefont {Gidney}, \citenamefont {Giustina},
  \citenamefont {Graff}, \citenamefont {Guerin}, \citenamefont {Habegger},
  \citenamefont {Harrigan}, \citenamefont {Hartmann}, \citenamefont {Ho},
  \citenamefont {Hoffmann}, \citenamefont {Huang}, \citenamefont {Humble},
  \citenamefont {Isakov}, \citenamefont {Jeffrey}, \citenamefont {Jiang},
  \citenamefont {Kafri}, \citenamefont {Kechedzhi}, \citenamefont {Kelly},
  \citenamefont {Klimov}, \citenamefont {Knysh}, \citenamefont {Korotkov},
  \citenamefont {Kostritsa}, \citenamefont {Landhuis}, \citenamefont
  {Lindmark}, \citenamefont {Lucero}, \citenamefont {Lyakh}, \citenamefont
  {Mandr{\`{a}}}, \citenamefont {McClean}, \citenamefont {McEwen},
  \citenamefont {Megrant}, \citenamefont {Mi}, \citenamefont {Michielsen},
  \citenamefont {Mohseni}, \citenamefont {Mutus}, \citenamefont {Naaman},
  \citenamefont {Neeley}, \citenamefont {Neill}, \citenamefont {Niu},
  \citenamefont {Ostby}, \citenamefont {Petukhov}, \citenamefont {Platt},
  \citenamefont {Quintana}, \citenamefont {Rieffel}, \citenamefont {Roushan},
  \citenamefont {Rubin}, \citenamefont {Sank}, \citenamefont {Satzinger},
  \citenamefont {Smelyanskiy}, \citenamefont {Sung}, \citenamefont
  {Trevithick}, \citenamefont {Vainsencher}, \citenamefont {Villalonga},
  \citenamefont {White}, \citenamefont {Yao}, \citenamefont {Yeh},
  \citenamefont {Zalcman}, \citenamefont {Neven},\ and\ \citenamefont
  {Martinis}}]{Arute_2019}%
  \BibitemOpen
  \bibfield  {author} {\bibinfo {author} {\bibfnamefont {F.}~\bibnamefont
  {Arute}}, \bibinfo {author} {\bibfnamefont {K.}~\bibnamefont {Arya}},
  \bibinfo {author} {\bibfnamefont {R.}~\bibnamefont {Babbush}}, \bibinfo
  {author} {\bibfnamefont {D.}~\bibnamefont {Bacon}}, \bibinfo {author}
  {\bibfnamefont {J.~C.}\ \bibnamefont {Bardin}}, \bibinfo {author}
  {\bibfnamefont {R.}~\bibnamefont {Barends}}, \bibinfo {author} {\bibfnamefont
  {R.}~\bibnamefont {Biswas}}, \bibinfo {author} {\bibfnamefont
  {S.}~\bibnamefont {Boixo}}, \bibinfo {author} {\bibfnamefont {F.~G. S.~L.}\
  \bibnamefont {Brandao}}, \bibinfo {author} {\bibfnamefont {D.~A.}\
  \bibnamefont {Buell}}, \bibinfo {author} {\bibfnamefont {B.}~\bibnamefont
  {Burkett}}, \bibinfo {author} {\bibfnamefont {Y.}~\bibnamefont {Chen}},
  \bibinfo {author} {\bibfnamefont {Z.}~\bibnamefont {Chen}}, \bibinfo {author}
  {\bibfnamefont {B.}~\bibnamefont {Chiaro}}, \bibinfo {author} {\bibfnamefont
  {R.}~\bibnamefont {Collins}}, \bibinfo {author} {\bibfnamefont
  {W.}~\bibnamefont {Courtney}}, \bibinfo {author} {\bibfnamefont
  {A.}~\bibnamefont {Dunsworth}}, \bibinfo {author} {\bibfnamefont
  {E.}~\bibnamefont {Farhi}}, \bibinfo {author} {\bibfnamefont
  {B.}~\bibnamefont {Foxen}}, \bibinfo {author} {\bibfnamefont
  {A.}~\bibnamefont {Fowler}}, \bibinfo {author} {\bibfnamefont
  {C.}~\bibnamefont {Gidney}}, \bibinfo {author} {\bibfnamefont
  {M.}~\bibnamefont {Giustina}}, \bibinfo {author} {\bibfnamefont
  {R.}~\bibnamefont {Graff}}, \bibinfo {author} {\bibfnamefont
  {K.}~\bibnamefont {Guerin}}, \bibinfo {author} {\bibfnamefont
  {S.}~\bibnamefont {Habegger}}, \bibinfo {author} {\bibfnamefont {M.~P.}\
  \bibnamefont {Harrigan}}, \bibinfo {author} {\bibfnamefont {M.~J.}\
  \bibnamefont {Hartmann}}, \bibinfo {author} {\bibfnamefont {A.}~\bibnamefont
  {Ho}}, \bibinfo {author} {\bibfnamefont {M.}~\bibnamefont {Hoffmann}},
  \bibinfo {author} {\bibfnamefont {T.}~\bibnamefont {Huang}}, \bibinfo
  {author} {\bibfnamefont {T.~S.}\ \bibnamefont {Humble}}, \bibinfo {author}
  {\bibfnamefont {S.~V.}\ \bibnamefont {Isakov}}, \bibinfo {author}
  {\bibfnamefont {E.}~\bibnamefont {Jeffrey}}, \bibinfo {author} {\bibfnamefont
  {Z.}~\bibnamefont {Jiang}}, \bibinfo {author} {\bibfnamefont
  {D.}~\bibnamefont {Kafri}}, \bibinfo {author} {\bibfnamefont
  {K.}~\bibnamefont {Kechedzhi}}, \bibinfo {author} {\bibfnamefont
  {J.}~\bibnamefont {Kelly}}, \bibinfo {author} {\bibfnamefont {P.~V.}\
  \bibnamefont {Klimov}}, \bibinfo {author} {\bibfnamefont {S.}~\bibnamefont
  {Knysh}}, \bibinfo {author} {\bibfnamefont {A.}~\bibnamefont {Korotkov}},
  \bibinfo {author} {\bibfnamefont {F.}~\bibnamefont {Kostritsa}}, \bibinfo
  {author} {\bibfnamefont {D.}~\bibnamefont {Landhuis}}, \bibinfo {author}
  {\bibfnamefont {M.}~\bibnamefont {Lindmark}}, \bibinfo {author}
  {\bibfnamefont {E.}~\bibnamefont {Lucero}}, \bibinfo {author} {\bibfnamefont
  {D.}~\bibnamefont {Lyakh}}, \bibinfo {author} {\bibfnamefont
  {S.}~\bibnamefont {Mandr{\`{a}}}}, \bibinfo {author} {\bibfnamefont {J.~R.}\
  \bibnamefont {McClean}}, \bibinfo {author} {\bibfnamefont {M.}~\bibnamefont
  {McEwen}}, \bibinfo {author} {\bibfnamefont {A.}~\bibnamefont {Megrant}},
  \bibinfo {author} {\bibfnamefont {X.}~\bibnamefont {Mi}}, \bibinfo {author}
  {\bibfnamefont {K.}~\bibnamefont {Michielsen}}, \bibinfo {author}
  {\bibfnamefont {M.}~\bibnamefont {Mohseni}}, \bibinfo {author} {\bibfnamefont
  {J.}~\bibnamefont {Mutus}}, \bibinfo {author} {\bibfnamefont
  {O.}~\bibnamefont {Naaman}}, \bibinfo {author} {\bibfnamefont
  {M.}~\bibnamefont {Neeley}}, \bibinfo {author} {\bibfnamefont
  {C.}~\bibnamefont {Neill}}, \bibinfo {author} {\bibfnamefont {M.~Y.}\
  \bibnamefont {Niu}}, \bibinfo {author} {\bibfnamefont {E.}~\bibnamefont
  {Ostby}}, \bibinfo {author} {\bibfnamefont {A.}~\bibnamefont {Petukhov}},
  \bibinfo {author} {\bibfnamefont {J.~C.}\ \bibnamefont {Platt}}, \bibinfo
  {author} {\bibfnamefont {C.}~\bibnamefont {Quintana}}, \bibinfo {author}
  {\bibfnamefont {E.~G.}\ \bibnamefont {Rieffel}}, \bibinfo {author}
  {\bibfnamefont {P.}~\bibnamefont {Roushan}}, \bibinfo {author} {\bibfnamefont
  {N.~C.}\ \bibnamefont {Rubin}}, \bibinfo {author} {\bibfnamefont
  {D.}~\bibnamefont {Sank}}, \bibinfo {author} {\bibfnamefont {K.~J.}\
  \bibnamefont {Satzinger}}, \bibinfo {author} {\bibfnamefont {V.}~\bibnamefont
  {Smelyanskiy}}, \bibinfo {author} {\bibfnamefont {K.~J.}\ \bibnamefont
  {Sung}}, \bibinfo {author} {\bibfnamefont {M.~D.}\ \bibnamefont
  {Trevithick}}, \bibinfo {author} {\bibfnamefont {A.}~\bibnamefont
  {Vainsencher}}, \bibinfo {author} {\bibfnamefont {B.}~\bibnamefont
  {Villalonga}}, \bibinfo {author} {\bibfnamefont {T.}~\bibnamefont {White}},
  \bibinfo {author} {\bibfnamefont {Z.~J.}\ \bibnamefont {Yao}}, \bibinfo
  {author} {\bibfnamefont {P.}~\bibnamefont {Yeh}}, \bibinfo {author}
  {\bibfnamefont {A.}~\bibnamefont {Zalcman}}, \bibinfo {author} {\bibfnamefont
  {H.}~\bibnamefont {Neven}},\ and\ \bibinfo {author} {\bibfnamefont {J.~M.}\
  \bibnamefont {Martinis}},\ }\href {https://doi.org/10.1038/s41586-019-1666-5}
  {\bibfield  {journal} {\bibinfo  {journal} {Nature}\ }\textbf {\bibinfo
  {volume} {574}},\ \bibinfo {pages} {505} (\bibinfo {year}
  {2019})}\BibitemShut {NoStop}%
\bibitem [{\citenamefont {Muralidharan}\ \emph {et~al.}(2016)\citenamefont
  {Muralidharan}, \citenamefont {Li}, \citenamefont {Kim}, \citenamefont
  {Lütkenhaus}, \citenamefont {Lukin},\ and\ \citenamefont
  {Jiang}}]{Muralidharan_2016}%
  \BibitemOpen
  \bibfield  {author} {\bibinfo {author} {\bibfnamefont {S.}~\bibnamefont
  {Muralidharan}}, \bibinfo {author} {\bibfnamefont {L.}~\bibnamefont {Li}},
  \bibinfo {author} {\bibfnamefont {J.}~\bibnamefont {Kim}}, \bibinfo {author}
  {\bibfnamefont {N.}~\bibnamefont {Lütkenhaus}}, \bibinfo {author}
  {\bibfnamefont {M.~D.}\ \bibnamefont {Lukin}},\ and\ \bibinfo {author}
  {\bibfnamefont {L.}~\bibnamefont {Jiang}},\ }\href
  {https://doi.org/10.1038/srep20463} {\bibfield  {journal} {\bibinfo
  {journal} {Scientific Reports}\ }\textbf {\bibinfo {volume} {6}},\ \bibinfo
  {pages} {20463} (\bibinfo {year} {2016})}\BibitemShut {NoStop}%
\bibitem [{\citenamefont {Awschalom}\ \emph {et~al.}(2021)\citenamefont
  {Awschalom}, \citenamefont {Berggren}, \citenamefont {Bernien}, \citenamefont
  {Bhave}, \citenamefont {Carr}, \citenamefont {Davids}, \citenamefont
  {Economou}, \citenamefont {Englund}, \citenamefont {Faraon}, \citenamefont
  {Fejer}, \citenamefont {Guha}, \citenamefont {Gustafsson}, \citenamefont
  {Hu}, \citenamefont {Jiang}, \citenamefont {Kim}, \citenamefont {Korzh},
  \citenamefont {Kumar}, \citenamefont {Kwiat}, \citenamefont {Lon{\v{c}}ar},
  \citenamefont {Lukin}, \citenamefont {Miller}, \citenamefont {Monroe},
  \citenamefont {Nam}, \citenamefont {Narang}, \citenamefont {Orcutt},
  \citenamefont {Raymer}, \citenamefont {Safavi-Naeini}, \citenamefont
  {Spiropulu}, \citenamefont {Srinivasan}, \citenamefont {Sun}, \citenamefont
  {Vu{\v{c}}kovi{\'{c}}}, \citenamefont {Waks}, \citenamefont {Walsworth},
  \citenamefont {Weiner},\ and\ \citenamefont {Zhang}}]{Awschalom_2021}%
  \BibitemOpen
  \bibfield  {author} {\bibinfo {author} {\bibfnamefont {D.}~\bibnamefont
  {Awschalom}}, \bibinfo {author} {\bibfnamefont {K.~K.}\ \bibnamefont
  {Berggren}}, \bibinfo {author} {\bibfnamefont {H.}~\bibnamefont {Bernien}},
  \bibinfo {author} {\bibfnamefont {S.}~\bibnamefont {Bhave}}, \bibinfo
  {author} {\bibfnamefont {L.~D.}\ \bibnamefont {Carr}}, \bibinfo {author}
  {\bibfnamefont {P.}~\bibnamefont {Davids}}, \bibinfo {author} {\bibfnamefont
  {S.~E.}\ \bibnamefont {Economou}}, \bibinfo {author} {\bibfnamefont
  {D.}~\bibnamefont {Englund}}, \bibinfo {author} {\bibfnamefont
  {A.}~\bibnamefont {Faraon}}, \bibinfo {author} {\bibfnamefont
  {M.}~\bibnamefont {Fejer}}, \bibinfo {author} {\bibfnamefont
  {S.}~\bibnamefont {Guha}}, \bibinfo {author} {\bibfnamefont {M.~V.}\
  \bibnamefont {Gustafsson}}, \bibinfo {author} {\bibfnamefont
  {E.}~\bibnamefont {Hu}}, \bibinfo {author} {\bibfnamefont {L.}~\bibnamefont
  {Jiang}}, \bibinfo {author} {\bibfnamefont {J.}~\bibnamefont {Kim}}, \bibinfo
  {author} {\bibfnamefont {B.}~\bibnamefont {Korzh}}, \bibinfo {author}
  {\bibfnamefont {P.}~\bibnamefont {Kumar}}, \bibinfo {author} {\bibfnamefont
  {P.~G.}\ \bibnamefont {Kwiat}}, \bibinfo {author} {\bibfnamefont
  {M.}~\bibnamefont {Lon{\v{c}}ar}}, \bibinfo {author} {\bibfnamefont {M.~D.}\
  \bibnamefont {Lukin}}, \bibinfo {author} {\bibfnamefont {D.~A.}\ \bibnamefont
  {Miller}}, \bibinfo {author} {\bibfnamefont {C.}~\bibnamefont {Monroe}},
  \bibinfo {author} {\bibfnamefont {S.~W.}\ \bibnamefont {Nam}}, \bibinfo
  {author} {\bibfnamefont {P.}~\bibnamefont {Narang}}, \bibinfo {author}
  {\bibfnamefont {J.~S.}\ \bibnamefont {Orcutt}}, \bibinfo {author}
  {\bibfnamefont {M.~G.}\ \bibnamefont {Raymer}}, \bibinfo {author}
  {\bibfnamefont {A.~H.}\ \bibnamefont {Safavi-Naeini}}, \bibinfo {author}
  {\bibfnamefont {M.}~\bibnamefont {Spiropulu}}, \bibinfo {author}
  {\bibfnamefont {K.}~\bibnamefont {Srinivasan}}, \bibinfo {author}
  {\bibfnamefont {S.}~\bibnamefont {Sun}}, \bibinfo {author} {\bibfnamefont
  {J.}~\bibnamefont {Vu{\v{c}}kovi{\'{c}}}}, \bibinfo {author} {\bibfnamefont
  {E.}~\bibnamefont {Waks}}, \bibinfo {author} {\bibfnamefont {R.}~\bibnamefont
  {Walsworth}}, \bibinfo {author} {\bibfnamefont {A.~M.}\ \bibnamefont
  {Weiner}},\ and\ \bibinfo {author} {\bibfnamefont {Z.}~\bibnamefont
  {Zhang}},\ }\href {https://doi.org/10.1103/prxquantum.2.017002} {\bibfield
  {journal} {\bibinfo  {journal} {{PRX} Quantum}\ }\textbf {\bibinfo {volume}
  {2}},\ \bibinfo {pages} {017002} (\bibinfo {year} {2021})}\BibitemShut
  {NoStop}%
\bibitem [{\citenamefont {Riechers}\ \emph {et~al.}(2022)\citenamefont
  {Riechers}, \citenamefont {Weiss}, \citenamefont {Walter}, \citenamefont
  {Carilli}, \citenamefont {Cox}, \citenamefont {Decarli},\ and\ \citenamefont
  {Neri}}]{Riechers_2022}%
  \BibitemOpen
  \bibfield  {author} {\bibinfo {author} {\bibfnamefont {D.~A.}\ \bibnamefont
  {Riechers}}, \bibinfo {author} {\bibfnamefont {A.}~\bibnamefont {Weiss}},
  \bibinfo {author} {\bibfnamefont {F.}~\bibnamefont {Walter}}, \bibinfo
  {author} {\bibfnamefont {C.~L.}\ \bibnamefont {Carilli}}, \bibinfo {author}
  {\bibfnamefont {P.}~\bibnamefont {Cox}}, \bibinfo {author} {\bibfnamefont
  {R.}~\bibnamefont {Decarli}},\ and\ \bibinfo {author} {\bibfnamefont
  {R.}~\bibnamefont {Neri}},\ }\href
  {https://doi.org/10.1038/s41586-021-04294-5} {\bibfield  {journal} {\bibinfo
  {journal} {Nature}\ }\textbf {\bibinfo {volume} {602}},\ \bibinfo {pages}
  {58} (\bibinfo {year} {2022})}\BibitemShut {NoStop}%
\bibitem [{\citenamefont {Pankratov}\ \emph {et~al.}(2022)\citenamefont
  {Pankratov}, \citenamefont {Revin}, \citenamefont {Gordeeva}, \citenamefont
  {Yablokov}, \citenamefont {Kuzmin},\ and\ \citenamefont
  {Il'ichev}}]{Pankratov_2022}%
  \BibitemOpen
  \bibfield  {author} {\bibinfo {author} {\bibfnamefont {A.~L.}\ \bibnamefont
  {Pankratov}}, \bibinfo {author} {\bibfnamefont {L.~S.}\ \bibnamefont
  {Revin}}, \bibinfo {author} {\bibfnamefont {A.~V.}\ \bibnamefont {Gordeeva}},
  \bibinfo {author} {\bibfnamefont {A.~A.}\ \bibnamefont {Yablokov}}, \bibinfo
  {author} {\bibfnamefont {L.~S.}\ \bibnamefont {Kuzmin}},\ and\ \bibinfo
  {author} {\bibfnamefont {E.}~\bibnamefont {Il'ichev}},\ }\href
  {https://doi.org/10.1038/s41534-022-00569-5} {\bibfield  {journal} {\bibinfo
  {journal} {npj Quantum Information}\ }\textbf {\bibinfo {volume} {8}},\
  \bibinfo {pages} {61} (\bibinfo {year} {2022})}\BibitemShut {NoStop}%
\bibitem [{\citenamefont {Komatsu}(2022)}]{Komatsu_2022}%
  \BibitemOpen
  \bibfield  {author} {\bibinfo {author} {\bibfnamefont {E.}~\bibnamefont
  {Komatsu}},\ }\href {https://doi.org/10.1038/s42254-022-00452-4} {\bibfield
  {journal} {\bibinfo  {journal} {Nature Reviews Physics}\ }\textbf {\bibinfo
  {volume} {4}},\ \bibinfo {pages} {452} (\bibinfo {year} {2022})}\BibitemShut
  {NoStop}%
\bibitem [{\citenamefont {Greaves}\ \emph {et~al.}(2018)\citenamefont
  {Greaves}, \citenamefont {Scaife}, \citenamefont {Frayer}, \citenamefont
  {Green}, \citenamefont {Mason},\ and\ \citenamefont {Smith}}]{Greaves_2018}%
  \BibitemOpen
  \bibfield  {author} {\bibinfo {author} {\bibfnamefont {J.~S.}\ \bibnamefont
  {Greaves}}, \bibinfo {author} {\bibfnamefont {A.~M.~M.}\ \bibnamefont
  {Scaife}}, \bibinfo {author} {\bibfnamefont {D.~T.}\ \bibnamefont {Frayer}},
  \bibinfo {author} {\bibfnamefont {D.~A.}\ \bibnamefont {Green}}, \bibinfo
  {author} {\bibfnamefont {B.~S.}\ \bibnamefont {Mason}},\ and\ \bibinfo
  {author} {\bibfnamefont {A.~M.~S.}\ \bibnamefont {Smith}},\ }\href
  {https://doi.org/10.1038/s41550-018-0495-z} {\bibfield  {journal} {\bibinfo
  {journal} {Nature Astronomy}\ }\textbf {\bibinfo {volume} {2}},\ \bibinfo
  {pages} {662} (\bibinfo {year} {2018})}\BibitemShut {NoStop}%
\bibitem [{\citenamefont {Gorski}\ \emph {et~al.}(2021)\citenamefont {Gorski},
  \citenamefont {Aalto}, \citenamefont {Mangum}, \citenamefont {Momjian},
  \citenamefont {Black}, \citenamefont {Falstad}, \citenamefont {Gullberg},
  \citenamefont {König}, \citenamefont {Onishi}, \citenamefont {Sato},\ and\
  \citenamefont {Stanley}}]{Gorski_2021}%
  \BibitemOpen
  \bibfield  {author} {\bibinfo {author} {\bibfnamefont {M.~D.}\ \bibnamefont
  {Gorski}}, \bibinfo {author} {\bibfnamefont {S.}~\bibnamefont {Aalto}},
  \bibinfo {author} {\bibfnamefont {J.}~\bibnamefont {Mangum}}, \bibinfo
  {author} {\bibfnamefont {E.}~\bibnamefont {Momjian}}, \bibinfo {author}
  {\bibfnamefont {J.~H.}\ \bibnamefont {Black}}, \bibinfo {author}
  {\bibfnamefont {N.}~\bibnamefont {Falstad}}, \bibinfo {author} {\bibfnamefont
  {B.}~\bibnamefont {Gullberg}}, \bibinfo {author} {\bibfnamefont
  {S.}~\bibnamefont {König}}, \bibinfo {author} {\bibfnamefont
  {K.}~\bibnamefont {Onishi}}, \bibinfo {author} {\bibfnamefont
  {M.}~\bibnamefont {Sato}},\ and\ \bibinfo {author} {\bibfnamefont
  {F.}~\bibnamefont {Stanley}},\ }\href
  {https://doi.org/10.1051/0004-6361/202141633} {\bibfield  {journal} {\bibinfo
   {journal} {Astronomy {\&} Astrophysics}\ }\textbf {\bibinfo {volume}
  {654}},\ \bibinfo {pages} {A110} (\bibinfo {year} {2021})}\BibitemShut
  {NoStop}%
\bibitem [{\citenamefont {Kou}\ \emph {et~al.}(2022)\citenamefont {Kou},
  \citenamefont {Cheng}, \citenamefont {Wang}, \citenamefont {Yu},
  \citenamefont {Chen}, \citenamefont {Kontar},\ and\ \citenamefont
  {Ding}}]{Kou_2022}%
  \BibitemOpen
  \bibfield  {author} {\bibinfo {author} {\bibfnamefont {Y.}~\bibnamefont
  {Kou}}, \bibinfo {author} {\bibfnamefont {X.}~\bibnamefont {Cheng}}, \bibinfo
  {author} {\bibfnamefont {Y.}~\bibnamefont {Wang}}, \bibinfo {author}
  {\bibfnamefont {S.}~\bibnamefont {Yu}}, \bibinfo {author} {\bibfnamefont
  {B.}~\bibnamefont {Chen}}, \bibinfo {author} {\bibfnamefont {E.~P.}\
  \bibnamefont {Kontar}},\ and\ \bibinfo {author} {\bibfnamefont
  {M.}~\bibnamefont {Ding}},\ }\href
  {https://doi.org/10.1038/s41467-022-35377-0} {\bibfield  {journal} {\bibinfo
  {journal} {Nature Communications}\ }\textbf {\bibinfo {volume} {13}},\
  \bibinfo {pages} {7680} (\bibinfo {year} {2022})}\BibitemShut {NoStop}%
\bibitem [{\citenamefont {Wade}\ \emph {et~al.}(2016)\citenamefont {Wade},
  \citenamefont {{\v{S}}ibali{\'{c}}}, \citenamefont {de~Melo}, \citenamefont
  {Kondo}, \citenamefont {Adams},\ and\ \citenamefont
  {Weatherill}}]{Wade_2016}%
  \BibitemOpen
  \bibfield  {author} {\bibinfo {author} {\bibfnamefont {C.~G.}\ \bibnamefont
  {Wade}}, \bibinfo {author} {\bibfnamefont {N.}~\bibnamefont
  {{\v{S}}ibali{\'{c}}}}, \bibinfo {author} {\bibfnamefont {N.~R.}\
  \bibnamefont {de~Melo}}, \bibinfo {author} {\bibfnamefont {J.~M.}\
  \bibnamefont {Kondo}}, \bibinfo {author} {\bibfnamefont {C.~S.}\ \bibnamefont
  {Adams}},\ and\ \bibinfo {author} {\bibfnamefont {K.~J.}\ \bibnamefont
  {Weatherill}},\ }\href {https://doi.org/10.1038/nphoton.2016.214} {\bibfield
  {journal} {\bibinfo  {journal} {Nature Photonics}\ }\textbf {\bibinfo
  {volume} {11}},\ \bibinfo {pages} {40} (\bibinfo {year} {2016})}\BibitemShut
  {NoStop}%
\bibitem [{\citenamefont {Xia}\ \emph {et~al.}(2020)\citenamefont {Xia},
  \citenamefont {Li}, \citenamefont {Clark}, \citenamefont {Hart},
  \citenamefont {Zhuang},\ and\ \citenamefont {Zhang}}]{Xia_2020}%
  \BibitemOpen
  \bibfield  {author} {\bibinfo {author} {\bibfnamefont {Y.}~\bibnamefont
  {Xia}}, \bibinfo {author} {\bibfnamefont {W.}~\bibnamefont {Li}}, \bibinfo
  {author} {\bibfnamefont {W.}~\bibnamefont {Clark}}, \bibinfo {author}
  {\bibfnamefont {D.}~\bibnamefont {Hart}}, \bibinfo {author} {\bibfnamefont
  {Q.}~\bibnamefont {Zhuang}},\ and\ \bibinfo {author} {\bibfnamefont
  {Z.}~\bibnamefont {Zhang}},\ }\href
  {https://doi.org/10.1103/physrevlett.124.150502} {\bibfield  {journal}
  {\bibinfo  {journal} {Physical Review Letters}\ }\textbf {\bibinfo {volume}
  {124}},\ \bibinfo {pages} {150502} (\bibinfo {year} {2020})}\BibitemShut
  {NoStop}%
\bibitem [{\citenamefont {Bochmann}\ \emph {et~al.}(2013)\citenamefont
  {Bochmann}, \citenamefont {Vainsencher}, \citenamefont {Awschalom},\ and\
  \citenamefont {Cleland}}]{Bochmann_2013}%
  \BibitemOpen
  \bibfield  {author} {\bibinfo {author} {\bibfnamefont {J.}~\bibnamefont
  {Bochmann}}, \bibinfo {author} {\bibfnamefont {A.}~\bibnamefont
  {Vainsencher}}, \bibinfo {author} {\bibfnamefont {D.~D.}\ \bibnamefont
  {Awschalom}},\ and\ \bibinfo {author} {\bibfnamefont {A.~N.}\ \bibnamefont
  {Cleland}},\ }\href {https://doi.org/10.1038/nphys2748} {\bibfield  {journal}
  {\bibinfo  {journal} {Nature Physics}\ }\textbf {\bibinfo {volume} {9}},\
  \bibinfo {pages} {712} (\bibinfo {year} {2013})}\BibitemShut {NoStop}%
\bibitem [{\citenamefont {Forsch}\ \emph {et~al.}(2019)\citenamefont {Forsch},
  \citenamefont {Stockill}, \citenamefont {Wallucks}, \citenamefont
  {Marinkovi{\'{c}}}, \citenamefont {Gärtner}, \citenamefont {Norte},
  \citenamefont {van Otten}, \citenamefont {Fiore}, \citenamefont
  {Srinivasan},\ and\ \citenamefont {Gröblacher}}]{Forsch_2019}%
  \BibitemOpen
  \bibfield  {author} {\bibinfo {author} {\bibfnamefont {M.}~\bibnamefont
  {Forsch}}, \bibinfo {author} {\bibfnamefont {R.}~\bibnamefont {Stockill}},
  \bibinfo {author} {\bibfnamefont {A.}~\bibnamefont {Wallucks}}, \bibinfo
  {author} {\bibfnamefont {I.}~\bibnamefont {Marinkovi{\'{c}}}}, \bibinfo
  {author} {\bibfnamefont {C.}~\bibnamefont {Gärtner}}, \bibinfo {author}
  {\bibfnamefont {R.~A.}\ \bibnamefont {Norte}}, \bibinfo {author}
  {\bibfnamefont {F.}~\bibnamefont {van Otten}}, \bibinfo {author}
  {\bibfnamefont {A.}~\bibnamefont {Fiore}}, \bibinfo {author} {\bibfnamefont
  {K.}~\bibnamefont {Srinivasan}},\ and\ \bibinfo {author} {\bibfnamefont
  {S.}~\bibnamefont {Gröblacher}},\ }\href
  {https://doi.org/10.1038/s41567-019-0673-7} {\bibfield  {journal} {\bibinfo
  {journal} {Nature Physics}\ }\textbf {\bibinfo {volume} {16}},\ \bibinfo
  {pages} {69} (\bibinfo {year} {2019})}\BibitemShut {NoStop}%
\bibitem [{\citenamefont {Jiang}\ \emph {et~al.}(2020)\citenamefont {Jiang},
  \citenamefont {Sarabalis}, \citenamefont {Dahmani}, \citenamefont {Patel},
  \citenamefont {Mayor}, \citenamefont {McKenna}, \citenamefont {Laer},\ and\
  \citenamefont {Safavi-Naeini}}]{Jiang_2020}%
  \BibitemOpen
  \bibfield  {author} {\bibinfo {author} {\bibfnamefont {W.}~\bibnamefont
  {Jiang}}, \bibinfo {author} {\bibfnamefont {C.~J.}\ \bibnamefont
  {Sarabalis}}, \bibinfo {author} {\bibfnamefont {Y.~D.}\ \bibnamefont
  {Dahmani}}, \bibinfo {author} {\bibfnamefont {R.~N.}\ \bibnamefont {Patel}},
  \bibinfo {author} {\bibfnamefont {F.~M.}\ \bibnamefont {Mayor}}, \bibinfo
  {author} {\bibfnamefont {T.~P.}\ \bibnamefont {McKenna}}, \bibinfo {author}
  {\bibfnamefont {R.~V.}\ \bibnamefont {Laer}},\ and\ \bibinfo {author}
  {\bibfnamefont {A.~H.}\ \bibnamefont {Safavi-Naeini}},\ }\href
  {https://doi.org/10.1038/s41467-020-14863-3} {\bibfield  {journal} {\bibinfo
  {journal} {Nature Communications}\ }\textbf {\bibinfo {volume} {11}},\
  \bibinfo {pages} {1166} (\bibinfo {year} {2020})}\BibitemShut {NoStop}%
\bibitem [{\citenamefont {Mirhosseini}\ \emph {et~al.}(2020)\citenamefont
  {Mirhosseini}, \citenamefont {Sipahigil}, \citenamefont {Kalaee},\ and\
  \citenamefont {Painter}}]{Mirhosseini_2020}%
  \BibitemOpen
  \bibfield  {author} {\bibinfo {author} {\bibfnamefont {M.}~\bibnamefont
  {Mirhosseini}}, \bibinfo {author} {\bibfnamefont {A.}~\bibnamefont
  {Sipahigil}}, \bibinfo {author} {\bibfnamefont {M.}~\bibnamefont {Kalaee}},\
  and\ \bibinfo {author} {\bibfnamefont {O.}~\bibnamefont {Painter}},\ }\href
  {https://doi.org/10.1038/s41586-020-3038-6} {\bibfield  {journal} {\bibinfo
  {journal} {Nature}\ }\textbf {\bibinfo {volume} {588}},\ \bibinfo {pages}
  {599} (\bibinfo {year} {2020})}\BibitemShut {NoStop}%
\bibitem [{\citenamefont {Hönl}\ \emph {et~al.}(2022)\citenamefont {Hönl},
  \citenamefont {Popoff}, \citenamefont {Caimi}, \citenamefont {Beccari},
  \citenamefont {Kippenberg},\ and\ \citenamefont {Seidler}}]{Honl_2022}%
  \BibitemOpen
  \bibfield  {author} {\bibinfo {author} {\bibfnamefont {S.}~\bibnamefont
  {Hönl}}, \bibinfo {author} {\bibfnamefont {Y.}~\bibnamefont {Popoff}},
  \bibinfo {author} {\bibfnamefont {D.}~\bibnamefont {Caimi}}, \bibinfo
  {author} {\bibfnamefont {A.}~\bibnamefont {Beccari}}, \bibinfo {author}
  {\bibfnamefont {T.~J.}\ \bibnamefont {Kippenberg}},\ and\ \bibinfo {author}
  {\bibfnamefont {P.}~\bibnamefont {Seidler}},\ }\href
  {https://doi.org/10.1038/s41467-022-28670-5} {\bibfield  {journal} {\bibinfo
  {journal} {Nature Communications}\ }\textbf {\bibinfo {volume} {13}},\
  \bibinfo {pages} {2065} (\bibinfo {year} {2022})}\BibitemShut {NoStop}%
\bibitem [{\citenamefont {Stockill}\ \emph {et~al.}(2022)\citenamefont
  {Stockill}, \citenamefont {Forsch}, \citenamefont {Hijazi}, \citenamefont
  {Beaudoin}, \citenamefont {Pantzas}, \citenamefont {Sagnes}, \citenamefont
  {Braive},\ and\ \citenamefont {Gröblacher}}]{Stockill_2022}%
  \BibitemOpen
  \bibfield  {author} {\bibinfo {author} {\bibfnamefont {R.}~\bibnamefont
  {Stockill}}, \bibinfo {author} {\bibfnamefont {M.}~\bibnamefont {Forsch}},
  \bibinfo {author} {\bibfnamefont {F.}~\bibnamefont {Hijazi}}, \bibinfo
  {author} {\bibfnamefont {G.}~\bibnamefont {Beaudoin}}, \bibinfo {author}
  {\bibfnamefont {K.}~\bibnamefont {Pantzas}}, \bibinfo {author} {\bibfnamefont
  {I.}~\bibnamefont {Sagnes}}, \bibinfo {author} {\bibfnamefont
  {R.}~\bibnamefont {Braive}},\ and\ \bibinfo {author} {\bibfnamefont
  {S.}~\bibnamefont {Gröblacher}},\ }\href
  {https://doi.org/10.1038/s41467-022-34338-x} {\bibfield  {journal} {\bibinfo
  {journal} {Nature Communications}\ }\textbf {\bibinfo {volume} {13}},\
  \bibinfo {pages} {6583} (\bibinfo {year} {2022})}\BibitemShut {NoStop}%
\bibitem [{\citenamefont {Andrews}\ \emph {et~al.}(2014)\citenamefont
  {Andrews}, \citenamefont {Peterson}, \citenamefont {Purdy}, \citenamefont
  {Cicak}, \citenamefont {Simmonds}, \citenamefont {Regal},\ and\ \citenamefont
  {Lehnert}}]{Andrews_2014}%
  \BibitemOpen
  \bibfield  {author} {\bibinfo {author} {\bibfnamefont {R.~W.}\ \bibnamefont
  {Andrews}}, \bibinfo {author} {\bibfnamefont {R.~W.}\ \bibnamefont
  {Peterson}}, \bibinfo {author} {\bibfnamefont {T.~P.}\ \bibnamefont {Purdy}},
  \bibinfo {author} {\bibfnamefont {K.}~\bibnamefont {Cicak}}, \bibinfo
  {author} {\bibfnamefont {R.~W.}\ \bibnamefont {Simmonds}}, \bibinfo {author}
  {\bibfnamefont {C.~A.}\ \bibnamefont {Regal}},\ and\ \bibinfo {author}
  {\bibfnamefont {K.~W.}\ \bibnamefont {Lehnert}},\ }\href
  {https://doi.org/10.1038/nphys2911} {\bibfield  {journal} {\bibinfo
  {journal} {Nature Physics}\ }\textbf {\bibinfo {volume} {10}},\ \bibinfo
  {pages} {321} (\bibinfo {year} {2014})}\BibitemShut {NoStop}%
\bibitem [{\citenamefont {Peairs}\ \emph {et~al.}(2020)\citenamefont {Peairs},
  \citenamefont {Chou}, \citenamefont {Bienfait}, \citenamefont {Chang},
  \citenamefont {Conner}, \citenamefont {Dumur}, \citenamefont {Grebel},
  \citenamefont {Povey}, \citenamefont {{\c{S}}ahin}, \citenamefont
  {Satzinger}, \citenamefont {Zhong},\ and\ \citenamefont
  {Cleland}}]{Peairs_2020}%
  \BibitemOpen
  \bibfield  {author} {\bibinfo {author} {\bibfnamefont {G.}~\bibnamefont
  {Peairs}}, \bibinfo {author} {\bibfnamefont {M.-H.}\ \bibnamefont {Chou}},
  \bibinfo {author} {\bibfnamefont {A.}~\bibnamefont {Bienfait}}, \bibinfo
  {author} {\bibfnamefont {H.-S.}\ \bibnamefont {Chang}}, \bibinfo {author}
  {\bibfnamefont {C.}~\bibnamefont {Conner}}, \bibinfo {author} {\bibfnamefont
  {{\'{E}}.}~\bibnamefont {Dumur}}, \bibinfo {author} {\bibfnamefont
  {J.}~\bibnamefont {Grebel}}, \bibinfo {author} {\bibfnamefont
  {R.}~\bibnamefont {Povey}}, \bibinfo {author} {\bibfnamefont
  {E.}~\bibnamefont {{\c{S}}ahin}}, \bibinfo {author} {\bibfnamefont
  {K.}~\bibnamefont {Satzinger}}, \bibinfo {author} {\bibfnamefont
  {Y.}~\bibnamefont {Zhong}},\ and\ \bibinfo {author} {\bibfnamefont
  {A.}~\bibnamefont {Cleland}},\ }\href
  {https://doi.org/10.1103/physrevapplied.14.061001} {\bibfield  {journal}
  {\bibinfo  {journal} {Physical Review Applied}\ }\textbf {\bibinfo {volume}
  {14}},\ \bibinfo {pages} {061001} (\bibinfo {year} {2020})}\BibitemShut
  {NoStop}%
\bibitem [{\citenamefont {Arnold}\ \emph {et~al.}(2020)\citenamefont {Arnold},
  \citenamefont {Wulf}, \citenamefont {Barzanjeh}, \citenamefont {Redchenko},
  \citenamefont {Rueda}, \citenamefont {Hease}, \citenamefont {Hassani},\ and\
  \citenamefont {Fink}}]{Arnold_2020}%
  \BibitemOpen
  \bibfield  {author} {\bibinfo {author} {\bibfnamefont {G.}~\bibnamefont
  {Arnold}}, \bibinfo {author} {\bibfnamefont {M.}~\bibnamefont {Wulf}},
  \bibinfo {author} {\bibfnamefont {S.}~\bibnamefont {Barzanjeh}}, \bibinfo
  {author} {\bibfnamefont {E.~S.}\ \bibnamefont {Redchenko}}, \bibinfo {author}
  {\bibfnamefont {A.}~\bibnamefont {Rueda}}, \bibinfo {author} {\bibfnamefont
  {W.~J.}\ \bibnamefont {Hease}}, \bibinfo {author} {\bibfnamefont
  {F.}~\bibnamefont {Hassani}},\ and\ \bibinfo {author} {\bibfnamefont {J.~M.}\
  \bibnamefont {Fink}},\ }\href {https://doi.org/10.1038/s41467-020-18269-z}
  {\bibfield  {journal} {\bibinfo  {journal} {Nature Communications}\ }\textbf
  {\bibinfo {volume} {11}},\ \bibinfo {pages} {4460} (\bibinfo {year}
  {2020})}\BibitemShut {NoStop}%
\bibitem [{\citenamefont {Delaney}\ \emph {et~al.}(2022)\citenamefont
  {Delaney}, \citenamefont {Urmey}, \citenamefont {Mittal}, \citenamefont
  {Brubaker}, \citenamefont {Kindem}, \citenamefont {Burns}, \citenamefont
  {Regal},\ and\ \citenamefont {Lehnert}}]{Delaney_2022}%
  \BibitemOpen
  \bibfield  {author} {\bibinfo {author} {\bibfnamefont {R.~D.}\ \bibnamefont
  {Delaney}}, \bibinfo {author} {\bibfnamefont {M.~D.}\ \bibnamefont {Urmey}},
  \bibinfo {author} {\bibfnamefont {S.}~\bibnamefont {Mittal}}, \bibinfo
  {author} {\bibfnamefont {B.~M.}\ \bibnamefont {Brubaker}}, \bibinfo {author}
  {\bibfnamefont {J.~M.}\ \bibnamefont {Kindem}}, \bibinfo {author}
  {\bibfnamefont {P.~S.}\ \bibnamefont {Burns}}, \bibinfo {author}
  {\bibfnamefont {C.~A.}\ \bibnamefont {Regal}},\ and\ \bibinfo {author}
  {\bibfnamefont {K.~W.}\ \bibnamefont {Lehnert}},\ }\href
  {https://doi.org/10.1038/s41586-022-04720-2} {\bibfield  {journal} {\bibinfo
  {journal} {Nature}\ }\textbf {\bibinfo {volume} {606}},\ \bibinfo {pages}
  {489} (\bibinfo {year} {2022})}\BibitemShut {NoStop}%
\bibitem [{\citenamefont {Hisatomi}\ \emph {et~al.}(2016)\citenamefont
  {Hisatomi}, \citenamefont {Osada}, \citenamefont {Tabuchi}, \citenamefont
  {Ishikawa}, \citenamefont {Noguchi}, \citenamefont {Yamazaki}, \citenamefont
  {Usami},\ and\ \citenamefont {Nakamura}}]{Hisatomi_2016}%
  \BibitemOpen
  \bibfield  {author} {\bibinfo {author} {\bibfnamefont {R.}~\bibnamefont
  {Hisatomi}}, \bibinfo {author} {\bibfnamefont {A.}~\bibnamefont {Osada}},
  \bibinfo {author} {\bibfnamefont {Y.}~\bibnamefont {Tabuchi}}, \bibinfo
  {author} {\bibfnamefont {T.}~\bibnamefont {Ishikawa}}, \bibinfo {author}
  {\bibfnamefont {A.}~\bibnamefont {Noguchi}}, \bibinfo {author} {\bibfnamefont
  {R.}~\bibnamefont {Yamazaki}}, \bibinfo {author} {\bibfnamefont
  {K.}~\bibnamefont {Usami}},\ and\ \bibinfo {author} {\bibfnamefont
  {Y.}~\bibnamefont {Nakamura}},\ }\href
  {https://doi.org/10.1103/physrevb.93.174427} {\bibfield  {journal} {\bibinfo
  {journal} {Physical Review B}\ }\textbf {\bibinfo {volume} {93}},\ \bibinfo
  {pages} {174427} (\bibinfo {year} {2016})}\BibitemShut {NoStop}%
\bibitem [{\citenamefont {Bartholomew}\ \emph {et~al.}(2020)\citenamefont
  {Bartholomew}, \citenamefont {Rochman}, \citenamefont {Xie}, \citenamefont
  {Kindem}, \citenamefont {Ruskuc}, \citenamefont {Craiciu}, \citenamefont
  {Lei},\ and\ \citenamefont {Faraon}}]{Bartholomew_2020}%
  \BibitemOpen
  \bibfield  {author} {\bibinfo {author} {\bibfnamefont {J.~G.}\ \bibnamefont
  {Bartholomew}}, \bibinfo {author} {\bibfnamefont {J.}~\bibnamefont
  {Rochman}}, \bibinfo {author} {\bibfnamefont {T.}~\bibnamefont {Xie}},
  \bibinfo {author} {\bibfnamefont {J.~M.}\ \bibnamefont {Kindem}}, \bibinfo
  {author} {\bibfnamefont {A.}~\bibnamefont {Ruskuc}}, \bibinfo {author}
  {\bibfnamefont {I.}~\bibnamefont {Craiciu}}, \bibinfo {author} {\bibfnamefont
  {M.}~\bibnamefont {Lei}},\ and\ \bibinfo {author} {\bibfnamefont
  {A.}~\bibnamefont {Faraon}},\ }\href
  {https://doi.org/10.1038/s41467-020-16996-x} {\bibfield  {journal} {\bibinfo
  {journal} {Nature Communications}\ }\textbf {\bibinfo {volume} {11}},\
  \bibinfo {pages} {3266} (\bibinfo {year} {2020})}\BibitemShut {NoStop}%
\bibitem [{\citenamefont {Zhu}\ \emph {et~al.}(2020)\citenamefont {Zhu},
  \citenamefont {Zhang}, \citenamefont {Han}, \citenamefont {Zou},
  \citenamefont {Zhong}, \citenamefont {Wang}, \citenamefont {Jiang},\ and\
  \citenamefont {Tang}}]{Zhu_2020}%
  \BibitemOpen
  \bibfield  {author} {\bibinfo {author} {\bibfnamefont {N.}~\bibnamefont
  {Zhu}}, \bibinfo {author} {\bibfnamefont {X.}~\bibnamefont {Zhang}}, \bibinfo
  {author} {\bibfnamefont {X.}~\bibnamefont {Han}}, \bibinfo {author}
  {\bibfnamefont {C.-L.}\ \bibnamefont {Zou}}, \bibinfo {author} {\bibfnamefont
  {C.}~\bibnamefont {Zhong}}, \bibinfo {author} {\bibfnamefont {C.-H.}\
  \bibnamefont {Wang}}, \bibinfo {author} {\bibfnamefont {L.}~\bibnamefont
  {Jiang}},\ and\ \bibinfo {author} {\bibfnamefont {H.~X.}\ \bibnamefont
  {Tang}},\ }\href {https://doi.org/10.1364/optica.397967} {\bibfield
  {journal} {\bibinfo  {journal} {Optica}\ }\textbf {\bibinfo {volume} {7}},\
  \bibinfo {pages} {1291} (\bibinfo {year} {2020})}\BibitemShut {NoStop}%
\bibitem [{\citenamefont {Rueda}\ \emph {et~al.}(2016)\citenamefont {Rueda},
  \citenamefont {Sedlmeir}, \citenamefont {Collodo}, \citenamefont {Vogl},
  \citenamefont {Stiller}, \citenamefont {Schunk}, \citenamefont {Strekalov},
  \citenamefont {Marquardt}, \citenamefont {Fink}, \citenamefont {Painter},
  \citenamefont {Leuchs},\ and\ \citenamefont {Schwefel}}]{Rueda_2016}%
  \BibitemOpen
  \bibfield  {author} {\bibinfo {author} {\bibfnamefont {A.}~\bibnamefont
  {Rueda}}, \bibinfo {author} {\bibfnamefont {F.}~\bibnamefont {Sedlmeir}},
  \bibinfo {author} {\bibfnamefont {M.~C.}\ \bibnamefont {Collodo}}, \bibinfo
  {author} {\bibfnamefont {U.}~\bibnamefont {Vogl}}, \bibinfo {author}
  {\bibfnamefont {B.}~\bibnamefont {Stiller}}, \bibinfo {author} {\bibfnamefont
  {G.}~\bibnamefont {Schunk}}, \bibinfo {author} {\bibfnamefont {D.~V.}\
  \bibnamefont {Strekalov}}, \bibinfo {author} {\bibfnamefont {C.}~\bibnamefont
  {Marquardt}}, \bibinfo {author} {\bibfnamefont {J.~M.}\ \bibnamefont {Fink}},
  \bibinfo {author} {\bibfnamefont {O.}~\bibnamefont {Painter}}, \bibinfo
  {author} {\bibfnamefont {G.}~\bibnamefont {Leuchs}},\ and\ \bibinfo {author}
  {\bibfnamefont {H.~G.~L.}\ \bibnamefont {Schwefel}},\ }\href
  {https://doi.org/10.1364/optica.3.000597} {\bibfield  {journal} {\bibinfo
  {journal} {Optica}\ }\textbf {\bibinfo {volume} {3}},\ \bibinfo {pages} {597}
  (\bibinfo {year} {2016})}\BibitemShut {NoStop}%
\bibitem [{\citenamefont {Witmer}\ \emph {et~al.}(2020)\citenamefont {Witmer},
  \citenamefont {McKenna}, \citenamefont {Arrangoiz-Arriola}, \citenamefont
  {Laer}, \citenamefont {Wollack}, \citenamefont {Lin}, \citenamefont {Jen},
  \citenamefont {Luo},\ and\ \citenamefont {Safavi-Naeini}}]{Witmer_2020}%
  \BibitemOpen
  \bibfield  {author} {\bibinfo {author} {\bibfnamefont {J.~D.}\ \bibnamefont
  {Witmer}}, \bibinfo {author} {\bibfnamefont {T.~P.}\ \bibnamefont {McKenna}},
  \bibinfo {author} {\bibfnamefont {P.}~\bibnamefont {Arrangoiz-Arriola}},
  \bibinfo {author} {\bibfnamefont {R.~V.}\ \bibnamefont {Laer}}, \bibinfo
  {author} {\bibfnamefont {E.~A.}\ \bibnamefont {Wollack}}, \bibinfo {author}
  {\bibfnamefont {F.}~\bibnamefont {Lin}}, \bibinfo {author} {\bibfnamefont
  {A.~K.-Y.}\ \bibnamefont {Jen}}, \bibinfo {author} {\bibfnamefont
  {J.}~\bibnamefont {Luo}},\ and\ \bibinfo {author} {\bibfnamefont {A.~H.}\
  \bibnamefont {Safavi-Naeini}},\ }\href
  {https://doi.org/10.1088/2058-9565/ab7eed} {\bibfield  {journal} {\bibinfo
  {journal} {Quantum Science and Technology}\ }\textbf {\bibinfo {volume}
  {5}},\ \bibinfo {pages} {034004} (\bibinfo {year} {2020})}\BibitemShut
  {NoStop}%
\bibitem [{\citenamefont {McKenna}\ \emph {et~al.}(2020)\citenamefont
  {McKenna}, \citenamefont {Witmer}, \citenamefont {Patel}, \citenamefont
  {Jiang}, \citenamefont {Laer}, \citenamefont {Arrangoiz-Arriola},
  \citenamefont {Wollack}, \citenamefont {Herrmann},\ and\ \citenamefont
  {Safavi-Naeini}}]{McKenna_2020}%
  \BibitemOpen
  \bibfield  {author} {\bibinfo {author} {\bibfnamefont {T.~P.}\ \bibnamefont
  {McKenna}}, \bibinfo {author} {\bibfnamefont {J.~D.}\ \bibnamefont {Witmer}},
  \bibinfo {author} {\bibfnamefont {R.~N.}\ \bibnamefont {Patel}}, \bibinfo
  {author} {\bibfnamefont {W.}~\bibnamefont {Jiang}}, \bibinfo {author}
  {\bibfnamefont {R.~V.}\ \bibnamefont {Laer}}, \bibinfo {author}
  {\bibfnamefont {P.}~\bibnamefont {Arrangoiz-Arriola}}, \bibinfo {author}
  {\bibfnamefont {E.~A.}\ \bibnamefont {Wollack}}, \bibinfo {author}
  {\bibfnamefont {J.~F.}\ \bibnamefont {Herrmann}},\ and\ \bibinfo {author}
  {\bibfnamefont {A.~H.}\ \bibnamefont {Safavi-Naeini}},\ }\href
  {https://doi.org/10.1364/optica.397235} {\bibfield  {journal} {\bibinfo
  {journal} {Optica}\ }\textbf {\bibinfo {volume} {7}},\ \bibinfo {pages}
  {1737} (\bibinfo {year} {2020})}\BibitemShut {NoStop}%
\bibitem [{\citenamefont {Xu}\ \emph {et~al.}(2021)\citenamefont {Xu},
  \citenamefont {Sayem}, \citenamefont {Fan}, \citenamefont {Zou},
  \citenamefont {Wang}, \citenamefont {Cheng}, \citenamefont {Fu},
  \citenamefont {Yang}, \citenamefont {Xu},\ and\ \citenamefont
  {Tang}}]{Xu_2021}%
  \BibitemOpen
  \bibfield  {author} {\bibinfo {author} {\bibfnamefont {Y.}~\bibnamefont
  {Xu}}, \bibinfo {author} {\bibfnamefont {A.~A.}\ \bibnamefont {Sayem}},
  \bibinfo {author} {\bibfnamefont {L.}~\bibnamefont {Fan}}, \bibinfo {author}
  {\bibfnamefont {C.-L.}\ \bibnamefont {Zou}}, \bibinfo {author} {\bibfnamefont
  {S.}~\bibnamefont {Wang}}, \bibinfo {author} {\bibfnamefont {R.}~\bibnamefont
  {Cheng}}, \bibinfo {author} {\bibfnamefont {W.}~\bibnamefont {Fu}}, \bibinfo
  {author} {\bibfnamefont {L.}~\bibnamefont {Yang}}, \bibinfo {author}
  {\bibfnamefont {M.}~\bibnamefont {Xu}},\ and\ \bibinfo {author}
  {\bibfnamefont {H.~X.}\ \bibnamefont {Tang}},\ }\href
  {https://doi.org/10.1038/s41467-021-24809-y} {\bibfield  {journal} {\bibinfo
  {journal} {Nature Communications}\ }\textbf {\bibinfo {volume} {12}},\
  \bibinfo {pages} {4453} (\bibinfo {year} {2021})}\BibitemShut {NoStop}%
\bibitem [{\citenamefont {Sahu}\ \emph {et~al.}(2022)\citenamefont {Sahu},
  \citenamefont {Hease}, \citenamefont {Rueda}, \citenamefont {Arnold},
  \citenamefont {Qiu},\ and\ \citenamefont {Fink}}]{Sahu_2022}%
  \BibitemOpen
  \bibfield  {author} {\bibinfo {author} {\bibfnamefont {R.}~\bibnamefont
  {Sahu}}, \bibinfo {author} {\bibfnamefont {W.}~\bibnamefont {Hease}},
  \bibinfo {author} {\bibfnamefont {A.}~\bibnamefont {Rueda}}, \bibinfo
  {author} {\bibfnamefont {G.}~\bibnamefont {Arnold}}, \bibinfo {author}
  {\bibfnamefont {L.}~\bibnamefont {Qiu}},\ and\ \bibinfo {author}
  {\bibfnamefont {J.~M.}\ \bibnamefont {Fink}},\ }\href
  {https://doi.org/10.1038/s41467-022-28924-2} {\bibfield  {journal} {\bibinfo
  {journal} {Nature Communications}\ }\textbf {\bibinfo {volume} {13}},\
  \bibinfo {pages} {1276} (\bibinfo {year} {2022})}\BibitemShut {NoStop}%
\bibitem [{\citenamefont {Wang}\ \emph {et~al.}(2022)\citenamefont {Wang},
  \citenamefont {Gonin}, \citenamefont {Grassellino}, \citenamefont {Kazakov},
  \citenamefont {Romanenko}, \citenamefont {Yakovlev},\ and\ \citenamefont
  {Zorzetti}}]{Wang_2022}%
  \BibitemOpen
  \bibfield  {author} {\bibinfo {author} {\bibfnamefont {C.}~\bibnamefont
  {Wang}}, \bibinfo {author} {\bibfnamefont {I.}~\bibnamefont {Gonin}},
  \bibinfo {author} {\bibfnamefont {A.}~\bibnamefont {Grassellino}}, \bibinfo
  {author} {\bibfnamefont {S.}~\bibnamefont {Kazakov}}, \bibinfo {author}
  {\bibfnamefont {A.}~\bibnamefont {Romanenko}}, \bibinfo {author}
  {\bibfnamefont {V.~P.}\ \bibnamefont {Yakovlev}},\ and\ \bibinfo {author}
  {\bibfnamefont {S.}~\bibnamefont {Zorzetti}},\ }\href
  {https://doi.org/10.1038/s41534-022-00664-7} {\bibfield  {journal} {\bibinfo
  {journal} {npj Quantum Information}\ }\textbf {\bibinfo {volume} {8}},\
  \bibinfo {pages} {149} (\bibinfo {year} {2022})}\BibitemShut {NoStop}%
\bibitem [{\citenamefont {Lekavicius}\ \emph {et~al.}(2017)\citenamefont
  {Lekavicius}, \citenamefont {Golter}, \citenamefont {Oo},\ and\ \citenamefont
  {Wang}}]{Lekavicius_2017}%
  \BibitemOpen
  \bibfield  {author} {\bibinfo {author} {\bibfnamefont {I.}~\bibnamefont
  {Lekavicius}}, \bibinfo {author} {\bibfnamefont {D.~A.}\ \bibnamefont
  {Golter}}, \bibinfo {author} {\bibfnamefont {T.}~\bibnamefont {Oo}},\ and\
  \bibinfo {author} {\bibfnamefont {H.}~\bibnamefont {Wang}},\ }\href
  {https://doi.org/10.1103/physrevlett.119.063601} {\bibfield  {journal}
  {\bibinfo  {journal} {Physical Review Letters}\ }\textbf {\bibinfo {volume}
  {119}},\ \bibinfo {pages} {063601} (\bibinfo {year} {2017})}\BibitemShut
  {NoStop}%
\bibitem [{\citenamefont {Gallagher}\ \emph {et~al.}(2022)\citenamefont
  {Gallagher}, \citenamefont {Rogers}, \citenamefont {Pritchett}, \citenamefont
  {Mistry}, \citenamefont {Pizzey}, \citenamefont {Adams}, \citenamefont
  {Jones}, \citenamefont {Grünwald}, \citenamefont {Walther}, \citenamefont
  {Hodges}, \citenamefont {Langbein},\ and\ \citenamefont
  {Lynch}}]{Gallagher_2022}%
  \BibitemOpen
  \bibfield  {author} {\bibinfo {author} {\bibfnamefont {L.~A.~P.}\
  \bibnamefont {Gallagher}}, \bibinfo {author} {\bibfnamefont {J.~P.}\
  \bibnamefont {Rogers}}, \bibinfo {author} {\bibfnamefont {J.~D.}\
  \bibnamefont {Pritchett}}, \bibinfo {author} {\bibfnamefont {R.~A.}\
  \bibnamefont {Mistry}}, \bibinfo {author} {\bibfnamefont {D.}~\bibnamefont
  {Pizzey}}, \bibinfo {author} {\bibfnamefont {C.~S.}\ \bibnamefont {Adams}},
  \bibinfo {author} {\bibfnamefont {M.~P.~A.}\ \bibnamefont {Jones}}, \bibinfo
  {author} {\bibfnamefont {P.}~\bibnamefont {Grünwald}}, \bibinfo {author}
  {\bibfnamefont {V.}~\bibnamefont {Walther}}, \bibinfo {author} {\bibfnamefont
  {C.}~\bibnamefont {Hodges}}, \bibinfo {author} {\bibfnamefont
  {W.}~\bibnamefont {Langbein}},\ and\ \bibinfo {author} {\bibfnamefont
  {S.~A.}\ \bibnamefont {Lynch}},\ }\href
  {https://doi.org/10.1103/physrevresearch.4.013031} {\bibfield  {journal}
  {\bibinfo  {journal} {Physical Review Research}\ }\textbf {\bibinfo {volume}
  {4}},\ \bibinfo {pages} {013031} (\bibinfo {year} {2022})}\BibitemShut
  {NoStop}%
\bibitem [{\citenamefont {Han}\ \emph {et~al.}(2018)\citenamefont {Han},
  \citenamefont {Vogt}, \citenamefont {Gross}, \citenamefont {Jaksch},
  \citenamefont {Kiffner},\ and\ \citenamefont {Li}}]{Han_2018}%
  \BibitemOpen
  \bibfield  {author} {\bibinfo {author} {\bibfnamefont {J.}~\bibnamefont
  {Han}}, \bibinfo {author} {\bibfnamefont {T.}~\bibnamefont {Vogt}}, \bibinfo
  {author} {\bibfnamefont {C.}~\bibnamefont {Gross}}, \bibinfo {author}
  {\bibfnamefont {D.}~\bibnamefont {Jaksch}}, \bibinfo {author} {\bibfnamefont
  {M.}~\bibnamefont {Kiffner}},\ and\ \bibinfo {author} {\bibfnamefont
  {W.}~\bibnamefont {Li}},\ }\href
  {https://doi.org/10.1103/physrevlett.120.093201} {\bibfield  {journal}
  {\bibinfo  {journal} {Physical Review Letters}\ }\textbf {\bibinfo {volume}
  {120}},\ \bibinfo {pages} {093201} (\bibinfo {year} {2018})}\BibitemShut
  {NoStop}%
\bibitem [{\citenamefont {Vogt}\ \emph {et~al.}(2019)\citenamefont {Vogt},
  \citenamefont {Gross}, \citenamefont {Han}, \citenamefont {Pal},
  \citenamefont {Lam}, \citenamefont {Kiffner},\ and\ \citenamefont
  {Li}}]{Vogt_2019}%
  \BibitemOpen
  \bibfield  {author} {\bibinfo {author} {\bibfnamefont {T.}~\bibnamefont
  {Vogt}}, \bibinfo {author} {\bibfnamefont {C.}~\bibnamefont {Gross}},
  \bibinfo {author} {\bibfnamefont {J.}~\bibnamefont {Han}}, \bibinfo {author}
  {\bibfnamefont {S.~B.}\ \bibnamefont {Pal}}, \bibinfo {author} {\bibfnamefont
  {M.}~\bibnamefont {Lam}}, \bibinfo {author} {\bibfnamefont {M.}~\bibnamefont
  {Kiffner}},\ and\ \bibinfo {author} {\bibfnamefont {W.}~\bibnamefont {Li}},\
  }\href {https://doi.org/10.1103/physreva.99.023832} {\bibfield  {journal}
  {\bibinfo  {journal} {Physical Review A}\ }\textbf {\bibinfo {volume} {99}},\
  \bibinfo {pages} {023832} (\bibinfo {year} {2019})}\BibitemShut {NoStop}%
\bibitem [{\citenamefont {Tu}\ \emph {et~al.}(2022)\citenamefont {Tu},
  \citenamefont {Liao}, \citenamefont {Zhang}, \citenamefont {Liu},
  \citenamefont {Zheng}, \citenamefont {Yang}, \citenamefont {Zhang},
  \citenamefont {Yan},\ and\ \citenamefont {Zhu}}]{Tu_2022}%
  \BibitemOpen
  \bibfield  {author} {\bibinfo {author} {\bibfnamefont {H.-T.}\ \bibnamefont
  {Tu}}, \bibinfo {author} {\bibfnamefont {K.-Y.}\ \bibnamefont {Liao}},
  \bibinfo {author} {\bibfnamefont {Z.-X.}\ \bibnamefont {Zhang}}, \bibinfo
  {author} {\bibfnamefont {X.-H.}\ \bibnamefont {Liu}}, \bibinfo {author}
  {\bibfnamefont {S.-Y.}\ \bibnamefont {Zheng}}, \bibinfo {author}
  {\bibfnamefont {S.-Z.}\ \bibnamefont {Yang}}, \bibinfo {author}
  {\bibfnamefont {X.-D.}\ \bibnamefont {Zhang}}, \bibinfo {author}
  {\bibfnamefont {H.}~\bibnamefont {Yan}},\ and\ \bibinfo {author}
  {\bibfnamefont {S.-L.}\ \bibnamefont {Zhu}},\ }\href
  {https://doi.org/10.1038/s41566-022-00959-3} {\bibfield  {journal} {\bibinfo
  {journal} {Nature Photonics}\ }\textbf {\bibinfo {volume} {16}},\ \bibinfo
  {pages} {291} (\bibinfo {year} {2022})}\BibitemShut {NoStop}%
\bibitem [{\citenamefont {Kumar}\ \emph {et~al.}(2023)\citenamefont {Kumar},
  \citenamefont {Suleymanzade}, \citenamefont {Stone}, \citenamefont {Taneja},
  \citenamefont {Anferov}, \citenamefont {Schuster},\ and\ \citenamefont
  {Simon}}]{Kumar_2022}%
  \BibitemOpen
  \bibfield  {author} {\bibinfo {author} {\bibfnamefont {A.}~\bibnamefont
  {Kumar}}, \bibinfo {author} {\bibfnamefont {A.}~\bibnamefont {Suleymanzade}},
  \bibinfo {author} {\bibfnamefont {M.}~\bibnamefont {Stone}}, \bibinfo
  {author} {\bibfnamefont {L.}~\bibnamefont {Taneja}}, \bibinfo {author}
  {\bibfnamefont {A.}~\bibnamefont {Anferov}}, \bibinfo {author} {\bibfnamefont
  {D.~I.}\ \bibnamefont {Schuster}},\ and\ \bibinfo {author} {\bibfnamefont
  {J.}~\bibnamefont {Simon}},\ }\href
  {https://doi.org/10.1038/s41586-023-05740-2} {\bibfield  {journal} {\bibinfo
  {journal} {Nature}\ }\textbf {\bibinfo {volume} {615}},\ \bibinfo {pages}
  {614} (\bibinfo {year} {2023})}\BibitemShut {NoStop}%
\bibitem [{\citenamefont {Goy}\ \emph {et~al.}(1980)\citenamefont {Goy},
  \citenamefont {Fabre}, \citenamefont {Gross},\ and\ \citenamefont
  {Haroche}}]{Goy_1980}%
  \BibitemOpen
  \bibfield  {author} {\bibinfo {author} {\bibfnamefont {P.}~\bibnamefont
  {Goy}}, \bibinfo {author} {\bibfnamefont {C.}~\bibnamefont {Fabre}}, \bibinfo
  {author} {\bibfnamefont {M.}~\bibnamefont {Gross}},\ and\ \bibinfo {author}
  {\bibfnamefont {S.}~\bibnamefont {Haroche}},\ }\href
  {https://doi.org/10.1088/0022-3700/13/3/001} {\bibfield  {journal} {\bibinfo
  {journal} {Journal of Physics B: Atomic and Molecular Physics}\ }\textbf
  {\bibinfo {volume} {13}},\ \bibinfo {pages} {L83} (\bibinfo {year}
  {1980})}\BibitemShut {NoStop}%
\bibitem [{\citenamefont {Raimond}\ \emph {et~al.}(1982)\citenamefont
  {Raimond}, \citenamefont {Goy}, \citenamefont {Gross}, \citenamefont
  {Fabre},\ and\ \citenamefont {Haroche}}]{Raimond_1982}%
  \BibitemOpen
  \bibfield  {author} {\bibinfo {author} {\bibfnamefont {J.~M.}\ \bibnamefont
  {Raimond}}, \bibinfo {author} {\bibfnamefont {P.}~\bibnamefont {Goy}},
  \bibinfo {author} {\bibfnamefont {M.}~\bibnamefont {Gross}}, \bibinfo
  {author} {\bibfnamefont {C.}~\bibnamefont {Fabre}},\ and\ \bibinfo {author}
  {\bibfnamefont {S.}~\bibnamefont {Haroche}},\ }\href
  {https://doi.org/10.1103/physrevlett.49.117} {\bibfield  {journal} {\bibinfo
  {journal} {Physical Review Letters}\ }\textbf {\bibinfo {volume} {49}},\
  \bibinfo {pages} {117} (\bibinfo {year} {1982})}\BibitemShut {NoStop}%
\bibitem [{\citenamefont {Meschede}\ \emph {et~al.}(1985)\citenamefont
  {Meschede}, \citenamefont {Walther},\ and\ \citenamefont
  {Müller}}]{Meschede_1985}%
  \BibitemOpen
  \bibfield  {author} {\bibinfo {author} {\bibfnamefont {D.}~\bibnamefont
  {Meschede}}, \bibinfo {author} {\bibfnamefont {H.}~\bibnamefont {Walther}},\
  and\ \bibinfo {author} {\bibfnamefont {G.}~\bibnamefont {Müller}},\ }\href
  {https://doi.org/10.1103/physrevlett.54.551} {\bibfield  {journal} {\bibinfo
  {journal} {Physical Review Letters}\ }\textbf {\bibinfo {volume} {54}},\
  \bibinfo {pages} {551} (\bibinfo {year} {1985})}\BibitemShut {NoStop}%
\bibitem [{\citenamefont {Finkelstein}\ \emph {et~al.}(2018)\citenamefont
  {Finkelstein}, \citenamefont {Poem}, \citenamefont {Michel}, \citenamefont
  {Lahad},\ and\ \citenamefont {Firstenberg}}]{Finkelstein_2018}%
  \BibitemOpen
  \bibfield  {author} {\bibinfo {author} {\bibfnamefont {R.}~\bibnamefont
  {Finkelstein}}, \bibinfo {author} {\bibfnamefont {E.}~\bibnamefont {Poem}},
  \bibinfo {author} {\bibfnamefont {O.}~\bibnamefont {Michel}}, \bibinfo
  {author} {\bibfnamefont {O.}~\bibnamefont {Lahad}},\ and\ \bibinfo {author}
  {\bibfnamefont {O.}~\bibnamefont {Firstenberg}},\ }\href
  {https://doi.org/10.1126/sciadv.aap8598} {\bibfield  {journal} {\bibinfo
  {journal} {Science Advances}\ }\textbf {\bibinfo {volume} {4}},\ \bibinfo
  {pages} {eaap8598} (\bibinfo {year} {2018})}\BibitemShut {NoStop}%
\bibitem [{\citenamefont {Kaczmarek}\ \emph {et~al.}(2018)\citenamefont
  {Kaczmarek}, \citenamefont {Ledingham}, \citenamefont {Brecht}, \citenamefont
  {Thomas}, \citenamefont {Thekkadath}, \citenamefont {Lazo-Arjona},
  \citenamefont {Munns}, \citenamefont {Poem}, \citenamefont {Feizpour},
  \citenamefont {Saunders}, \citenamefont {Nunn},\ and\ \citenamefont
  {Walmsley}}]{Kaczmarek_2018}%
  \BibitemOpen
  \bibfield  {author} {\bibinfo {author} {\bibfnamefont {K.~T.}\ \bibnamefont
  {Kaczmarek}}, \bibinfo {author} {\bibfnamefont {P.~M.}\ \bibnamefont
  {Ledingham}}, \bibinfo {author} {\bibfnamefont {B.}~\bibnamefont {Brecht}},
  \bibinfo {author} {\bibfnamefont {S.~E.}\ \bibnamefont {Thomas}}, \bibinfo
  {author} {\bibfnamefont {G.~S.}\ \bibnamefont {Thekkadath}}, \bibinfo
  {author} {\bibfnamefont {O.}~\bibnamefont {Lazo-Arjona}}, \bibinfo {author}
  {\bibfnamefont {J.~H.~D.}\ \bibnamefont {Munns}}, \bibinfo {author}
  {\bibfnamefont {E.}~\bibnamefont {Poem}}, \bibinfo {author} {\bibfnamefont
  {A.}~\bibnamefont {Feizpour}}, \bibinfo {author} {\bibfnamefont {D.~J.}\
  \bibnamefont {Saunders}}, \bibinfo {author} {\bibfnamefont {J.}~\bibnamefont
  {Nunn}},\ and\ \bibinfo {author} {\bibfnamefont {I.~A.}\ \bibnamefont
  {Walmsley}},\ }\href {https://doi.org/10.1103/physreva.97.042316} {\bibfield
  {journal} {\bibinfo  {journal} {Physical Review A}\ }\textbf {\bibinfo
  {volume} {97}},\ \bibinfo {pages} {042316} (\bibinfo {year}
  {2018})}\BibitemShut {NoStop}%
\bibitem [{\citenamefont {Ripka}\ \emph {et~al.}(2018)\citenamefont {Ripka},
  \citenamefont {Kübler}, \citenamefont {Löw},\ and\ \citenamefont
  {Pfau}}]{Ripka_2018}%
  \BibitemOpen
  \bibfield  {author} {\bibinfo {author} {\bibfnamefont {F.}~\bibnamefont
  {Ripka}}, \bibinfo {author} {\bibfnamefont {H.}~\bibnamefont {Kübler}},
  \bibinfo {author} {\bibfnamefont {R.}~\bibnamefont {Löw}},\ and\ \bibinfo
  {author} {\bibfnamefont {T.}~\bibnamefont {Pfau}},\ }\href
  {https://doi.org/10.1126/science.aau1949} {\bibfield  {journal} {\bibinfo
  {journal} {Science}\ }\textbf {\bibinfo {volume} {362}},\ \bibinfo {pages}
  {446} (\bibinfo {year} {2018})}\BibitemShut {NoStop}%
\bibitem [{\citenamefont {Dideriksen}\ \emph {et~al.}(2021)\citenamefont
  {Dideriksen}, \citenamefont {Schmieg}, \citenamefont {Zugenmaier},\ and\
  \citenamefont {Polzik}}]{Dideriksen_2021}%
  \BibitemOpen
  \bibfield  {author} {\bibinfo {author} {\bibfnamefont {K.~B.}\ \bibnamefont
  {Dideriksen}}, \bibinfo {author} {\bibfnamefont {R.}~\bibnamefont {Schmieg}},
  \bibinfo {author} {\bibfnamefont {M.}~\bibnamefont {Zugenmaier}},\ and\
  \bibinfo {author} {\bibfnamefont {E.~S.}\ \bibnamefont {Polzik}},\ }\href
  {https://doi.org/10.1038/s41467-021-24033-8} {\bibfield  {journal} {\bibinfo
  {journal} {Nature Communications}\ }\textbf {\bibinfo {volume} {12}},\
  \bibinfo {pages} {3699} (\bibinfo {year} {2021})}\BibitemShut {NoStop}%
\bibitem [{\citenamefont {Dong}\ \emph {et~al.}(2021)\citenamefont {Dong},
  \citenamefont {Xia}, \citenamefont {Zhang}, \citenamefont {Yu}, \citenamefont
  {Ye}, \citenamefont {Li}, \citenamefont {Zeng}, \citenamefont {Ding},
  \citenamefont {Shi}, \citenamefont {Guo},\ and\ \citenamefont
  {Nori}}]{Dong_2021}%
  \BibitemOpen
  \bibfield  {author} {\bibinfo {author} {\bibfnamefont {M.-X.}\ \bibnamefont
  {Dong}}, \bibinfo {author} {\bibfnamefont {K.-Y.}\ \bibnamefont {Xia}},
  \bibinfo {author} {\bibfnamefont {W.-H.}\ \bibnamefont {Zhang}}, \bibinfo
  {author} {\bibfnamefont {Y.-C.}\ \bibnamefont {Yu}}, \bibinfo {author}
  {\bibfnamefont {Y.-H.}\ \bibnamefont {Ye}}, \bibinfo {author} {\bibfnamefont
  {E.-Z.}\ \bibnamefont {Li}}, \bibinfo {author} {\bibfnamefont
  {L.}~\bibnamefont {Zeng}}, \bibinfo {author} {\bibfnamefont {D.-S.}\
  \bibnamefont {Ding}}, \bibinfo {author} {\bibfnamefont {B.-S.}\ \bibnamefont
  {Shi}}, \bibinfo {author} {\bibfnamefont {G.-C.}\ \bibnamefont {Guo}},\ and\
  \bibinfo {author} {\bibfnamefont {F.}~\bibnamefont {Nori}},\ }\href
  {https://doi.org/10.1126/sciadv.abe8924} {\bibfield  {journal} {\bibinfo
  {journal} {Science Advances}\ }\textbf {\bibinfo {volume} {7}},\ \bibinfo
  {pages} {eabe8924} (\bibinfo {year} {2021})}\BibitemShut {NoStop}%
\bibitem [{\citenamefont {Steigenberger}\ \emph {et~al.}(2015)\citenamefont
  {Steigenberger}, \citenamefont {Hugentobler}, \citenamefont {Loyer},
  \citenamefont {Perosanz}, \citenamefont {Prange}, \citenamefont {Dach},
  \citenamefont {Uhlemann}, \citenamefont {Gendt},\ and\ \citenamefont
  {Montenbruck}}]{Steigenberger_2015}%
  \BibitemOpen
  \bibfield  {author} {\bibinfo {author} {\bibfnamefont {P.}~\bibnamefont
  {Steigenberger}}, \bibinfo {author} {\bibfnamefont {U.}~\bibnamefont
  {Hugentobler}}, \bibinfo {author} {\bibfnamefont {S.}~\bibnamefont {Loyer}},
  \bibinfo {author} {\bibfnamefont {F.}~\bibnamefont {Perosanz}}, \bibinfo
  {author} {\bibfnamefont {L.}~\bibnamefont {Prange}}, \bibinfo {author}
  {\bibfnamefont {R.}~\bibnamefont {Dach}}, \bibinfo {author} {\bibfnamefont
  {M.}~\bibnamefont {Uhlemann}}, \bibinfo {author} {\bibfnamefont
  {G.}~\bibnamefont {Gendt}},\ and\ \bibinfo {author} {\bibfnamefont
  {O.}~\bibnamefont {Montenbruck}},\ }\href
  {https://doi.org/10.1016/j.asr.2014.06.030} {\bibfield  {journal} {\bibinfo
  {journal} {Advances in Space Research}\ }\textbf {\bibinfo {volume} {55}},\
  \bibinfo {pages} {269} (\bibinfo {year} {2015})}\BibitemShut {NoStop}%
\bibitem [{\citenamefont {Sedlacek}\ \emph {et~al.}(2012)\citenamefont
  {Sedlacek}, \citenamefont {Schwettmann}, \citenamefont {Kübler},
  \citenamefont {Löw}, \citenamefont {Pfau},\ and\ \citenamefont
  {Shaffer}}]{Sedlacek_2012}%
  \BibitemOpen
  \bibfield  {author} {\bibinfo {author} {\bibfnamefont {J.~A.}\ \bibnamefont
  {Sedlacek}}, \bibinfo {author} {\bibfnamefont {A.}~\bibnamefont
  {Schwettmann}}, \bibinfo {author} {\bibfnamefont {H.}~\bibnamefont
  {Kübler}}, \bibinfo {author} {\bibfnamefont {R.}~\bibnamefont {Löw}},
  \bibinfo {author} {\bibfnamefont {T.}~\bibnamefont {Pfau}},\ and\ \bibinfo
  {author} {\bibfnamefont {J.~P.}\ \bibnamefont {Shaffer}},\ }\href
  {https://doi.org/10.1038/nphys2423} {\bibfield  {journal} {\bibinfo
  {journal} {Nature Physics}\ }\textbf {\bibinfo {volume} {8}},\ \bibinfo
  {pages} {819} (\bibinfo {year} {2012})}\BibitemShut {NoStop}%
\bibitem [{\citenamefont {Jing}\ \emph {et~al.}(2020)\citenamefont {Jing},
  \citenamefont {Hu}, \citenamefont {Ma}, \citenamefont {Zhang}, \citenamefont
  {Zhang}, \citenamefont {Xiao},\ and\ \citenamefont {Jia}}]{Jing_2020}%
  \BibitemOpen
  \bibfield  {author} {\bibinfo {author} {\bibfnamefont {M.}~\bibnamefont
  {Jing}}, \bibinfo {author} {\bibfnamefont {Y.}~\bibnamefont {Hu}}, \bibinfo
  {author} {\bibfnamefont {J.}~\bibnamefont {Ma}}, \bibinfo {author}
  {\bibfnamefont {H.}~\bibnamefont {Zhang}}, \bibinfo {author} {\bibfnamefont
  {L.}~\bibnamefont {Zhang}}, \bibinfo {author} {\bibfnamefont
  {L.}~\bibnamefont {Xiao}},\ and\ \bibinfo {author} {\bibfnamefont
  {S.}~\bibnamefont {Jia}},\ }\href {https://doi.org/10.1038/s41567-020-0918-5}
  {\bibfield  {journal} {\bibinfo  {journal} {Nature Physics}\ }\textbf
  {\bibinfo {volume} {16}},\ \bibinfo {pages} {911} (\bibinfo {year}
  {2020})}\BibitemShut {NoStop}%
\bibitem [{\citenamefont {Deb}\ and\ \citenamefont
  {Kj{\ae}rgaard}(2018)}]{Deb_2018}%
  \BibitemOpen
  \bibfield  {author} {\bibinfo {author} {\bibfnamefont {A.~B.}\ \bibnamefont
  {Deb}}\ and\ \bibinfo {author} {\bibfnamefont {N.}~\bibnamefont
  {Kj{\ae}rgaard}},\ }\href {https://doi.org/10.1063/1.5031033} {\bibfield
  {journal} {\bibinfo  {journal} {Applied Physics Letters}\ }\textbf {\bibinfo
  {volume} {112}},\ \bibinfo {pages} {211106} (\bibinfo {year}
  {2018})}\BibitemShut {NoStop}%
\bibitem [{\citenamefont {Meyer}\ \emph {et~al.}(2018)\citenamefont {Meyer},
  \citenamefont {Cox}, \citenamefont {Fatemi},\ and\ \citenamefont
  {Kunz}}]{Meyer_2018}%
  \BibitemOpen
  \bibfield  {author} {\bibinfo {author} {\bibfnamefont {D.~H.}\ \bibnamefont
  {Meyer}}, \bibinfo {author} {\bibfnamefont {K.~C.}\ \bibnamefont {Cox}},
  \bibinfo {author} {\bibfnamefont {F.~K.}\ \bibnamefont {Fatemi}},\ and\
  \bibinfo {author} {\bibfnamefont {P.~D.}\ \bibnamefont {Kunz}},\ }\href
  {https://doi.org/10.1063/1.5028357} {\bibfield  {journal} {\bibinfo
  {journal} {Applied Physics Letters}\ }\textbf {\bibinfo {volume} {112}},\
  \bibinfo {pages} {211108} (\bibinfo {year} {2018})}\BibitemShut {NoStop}%
\bibitem [{\citenamefont {Bor{\'{o}}wka}\ \emph {et~al.}(2022)\citenamefont
  {Bor{\'{o}}wka}, \citenamefont {Pylypenko}, \citenamefont {Mazelanik},\ and\
  \citenamefont {Parniak}}]{Bor_wka_2022}%
  \BibitemOpen
  \bibfield  {author} {\bibinfo {author} {\bibfnamefont {S.}~\bibnamefont
  {Bor{\'{o}}wka}}, \bibinfo {author} {\bibfnamefont {U.}~\bibnamefont
  {Pylypenko}}, \bibinfo {author} {\bibfnamefont {M.}~\bibnamefont
  {Mazelanik}},\ and\ \bibinfo {author} {\bibfnamefont {M.}~\bibnamefont
  {Parniak}},\ }\href {https://doi.org/10.1364/ao.472295} {\bibfield  {journal}
  {\bibinfo  {journal} {Applied Optics}\ }\textbf {\bibinfo {volume} {61}},\
  \bibinfo {pages} {8806} (\bibinfo {year} {2022})}\BibitemShut {NoStop}%
\bibitem [{\citenamefont {Chen}\ \emph {et~al.}(2011)\citenamefont {Chen},
  \citenamefont {Hover}, \citenamefont {Sendelbach}, \citenamefont {Maurer},
  \citenamefont {Merkel}, \citenamefont {Pritchett}, \citenamefont {Wilhelm},\
  and\ \citenamefont {McDermott}}]{PhysRevLett.107.217401}%
  \BibitemOpen
  \bibfield  {author} {\bibinfo {author} {\bibfnamefont {Y.-F.}\ \bibnamefont
  {Chen}}, \bibinfo {author} {\bibfnamefont {D.}~\bibnamefont {Hover}},
  \bibinfo {author} {\bibfnamefont {S.}~\bibnamefont {Sendelbach}}, \bibinfo
  {author} {\bibfnamefont {L.}~\bibnamefont {Maurer}}, \bibinfo {author}
  {\bibfnamefont {S.~T.}\ \bibnamefont {Merkel}}, \bibinfo {author}
  {\bibfnamefont {E.~J.}\ \bibnamefont {Pritchett}}, \bibinfo {author}
  {\bibfnamefont {F.~K.}\ \bibnamefont {Wilhelm}},\ and\ \bibinfo {author}
  {\bibfnamefont {R.}~\bibnamefont {McDermott}},\ }\href
  {https://doi.org/10.1103/PhysRevLett.107.217401} {\bibfield  {journal}
  {\bibinfo  {journal} {Phys. Rev. Lett.}\ }\textbf {\bibinfo {volume} {107}},\
  \bibinfo {pages} {217401} (\bibinfo {year} {2011})}\BibitemShut {NoStop}%
\bibitem [{\citenamefont {Kitching}(2018)}]{Kitching_2018}%
  \BibitemOpen
  \bibfield  {author} {\bibinfo {author} {\bibfnamefont {J.}~\bibnamefont
  {Kitching}},\ }\href {https://doi.org/10.1063/1.5026238} {\bibfield
  {journal} {\bibinfo  {journal} {Applied Physics Reviews}\ }\textbf {\bibinfo
  {volume} {5}},\ \bibinfo {pages} {031302} (\bibinfo {year}
  {2018})}\BibitemShut {NoStop}%
\bibitem [{\citenamefont {Cutler}\ \emph {et~al.}(2020)\citenamefont {Cutler},
  \citenamefont {Hamlyn}, \citenamefont {Renger}, \citenamefont {Whittaker},
  \citenamefont {Pizzey}, \citenamefont {Hughes}, \citenamefont {Sandoghdar},\
  and\ \citenamefont {Adams}}]{Cutler_2020}%
  \BibitemOpen
  \bibfield  {author} {\bibinfo {author} {\bibfnamefont {T.}~\bibnamefont
  {Cutler}}, \bibinfo {author} {\bibfnamefont {W.}~\bibnamefont {Hamlyn}},
  \bibinfo {author} {\bibfnamefont {J.}~\bibnamefont {Renger}}, \bibinfo
  {author} {\bibfnamefont {K.}~\bibnamefont {Whittaker}}, \bibinfo {author}
  {\bibfnamefont {D.}~\bibnamefont {Pizzey}}, \bibinfo {author} {\bibfnamefont
  {I.}~\bibnamefont {Hughes}}, \bibinfo {author} {\bibfnamefont
  {V.}~\bibnamefont {Sandoghdar}},\ and\ \bibinfo {author} {\bibfnamefont
  {C.}~\bibnamefont {Adams}},\ }\href
  {https://doi.org/10.1103/physrevapplied.14.034054} {\bibfield  {journal}
  {\bibinfo  {journal} {Physical Review Applied}\ }\textbf {\bibinfo {volume}
  {14}},\ \bibinfo {pages} {034054} (\bibinfo {year} {2020})}\BibitemShut
  {NoStop}%
\bibitem [{\citenamefont {Lucivero}\ \emph {et~al.}(2022)\citenamefont
  {Lucivero}, \citenamefont {Zanoni}, \citenamefont {Corrielli}, \citenamefont
  {Osellame},\ and\ \citenamefont {Mitchell}}]{Lucivero_2022}%
  \BibitemOpen
  \bibfield  {author} {\bibinfo {author} {\bibfnamefont {V.~G.}\ \bibnamefont
  {Lucivero}}, \bibinfo {author} {\bibfnamefont {A.}~\bibnamefont {Zanoni}},
  \bibinfo {author} {\bibfnamefont {G.}~\bibnamefont {Corrielli}}, \bibinfo
  {author} {\bibfnamefont {R.}~\bibnamefont {Osellame}},\ and\ \bibinfo
  {author} {\bibfnamefont {M.~W.}\ \bibnamefont {Mitchell}},\ }\href
  {https://doi.org/10.1364/oe.469296} {\bibfield  {journal} {\bibinfo
  {journal} {Optics Express}\ }\textbf {\bibinfo {volume} {30}},\ \bibinfo
  {pages} {27149} (\bibinfo {year} {2022})}\BibitemShut {NoStop}%
\bibitem [{\citenamefont {Peters}\ \emph {et~al.}(2020)\citenamefont {Peters},
  \citenamefont {Wang}, \citenamefont {Neumann}, \citenamefont {Simeonov},\
  and\ \citenamefont {Halfmann}}]{Peters_2020}%
  \BibitemOpen
  \bibfield  {author} {\bibinfo {author} {\bibfnamefont {T.}~\bibnamefont
  {Peters}}, \bibinfo {author} {\bibfnamefont {T.-P.}\ \bibnamefont {Wang}},
  \bibinfo {author} {\bibfnamefont {A.}~\bibnamefont {Neumann}}, \bibinfo
  {author} {\bibfnamefont {L.~S.}\ \bibnamefont {Simeonov}},\ and\ \bibinfo
  {author} {\bibfnamefont {T.}~\bibnamefont {Halfmann}},\ }\href
  {https://doi.org/10.1364/oe.383999} {\bibfield  {journal} {\bibinfo
  {journal} {Optics Express}\ }\textbf {\bibinfo {volume} {28}},\ \bibinfo
  {pages} {5340} (\bibinfo {year} {2020})}\BibitemShut {NoStop}%
\bibitem [{\citenamefont {Kr{\"a}mer}\ \emph {et~al.}(2018)\citenamefont
  {Kr{\"a}mer}, \citenamefont {Plankensteiner}, \citenamefont {Ostermann},\
  and\ \citenamefont {Ritsch}}]{kramer2018quantumoptics}%
  \BibitemOpen
  \bibfield  {author} {\bibinfo {author} {\bibfnamefont {S.}~\bibnamefont
  {Kr{\"a}mer}}, \bibinfo {author} {\bibfnamefont {D.}~\bibnamefont
  {Plankensteiner}}, \bibinfo {author} {\bibfnamefont {L.}~\bibnamefont
  {Ostermann}},\ and\ \bibinfo {author} {\bibfnamefont {H.}~\bibnamefont
  {Ritsch}},\ }\href
  {https://doi.org/https://doi.org/10.1016/j.cpc.2018.02.004} {\bibfield
  {journal} {\bibinfo  {journal} {Computer Physics Communications}\ }\textbf
  {\bibinfo {volume} {227}},\ \bibinfo {pages} {109} (\bibinfo {year}
  {2018})}\BibitemShut {NoStop}%
\bibitem [{\citenamefont {Marian}\ and\ \citenamefont
  {Marian}(1993)}]{Marian_1993}%
  \BibitemOpen
  \bibfield  {author} {\bibinfo {author} {\bibfnamefont {P.}~\bibnamefont
  {Marian}}\ and\ \bibinfo {author} {\bibfnamefont {T.~A.}\ \bibnamefont
  {Marian}},\ }\href {https://doi.org/10.1103/physreva.47.4474} {\bibfield
  {journal} {\bibinfo  {journal} {Physical Review A}\ }\textbf {\bibinfo
  {volume} {47}},\ \bibinfo {pages} {4474} (\bibinfo {year}
  {1993})}\BibitemShut {NoStop}%
\bibitem [{\citenamefont {Borówka}\ \emph {et~al.}(2023)\citenamefont
  {Borówka}, \citenamefont {Pylypenko}, \citenamefont {Mazelanik},\ and\
  \citenamefont {Parniak}}]{Data23}%
  \BibitemOpen
  \bibfield  {author} {\bibinfo {author} {\bibfnamefont {S.}~\bibnamefont
  {Borówka}}, \bibinfo {author} {\bibfnamefont {U.}~\bibnamefont {Pylypenko}},
  \bibinfo {author} {\bibfnamefont {M.}~\bibnamefont {Mazelanik}},\ and\
  \bibinfo {author} {\bibfnamefont {M.}~\bibnamefont {Parniak}},\ }\href
  {https://doi.org/10.7910/DVN/W7XZXB} {\bibinfo {title} {Replication data for:
  Continuous wideband microwave-to-optical converter based on room-temperature
  rydberg atoms}},\ \bibinfo {howpublished} {Harvard Dataverse} (\bibinfo
  {year} {2023})\BibitemShut {NoStop}%
\end{thebibliography}%

\onecolumngrid
\clearpage
\vspace{0.8cm}
\begin{center}
	\textbf{\large Supplementary information for: "Continuous wideband microwave-to-optical converter based on room-temperature Rydberg atoms"}
\end{center}
\vspace{0.8cm}

\addto\captionsenglish{\renewcommand{\figurename}{Supplementary Figure}}
\addto\captionsenglish{\renewcommand{\tablename}{Supplementary Table}}

\setcounter{equation}{0}
\setcounter{figure}{0}
\setcounter{table}{0}
\makeatletter

	\section*{S.1 Supplementary discussion on the efficiency of conversion}
	The estimation of the efficiency of the microwave-to-optical conversion appears to be a complex topic if considered in free space. Most simulations predict optical photon rates $[\mathrm{phot/s}]$ for given MW intensities $[\mathrm{W/m^2}]$ (or equivalently $[\mathrm{phot/s/m^2}]$). These are also the quantities directly available in the experiment, where the intensity can be obtained from the Autler-Townes splitting calibration, and the photon rates from the single photon counting. To arrive with the efficiency of the conversion, $\eta$ $[\mathrm{phot/phot}]$, one has to perform a full 3D near-field simulation of the conversion process.
	
	A simpler approach is to consider an effective aperture $[\mathrm{m^2}]$, where the conversion occurs. This is completely correct in the case of a large (larger than the wavelength) conversion medium. However, we operate in the subwavelength regime, $w_0 = 100\ \mathrm{\mu m} \ll 21.5\ \mathrm{mm} = \lambda_{\mathrm{MW}}$, where this approach may lead to errors and a full near-field simulation would be more appropriate, especially in the case of significant conversion efficiency. In the case of great efficiency, the field is (by assumption) strongly affected by the ensemble, therefore significant diffraction may occur and thus the resulting effective area is larger than assumed. Nevertheless, the efficiency obtained from this approach will be the true photon-to-photon efficiency if the single-mode MW field is confined to the atomic medium volume, for example using a traveling-wave resonator/waveguide, but also with a cavity. In free space however, one needs to remember that a radiation mode cannot be confined to such a small volume, and therefore the given efficiency does not correspond to any free-space mode.
	
	To achieve a proper comparison with the results of other groups, especially Ref.~[33-35], we would like to perform a similar estimation of an effective aperture. However, it is important to note that our implementation differs from the ones in Ref.~[33-35]. In these cases the interaction area is determined by cold atoms trapped in MOT, whose shape and optical density can be straightforwardly measured e.g.~with camera imaging. On the other hand, in our case the interaction area is defined by three optical laser beams (Gaussian) interacting with atomic gas freely entering and exiting the area. Additionally, the conversion efficiency is in fact nontrivially position-dependent due to the beams' profiles, induced light shifts, and saturation effects.
	
	\subsection*{Effective aperture}
	We undertake a simple approach similar to the Ref.~[33-35] to estimate the effective interaction area, while stressing that as our overall efficiency is relatively small, we can neglect the diffraction effects along the atomic ensemble. In case of a single Gaussian beam with waist $w_0$, assuming conversion's linear dependence on the intensity of an optical field, $I$, one would arrive with effective area estimate at the focus of the beam
	\begin{equation}
		2 \pi \int_0^\infty \exp(-\frac{2 r^2}{w_0^2}) r \mathrm{d} r = \frac{1}{2} \pi w_0^2.
	\end{equation}
	
	In our case, however, the process is driven by four transitions in Gaussian beams (the probe transition considered twice as in the Fig.~1\textbf{b}) of the same waist. In this case the effective area of relevant Gaussian product is
	\begin{equation}
		2 \pi \int_0^\infty \left(\exp(-\frac{2 r^2}{w_0^2})\right)^4 r \mathrm{d} r = \frac{1}{8} \pi w_0^2.
	\end{equation}
	
	Taking that estimate in the $w_0 = 100\ \mathrm{\mu m}$ case, with the measured peak results -- photon rate $(4.44 {\pm} 0.08) {\cdot} 10^5\ \mathrm{phot/s}$ for MW intensity $(8.1 {\pm} 0.9) {\cdot} 10^{-8}\ \mathrm{W/m^2}$ -- leads us to the overall achieved peak conversion efficiency of $1.3 {\pm} 0.1 \%$.
	
	\subsection*{Losses in the experimental setup}
	The implemented setup is subject to losses in the output of the converter. We put the consideration of all the losses in the Sup.~Tab.~\ref{loss}. The measurements were performed with an assisting $776\ \mathrm{nm}$ field propagating along the expected conversion's direction. Overall losses account for $58\%$.
	
	\begin{table}[]
		\begin{tabular}{|l|l|}
			\hline
			\textbf{Source of loss}                  & \textbf{Loss percentage} \\ \hline
			exit window of the (uncoated) vapor cell & 10\%                     \\ \hline
			optical filtering setup                  & 19\%                     \\ \hline
			fiber coupling                           & 16\%                     \\ \hline
			fiber-fiber connector                    & 20\%                     \\ \hline
			photon detector                          & 15\%                     \\ \hline
			\textbf{overall loss}                    & 58\%                     \\ \hline
		\end{tabular}
		\caption{Analysis of losses at the output of the converter in the experimental setup. \label{loss}}
	\end{table}
	
	Accounting for losses in the setup leads us to an estimate of atomic peak conversion efficiency of $3.1 {\pm} 0.4 \%$.
	
	\subsection*{Theoretical estimate}
	To obtain a theoretical estimate of the efficiency of conversion, we perform numerical simulations in the context of the theory presented in the manuscript. Thus we interrogate the results of our full atomic simulation, as given by Eq.~(1) in the $\partial_t \hat{\rho} (t) = 0$ case. Within the linear regime (as in the Fig.~2\textbf{a}) we check that at peak conversion we obtain the mapping from incident microwave Rabi frequency onto an output optical atomic coherence $\rho_{s}$, observing a slope $\beta=\mathrm{d}\rho_s/\mathrm{d}\Omega_{\mathrm{MW}} = 17\ (\mathrm{rad/s})^{-1} \approx 1.1\times 10^{-4} (2\pi \mathrm{MHz})^{-1}$ (quantyfing amount of atomic coherence per unit of Rabi frequency). This allows us estimate an intensity-to-intensity efficiency as the ratio of microwave and optical field intensities normalized by the photon energy ratio. The expression is following, and in the linear regime:
	\begin{equation}
		\eta = \frac{\varepsilon_0 (\Omega_{\mathrm{MW}} \hbar/d_{\mathrm{MW}})^2}{\varepsilon_0|\frac{ik_{s}}{2\varepsilon_0}n L_\mathrm{eff}d_{s}\rho_{s}|^2} \frac{\omega_{\mathrm{MW}}}{\omega_{s}} =  \frac{ ( \hbar/d_{\mathrm{MW}})^2}{|\frac{ik_{s}}{2\varepsilon_0}n L_\mathrm{eff}d_{s}|^2} \beta^{-2} \frac{\omega_{\mathrm{MW}}}{\omega_{s}}
		\nonumber,
	\end{equation}
	where the variables follow the convention introduced in the main text, and additionally $d_{\mathrm{MW}} = 2500 a_0 e$ and $d_{s} = 0.95 a_0 e$ are electric dipole moments of MW and \emph{signal} transitions respectively ($a_0$ being Bohr radius and $e$ being elementary charge), $n \approx 7{\cdot}10^{16}\ \mathrm{1/m^3}$ is atomic number density and $L_\mathrm{eff}\approx2.5\ \mathrm{cm}$ is an effective interaction length.
	In the numerator we have the microwave field intensity obtained from the calibrated microwave Rabi frequency.
	In the denumerator we have the intensity of the optical field, under the assumption of uniform generation along the ensemble and no input signal depletion (which remains valid for low efficiencies). There are several parameters here that bear significant uncertainties, which leads to a rough estimate of $\eta=2.8\pm1.6\%$, which agrees with the experimental result.

	\section*{S.2 Photon rates, noise levels and noise-equivalent temperature}
	\subsection*{Cross-correlation analysis of laser fields}
\begin{figure*}
		\includegraphics[width=0.7\textwidth]{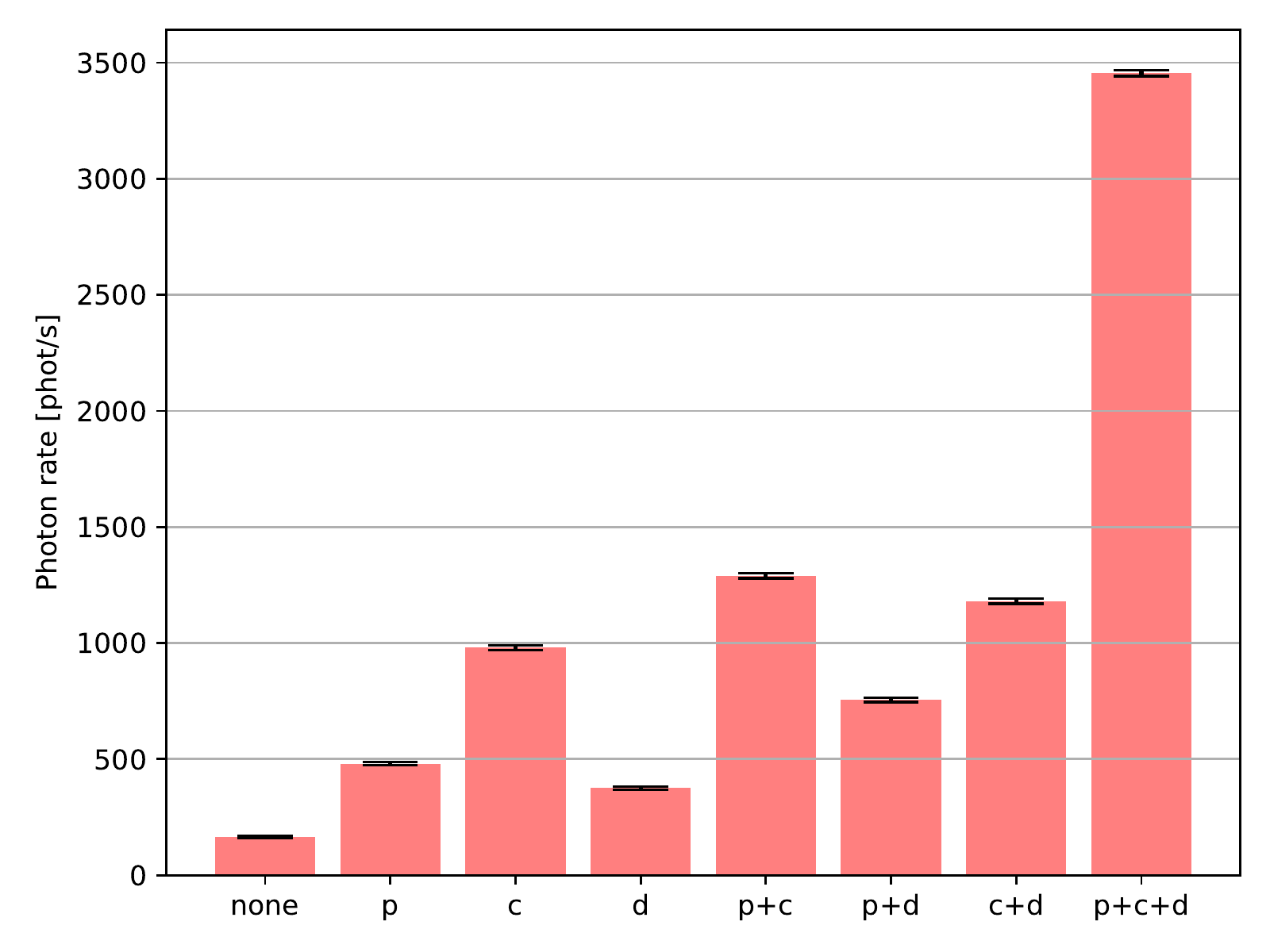}
		\caption{Photon rates registered for each combination of fields before introducing spectral filters. \emph{p} -- \emph{probe} ($780\ \mathrm{nm}$), \emph{c} -- \emph{coupling} ($480\ \mathrm{nm}$), \emph{d} -- \emph{decoupling} ($1258\ \mathrm{nm}$). The MW field has no effect on photon rates, apart from the case with all the other fields on. The error bars are calculated from photon-counting standard deviation, $\sqrt{\bar{n}}$, where in each case $\bar{n}$ is explicitly noted as photon rates on the Y axis.}
		\label{precor}
	\end{figure*}

\begin{figure*}
		\includegraphics[width=0.7\textwidth]{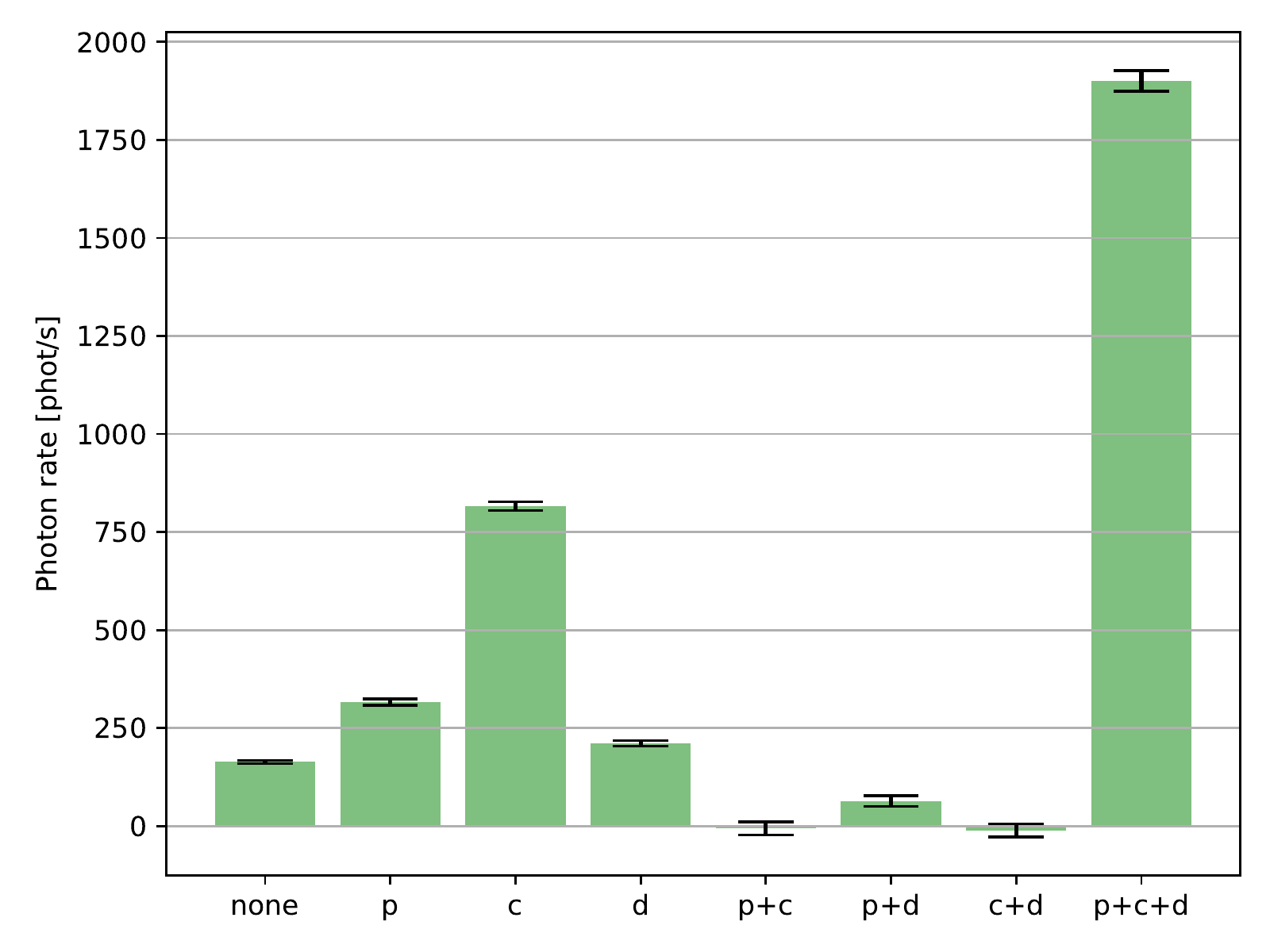}
		\caption{Photon rate contributions for each combination of fields before introducing spectral filters. \emph{p} -- \emph{probe} ($780\ \mathrm{nm}$), \emph{c} -- \emph{coupling} ($480\ \mathrm{nm}$), \emph{d} -- \emph{decoupling} ($1258\ \mathrm{nm}$). The MW field has no effect on photon rates, apart from the case with all the other fields on. The error bars are calculated from the propagation of the standard deviation from the Sup.~Fig.~\ref{precor}, where in each case $\bar{n}$ is explicitly noted as photon rates on the Y axis.}
		\label{postcor}
	\end{figure*}

	The values of various components of the observed optical signal have been obtained by a cross-correlation analysis based on measuring the output photon rate for all possible combinations of one or more lasers switched off. In the first step, the preliminary measurements identified the cross-correlation between all the introduced optical fields. Photons rates for all of the combinations of laser fields are presented in the Sup.~Fig.~\ref{precor}. It was observed that the MW field had no effect on photon rates, apart from the obvious case with all the other fields on, where we observe the conversion of coherent state.
	To extract the information about each field contributing to the overall registered signal (i.e.~in the p+c+d case), we treat the measured values (in the order from the Sup.~Fig.~\ref{precor}) as a vector and perform a matrix operation on it with the inverse of the correlation matrix:
	\begin{equation}
		\begin{pmatrix}
			1 & 0 & 0 & 0 & 0 & 0 & 0 & 0 \\
			1 & 1 & 0 & 0 & 0 & 0 & 0 & 0 \\
			1 & 0 & 1 & 0 & 0 & 0 & 0 & 0 \\
			1 & 0 & 0 & 1 & 0 & 0 & 0 & 0 \\
			1 & 1 & 1 & 0 & 1 & 0 & 0 & 0 \\
			1 & 1 & 0 & 1 & 0 & 1 & 0 & 0 \\
			1 & 0 & 1 & 1 & 0 & 0 & 1 & 0 \\
			1 & 1 & 1 & 1 & 1 & 1 & 1 & 1 
		\end{pmatrix}^{-1} = \begin{pmatrix}
			1 & 0 & 0 & 0 & 0 & 0 & 0 & 0 \\
			-1 & 1 & 0 & 0 & 0 & 0 & 0 & 0 \\
			-1 & 0 & 1 & 0 & 0 & 0 & 0 & 0 \\
			-1 & 0 & 0 & 1 & 0 & 0 & 0 & 0 \\
			1 & -1 & -1 & 0 & 1 & 0 & 0 & 0 \\
			1 & -1 & 0 & -1 & 0 & 1 & 0 & 0 \\
			1 & 0 & -1 & -1 & 0 & 0 & 1 & 0 \\
			-1 & 1 & 1 & 1 & -1 & -1 & -1 & 1 
		\end{pmatrix}.
	\end{equation}
	This results in photon rate contributions for each combination of fields, which we present in the Sup.~Fig.~\ref{postcor}.
	We note that the contributions of two-field combinations are near-zero and thus we neglect them in further considerations. Furthermore, we interpret the p+c+d case as the thermal radiation coupling to the converter.
	
\begin{figure*}
		\includegraphics[width=0.7\textwidth]{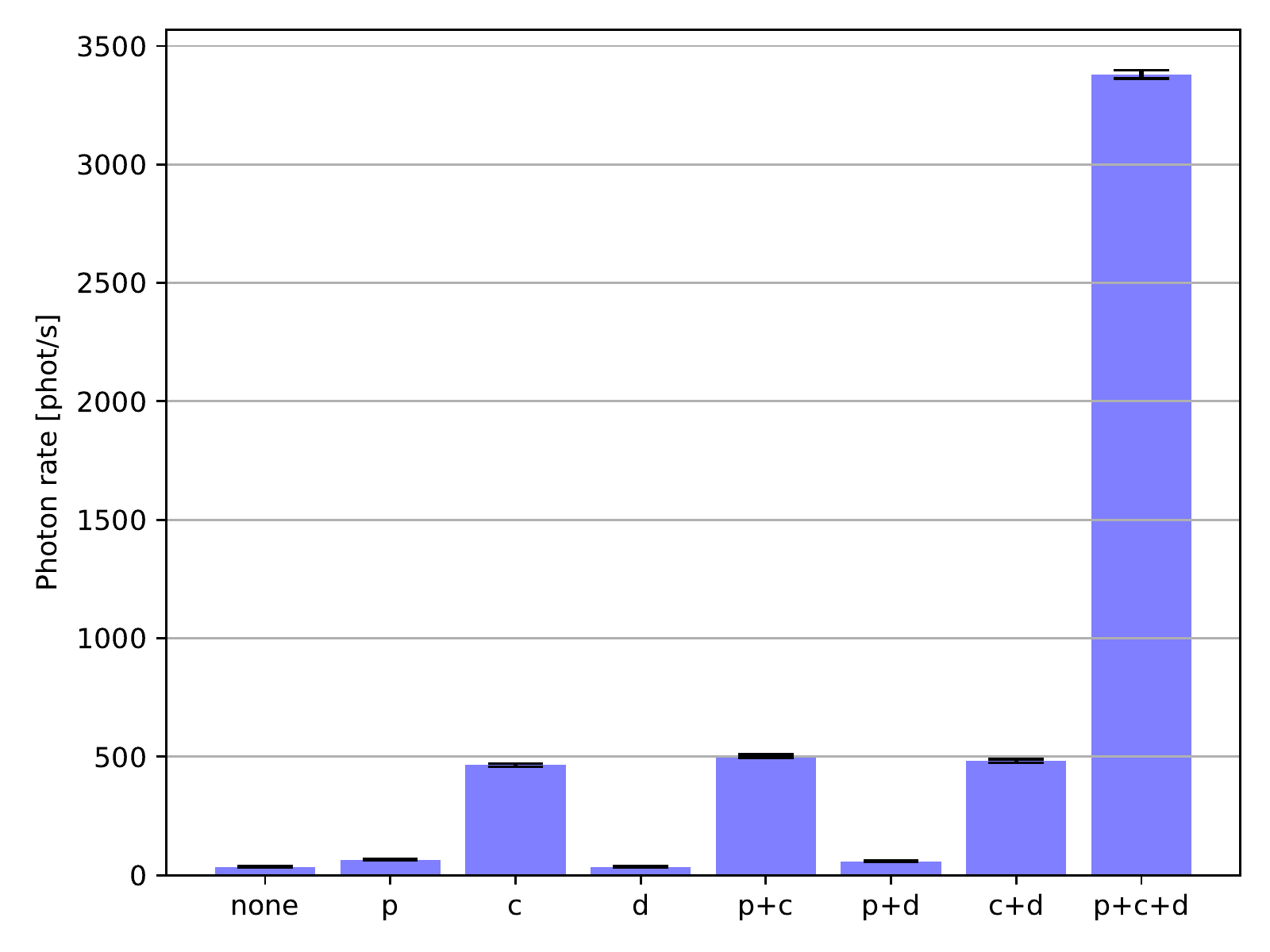}
		\caption{Photon rates registered for each combination of fields after introducing spectral filters. \emph{p} -- \emph{probe} ($780\ \mathrm{nm}$), \emph{c} -- \emph{coupling} ($480\ \mathrm{nm}$), \emph{d} -- \emph{decoupling} ($1258\ \mathrm{nm}$). Again, the MW field has no effect on photon rates, apart from the case with all the other fields on. The error bars are calculated from photon-counting standard deviation, $\sqrt{\bar{n}}$, where in each case $\bar{n}$ is explicitly noted as photon rates on the Y axis.}
		\label{precorfil}
	\end{figure*}

\begin{figure*}
		\includegraphics[width=0.7\textwidth]{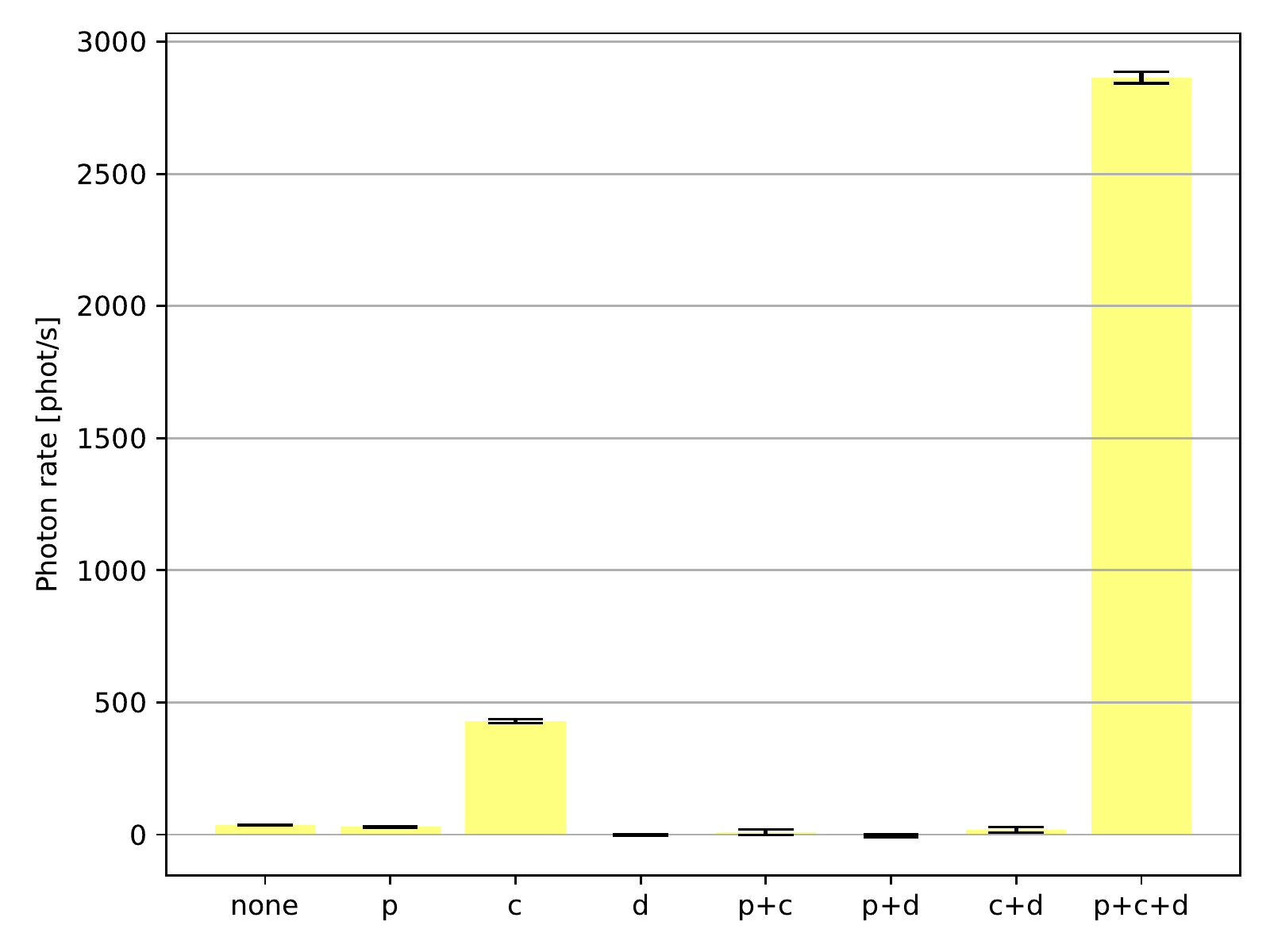}
		\caption{Photon rate contributions for each field after introducing spectral filters. \emph{p} -- \emph{probe} ($780\ \mathrm{nm}$), \emph{c} -- \emph{coupling} ($480\ \mathrm{nm}$), \emph{d} -- \emph{decoupling} ($1258\ \mathrm{nm}$). Again, the MW field has no effect on photon rates, apart from the case with all the other fields on. The error bars are calculated from the propagation of the standard deviation from the Sup.~Fig.~\ref{precorfil}, where in each case $\bar{n}$ is explicitly noted as photon rates on the Y axis.}
		\label{postcorfil}
	\end{figure*}

	In the second step we modified the optical filtering setup of the converted signal ($776\ \mathrm{nm}$) in order to eliminate the effects of other lasers disturbing the photon counting measurements (cases p, c and d). Namely, we replaced a $10\ \mathrm{nm}$ bandwidth filter used in the preliminary measurements with the set of filters described in the Methods (highpass, lowpass, $1.2\ \mathrm{nm}$ bandpass). This led to the registered rates in the Sup.~Fig.~\ref{precorfil} and the analysis has shown the elimination of photon rate contributions in the cases p and d, as shown in the Sup.~Fig.~\ref{postcorfil}. However, photon rate contribution in the case c could not be fully eliminated -- we identify this effect as wideband fluorescence of glass elements induced by the $480\ \mathrm{nm}$ field. From now on to assess the fluorescence (and other) noise level, we turned off the $1258\ \mathrm{nm}$ field only. To assess the other noise level, we turned off the $480\ \mathrm{nm}$ field only.
	
	In the third step, when measuring the dynamic range from the Fig.~2\textbf{a}, we measured photon rates of overall, thermal, fluorescence and other noise levels, as a reference. The results, consistent with the measurements in the main text, are presented in the Sup.~Tab.~\ref{fields}.
	
	\begin{table}[]
		\begin{tabular}{|l|l|l|l|l|l|l|l|l|}
			\hline
			\textbf{Source of signal}                                                                                   & \textbf{\begin{tabular}[c]{@{}l@{}}Overall\\ noise\end{tabular}} & \textbf{\begin{tabular}[c]{@{}l@{}}Thermal\\ noise\end{tabular}} & \textbf{\begin{tabular}[c]{@{}l@{}}Non-\\ thermal\\ noise\end{tabular}} & \textbf{\begin{tabular}[c]{@{}l@{}}Glass\\ fluorescence\\ noise\end{tabular}} & \textbf{\begin{tabular}[c]{@{}l@{}}Other\\ noise\end{tabular}} & \textbf{\begin{tabular}[c]{@{}l@{}}Thermal\\ level\\ (theory)\end{tabular}} & \textbf{\begin{tabular}[c]{@{}l@{}}Adjusted\\ thermal\\ level\\ (theory)\end{tabular}} & \textbf{\begin{tabular}[c]{@{}l@{}}Thermal\\ level\\ (experimental)\end{tabular}} \\ \hline
			\textbf{\begin{tabular}[c]{@{}l@{}}photon rate\\ $[\mathrm{phot/s}]$\end{tabular}}                          & 2050                                                             & 1740                                                             & 310                                                                     & 270                                                                           & 40                                                             & n/a                                                                         & n/a                                                                                    & n/a                                                                               \\ \hline
			\textbf{\begin{tabular}[c]{@{}l@{}}equivalent MW\\ intensity $[\mathrm{p W/m^2}]$\end{tabular}}             & 400                                                              & 340                                                              & 61                                                                      & 53                                                                            & 8                                                              & n/a                                                                         & 360                                                                                    & n/a                                                                               \\ \hline
			\textbf{\begin{tabular}[c]{@{}l@{}}equivalent MW\\ field $[\mathrm{\mu V / cm}]$\end{tabular}}              & 5.50                                                             & 5.07                                                             & 2.13                                                                    & 1.98                                                                          & 0.78                                                           & n/a                                                                         & 5.23                                                                                   & n/a                                                                               \\ \hline
			\textbf{\begin{tabular}[c]{@{}l@{}}MW field density\\ $[\mathrm{n V cm^{-1} (rad/s)^{-1/2}}]$\end{tabular}} & n/a                                                              & \textbf{0.48}                                                    & n/a                                                                     & n/a                                                                           & n/a                                                            & \textbf{1.64}                                                               & \textbf{0.495}                                                                         & \textbf{1.59}                                                                     \\ \hline
			\textbf{\begin{tabular}[c]{@{}l@{}}noise-equivalent\\ temperature $[\mathrm{K}]$\end{tabular}}              & 353                                                              & 300                                                              & \textbf{53}                                                             & 46                                                                            & 7                                                              & n/a                                                                         & n/a                                                                                    & n/a                                                                               \\ \hline
			\textbf{\begin{tabular}[c]{@{}l@{}}part of overall\\ noise\end{tabular}}                                    & 100\%                                                            & 85\%                                                             & 15\%                                                                    & 13\%                                                                          & 2\%                                                            & n/a                                                                         & n/a                                                                                    & n/a                                                                               \\ \hline
		\end{tabular}
		\caption{Reference for registered noise components and derived thermal level. \label{fields} }
	\end{table}
	
	\subsection*{Noise-equivalent temperatures}
	The noise level (thermal and non-thermal) was measured in the $T = 300 \mathrm{K}$ environment. We came to the conclusion that the ratio between thermal noise (measured if and only if every field was switched on) and non-thermal noise is $85\%$ to $15\%$, as shown in the Sup.~Fig.~\ref{postcorfil} and in the Sup.~Tab.~\ref{fields}. From that we deduce that the thermal noise corresponds to the measured temperature of black-body (MW absorbing foam) $T = 300 \mathrm{K}$ and non-thermal noise corresponds to noise-equivalent temperature of $T_{\mathrm{NE}} = 53\ \mathrm{K}$. In this approach we consider the thermal signal as the proper result (fundamentally guaranteed by the theory of black-body radiation) and the non-thermal noise as a noise added by our device. However, if we were to measure another signal in relation to the thermal, the noise-equivalent temperature in the presented environment would be $T_{\mathrm{NE}} = 353\ \mathrm{K}$, as presented in the Sup.~Tab.~\ref{fields}, along with other values.
	
	The lowest limit of $3.8 \mathrm{K}$ was obtained with an additional filtering incorporating an optical cavity, that on the one hand allowed for the conversion of thermal radiation from the whole bandwidth (the bandwidth of the cavity was $160\ \mathrm{MHz}$ versus $16\ \mathrm{MHz}$ of conversion bandwidth), and on the other hand allowed for successful filtration of wideband noise -- the initial bandwidth that could couple to the signal fiber after passing through filters, at the order of $500\ \mathrm{GHz}$ ($1.2\ \mathrm{nm}$), was narrowed down to $160\ \mathrm{MHz}$. This allowed for the elimination of noise, which we identified as 480nm-induced glass fluorescence, and led us to reducing noise-equivalent temperature of our device from $53 \mathrm{K}$ to $3.8 \mathrm{K}$ by improving the thermal/non-thermal noise ratio. This is a demonstration of sensitivity of our device to photon-based MW measurements. One could imagine approaching single-photon operations e.g.~on cosmic microwave background with such a device (still noisy but not impossible, and subject to further optimizations).
	
	The values obtained for the measurements with additional cavity filtering were $325 {\pm} 8\ \mathrm{phot/s}$ for thermal photons and $4.2 {\pm} 0.5 \mathrm{phot/s}$ for other noise photons, resulting in overall $T_{\mathrm{NE}} = 3.8 {\pm} 0.5\ \mathrm{K}$ noise-equivalent temperature.
	
	\subsection*{Photon rates and photon pair rates}
	In the Fig.~3\textbf{c} (and similarly in the Fig.~3\textbf{b}) the value of total registered photon rate with antenna turned off is $\bar{n}_{\mathrm{th}} + \bar{n}_{\mathrm{noise}} = 1550\ \mathrm{phot/s}$, thus from the $85\%$ proportion, the thermal photon rate is $\bar{n}_{\mathrm{th}} = 1320\ \mathrm{phot/s}$. The source of losses in this case (comparing to $1740\ \mathrm{phot/s}$ at the Fig.~2) is different measurement scheme compared to that in the Fig.~2. Namely, we utilise two channels of single-photon detector to measure correlation between them, so we use 50:50 fiber splitter -- thus there is fiber -- splitter -- two fibers connection versus fiber -- fiber connection. Actually, there is a slight asymmetry introduced by the splitter, where we register $660\ \mathrm{phot/s}$ on one channel of the detector and $890\ \mathrm{phot/s}$ on the other in case of the antenna turned off. However, these rates are for single photons on each channel. The rate of correlation-measured excess bunched photon pairs (contributing to the peak e.g.~in the Fig.~3\textbf{b}) in thermal radiation is around $20\ \mathrm{pairs/h}$.
	
	All of the values of $\bar{n}_{\mathrm{th}}$ and $\bar{n}_{\mathrm{coh}}$ were measured by measuring photon rates on the two channels of the photon detector. Actually, the measured quality is $\bar{n}_{\mathrm{th}} + \bar{n}_{\mathrm{coh}} + \bar{n}_{\mathrm{noise}}$ but individual rates can be extracted from a separate measurement with antenna turned off (thus $\bar{n}_{\mathrm{coh}} = 0$) and from the knowledge of thermal/non-thermal noise proportion.

	\section*{S.3 Derivation of the field spectral density of the measured thermal radiation}
	As for the derivation of $480\ \mathrm{pV}\mathrm{cm}^{-1}(\mathrm{rad}/\mathrm{s})^{-1/2}$ from the Fig.~2\textbf{a}, the registered signal with antenna turned off is $2050\ \mathrm{phot/s}$, as demonstrated in the Sup.~Tab.~\ref{fields}. From this we deduce that as thermal noise corresponds to $85\%$ of the signal, it is at the level of $1740\ \mathrm{phot/s}$. This consequently corresponds to a measured field of $5.07\ \mathrm{\mu V/cm}$, as we describe in the Methods. However, this is thermal field integrated over whole conversion bandwidth, so we divide by the square root of integrally measured bandwidth ($17.8 {\cdot} 2 \pi\ \mathrm{MHz}$, as described in the Methods) and arrive at the $480\ \mathrm{pV}\mathrm{cm}^{-1}(\mathrm{rad}/\mathrm{s})^{-1/2}$.
	
	To achieve consistency between measured $480\ \mathrm{pV}\mathrm{cm}^{-1}(\mathrm{rad}/\mathrm{s})^{-1/2}$ and theoretically predicted $1.64\ \mathrm{\mu V}\mathrm{cm}^{-1}(\mathrm{rad}/\mathrm{s})^{-1/2}$, we notice that the theory predicts thermal spectral density propagating in all directions, in both polarisations. However, in the \emph{Phase matching} section in the Methods, and in the Extended Data Fig.~2, we show that our converter is sensitive mainly to one circular polarisation and quite directional -- only $18\%$ of whole spherical distribution is coupled to the device. In consequence only around $9\%$ of all thermal radiation incident can couple to the converter. Adjusting for that property, we arrive at the value $1.59\ \mathrm{\mu V}\mathrm{cm}^{-1}(\mathrm{rad}/\mathrm{s})^{-1/2}$, from experimental data, which is very near the theoretical prediction of $1.64\ \mathrm{\mu V}\mathrm{cm}^{-1}(\mathrm{rad}/\mathrm{s})^{-1/2}$.

	\section*{S.4 Supplementary discussion on the coherence of the conversion process}
	Unaffected coherent conversion on long timescales can be demonstrated with introducing two MW fields of slightly different frequencies at the input of the converter. The converted signal for both MW fields would in this case experience the same fluctuations and thus the phase difference between the two signals will be preserved. 
	
\begin{figure*}
		\includegraphics[width=0.7\textwidth]{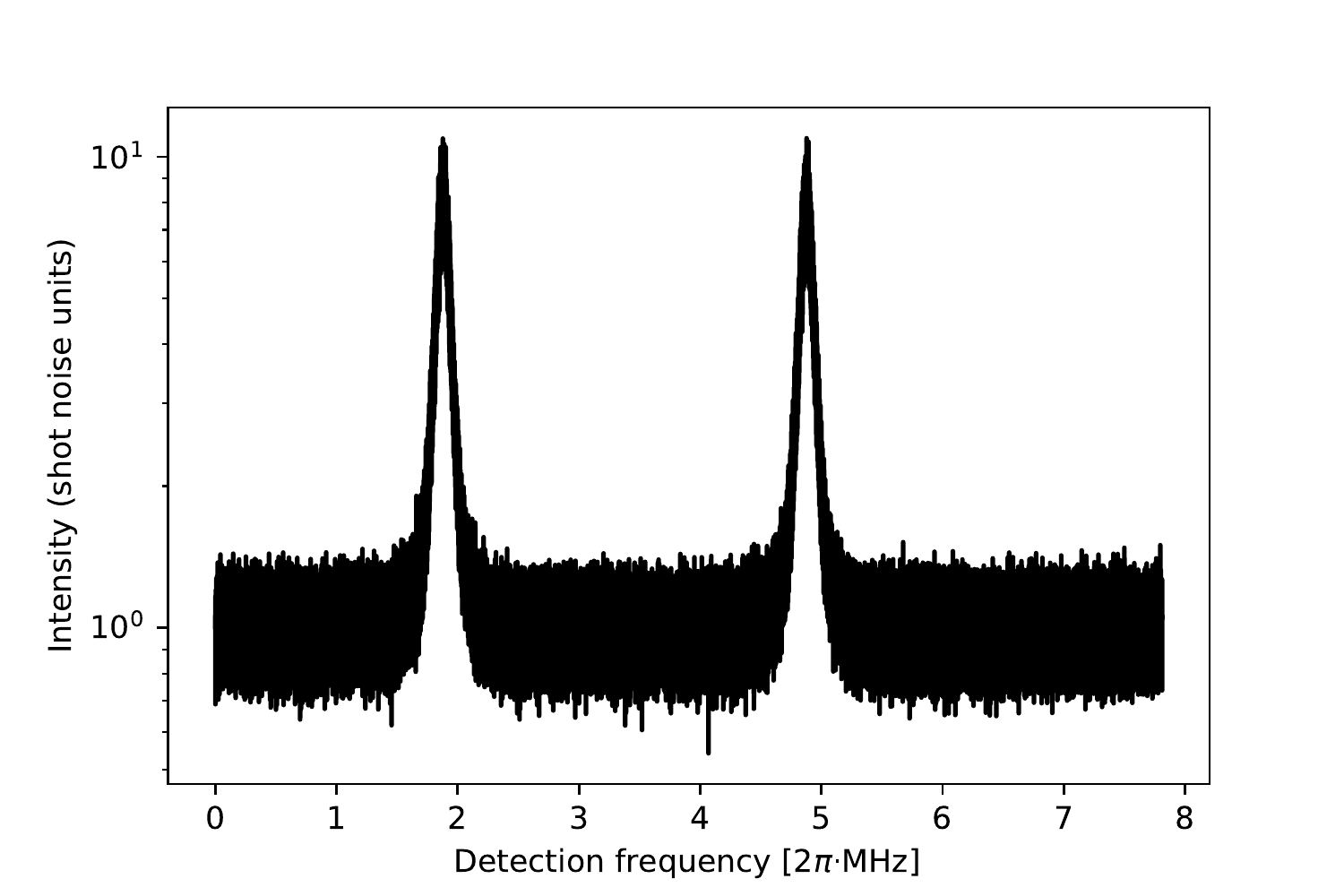}
		\caption{The spectrum of two simultaneously converted MW signals in the heterodyne dection setup. The spectrum is averaged over 100 measurements. The intensity is normalized to the shot noise of the heterodyne. The signal-to-noise ratio is measured as $\mathrm{SNR} = 8.2$ ($9.2\ \mathrm{dB}$).}
		\label{hetmeas}
	\end{figure*}

\begin{figure*}
		\includegraphics[width=0.7\textwidth]{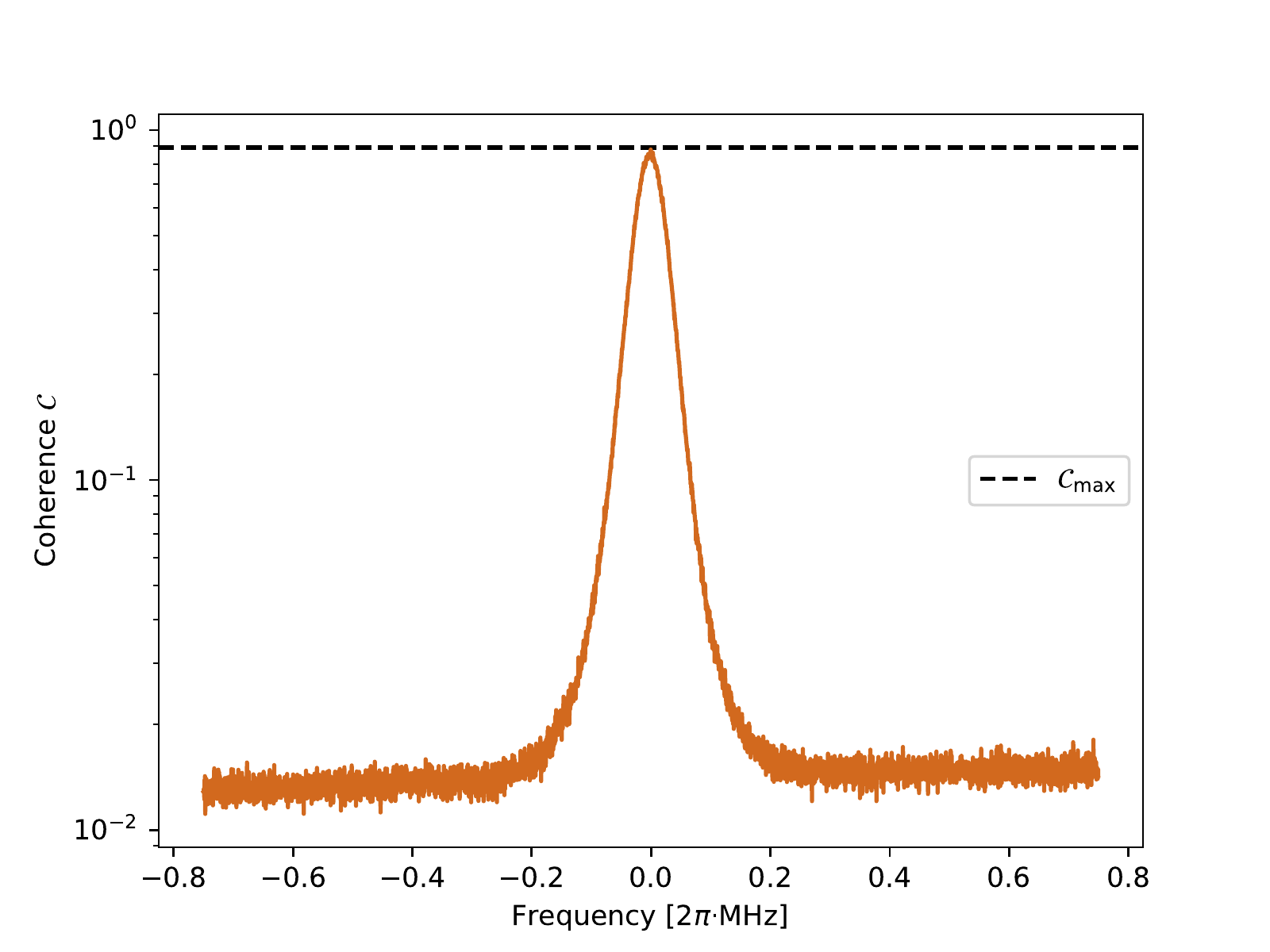}
		\caption{The coherence of the conversion process, being the result of normalized cross-spectral density measurement performed on two MW signals converted simultaneously, in relation to the limit based on their signal-to-noise ratio. The result is averaged over 100 measurements.}
		\label{csd}
	\end{figure*}

	To demonstrate the coherence of the conversion process, we apply simultaneously two near-optimal MW fields (generated from different VCOs synchronized with the same clock) of similar strengths, detuned by $3{\cdot}2 \pi\ \mathrm{MHz}$, to the antenna. The spectrum of the conversion signal registered with heterodyne detection (Extended Data Fig.~1\textbf{g}) is presented in the Sup.~Fig.~\ref{hetmeas}. We then take parts of the spectrum corresponding to the registered signals and perform an estimation of cross-spectral density between them using Welch's method. Then by using a standard normalization to the geometrical mean of both signals' power spectral densities obtained with the same method, we arrive with coherence $\mathcal{C}$. The results are presented in the Sup.~Fig.~\ref{csd}.
	
	The coherence measurement ($\mathcal{C} = 0.88$ at zero frequency) corresponds very well with the theoretical limit based on the signal-to-noise ratio
	\begin{equation}
		\mathcal{C}_{\mathrm{max}} = \frac{\mathrm{SNR}}{1+\mathrm{SNR}} \approx 0.89,
	\end{equation}
	constituting a proof of nearly perfect coherence of the presented microwave-to-optical conversion process.

	\section*{S.5 Microwave converter acting as a microwave detector}
	The proposed scheme could work as a MW detector instead of converter, for example by employing shot-noise limited heterodyne detection with a local oscillator at the converted signal's wavelength. We present an exemplary result of this kind as a spectral characteristics of the output signal in the Extended Data Fig.~1\textbf{e} in the Methods. There we show the readout of a signal of around $2.9\ \mathrm{mV/cm}$ with SNR of $13.5\ \mathrm{dB}$. As the signal have a bandwidth of $86\ \mathrm{kHz}$, we may assume a phase noise compensation mechanism and that would yield us a sensitivity of around $2.8\ \mathrm{\mu V {cm}^{-1} {Hz}^{-1/2}}$. This result would be comparable with standard Rydberg electrometric sensitivity but well below the superheterodyne state-of-the-art sensitivity.
	
	However, photon-counting detection, for which the proposed scheme is suitable (and standard electrometric setups are not due to poor spectral separation of optical fields), entails new type of photonic operations and measurements that are not possible elsewhere, e.g.~the presented autocorrelation measurement. These new types of measurements are not straightforwardly comparable to the heterodyne measurement, as here we are not subjected to the shot noise but, as we show, we hit the thermal noise limit (that has not been approached yet in heterodyne detection). In photon-counting detection the thermal limit is a limit not on the spectral sensitivity of the detection in $[\mathrm{\mu V {cm}^{-1} {Hz}^{-1/2}}]$ (as is the case for the shot noise limit in heterodyne detection) but for minimal detectable field in $[\mathrm{\mu V / cm}]$. If we wanted to measure electric field using that method, we would arrive at minimal detectable field of $5.5 \ \mathrm{\mu V / cm}$, as presented in the Sup.~Tab.~\ref{fields}, with SNR of 1 in relation to thermal (and other) noise.
	
	The wide operational bandwidth of the converter is the cause of high level of thermal noise that limits the sensitivity of the converter employed as an electrometric detector. We could imagine trading the bandwidth's figure of merit for the minimal detectable field, e.g.~by directing the converted signal via a narrowband ($20\ \mathrm{kHz}$) tunable cavity -- that would lead us to $170 \ \mathrm{nV {cm}^{-1}}$ of thermal field. However, at this level of field we would expect count rate of only $2\ \mathrm{phot/s}$, so it would require special care of other sources of noise. As the non-thermal noise registered with additional filtering translates to the level of $22\ \mathrm{phot/s}$ (as shown in the Fig.~2\textbf{a}), this would be the expected limit approached in this realization -- resulting in $600 \ \mathrm{nV {cm}^{-1}}$ of minimal detectable field.
	
	A more concrete comparison, e.g.~with the Ref.~[47] is not straightforward, as the detection method is fundamentally different. In Ref.~[47] a heterodyne detection is employed and the limit of the sensitivity -- the noise floor -- is given by atomic and technical noises, as claimed by the authors (we think that in most of the realizations the optical shot noise could also be an important contributing factor there), not the thermal radiation as in our case. Overall, this means that in homodyne/heterodyne detection quadrature operators of the field are measured ($\hat{X},\ \hat{P}$), while in our work we use the photon-number operator $\hat{a}^\dagger\hat{a}$. We use photon counting, which is independent of the converted light frequency -- registered photon counts from the thermal radiation correspond to the signal from the whole conversion bandwidth. This means that, when employing the converter as a field detector, the thermal limit is not a limit on sensitivity (in $[\mathrm{\mu V {cm}^{-1} {Hz}^{-1/2}}]$) but rather on a minimal detectable field (in $[\mathrm{\mu V / cm}]$), as the thermal radiation is integrated over the whole conversion bandwidth.
	
	Regarding the operational dynamic range, e.g.~in comparison to Ref.~[47], in our setup we claim the lower limit of the converter's working range as the minimal detectable field given by the thermal radiation noise floor ($4{\cdot}10^{-10}\ \mathrm{W/m^2}$ in the Fig.~2), and the higher limit as ${>}0.5$ relative conversion efficiency ($2{\cdot}10^{-4}\ \mathrm{W/m^2}$) -- this is $10\ \mathrm{dB}$ below the atomic saturation ($2{\cdot}10^{-3}\ \mathrm{W/m^2}$). The lower limit is dependent on the effective detection area, as well as the bandwidth of the conversion -- the effective integration bandwidth of thermal radiation spectral density. We can envisage the filtering of the converted signal through a narrowband (e.g.~$20\ \mathrm{kHz}$) cavity. In that realization we would expect the detected thermal radiation to be at around $2\ \mathrm{phot/s}$. Then other non-thermal noise at $22\ \mathrm{phot/s}$ (as shown in the Fig.~2\textbf{a}) would dominate the process, setting the minimal detectable field at $8{\cdot}10^{-12}\ \mathrm{W/m^2}$. That would result in $84\ \mathrm{dB}$ dynamic range (from the minimal detectable field to atomic saturation) and this result enables a fair comparison with the heterodyne detection device presented in the Ref.~[47]. Though, in this case the instantaneous bandwidth of the conversion would of course be limited by the bandwidth of the mentioned narrowband cavity.
	
	\section*{S.6 Supplementary discussion on the optical-to-microwave conversion}
	Demonstration of the reverse process of optical-to-microwave conversion is an important topic of research that has yielded solutions in different systems, Ref.~[14, 18, 22, 28, 29, 31]. In our case the reverse conversion should also be possible, as confirmed by the model [Eq.~(1)]. However, as the spontaneous emission branching ratios from the Rydberg states are not favorable, the emitted MW field would be very weak and difficult to detect in free space. A clear solution is to couple the atoms to a resonator which confines the MW field and enhances the decay rate on the MW transition.

\end{document}